\documentclass[12pt]{article}

\usepackage{amsfonts,amssymb,amstext,amsmath}
\usepackage{mathtools}

\usepackage[normal]{subfigure}
\usepackage{epsfig}
\usepackage{graphicx}
\usepackage{enumerate}
\usepackage{makeidx}
\usepackage[
linkcolor=blue,urlcolor=blue,citecolor=blue,
bookmarks=true,bookmarksopenlevel=2]
{hyperref}

\usepackage{color}
\usepackage[normalem]{ulem}
\usepackage{float}
\usepackage{amsmath}
\usepackage{amsthm}

\usepackage[overload]{empheq}
\usepackage{relsize}

\usepackage{cancel}
\usepackage{soul}
\usepackage{ulem}

\usepackage[titletoc,title]{appendix}

\allowdisplaybreaks

\input epsf.tex

\newcommand{\wh}[1]{\widehat{#1}}

\newcommand{\aum}{\alpha+\frac{1}{2}}

\newcommand{\abs}[1]{\left\vert #1 \right\vert}

\newtheorem{remark}{Remark}

\definecolor{greenn}{rgb}{0,0.75,0}

\def\Q{  {{\mathcal U}_{\mathcal Z}}}
\def\QH{  { {\mathcal U}_{\mathcal Z}^H}}

\def\Qs{  {{\mathcal V}_{\mathcal Z}}}
\def\QsH{  { {\mathcal V}_{\mathcal Z}^H}}
\def\QsV{  { {\mathcal V}_{\mathcal Z}^V}}

\def\R{\mathbb{R}}

\def\vu{\boldsymbol{u}}

\def\vv{\boldsymbol{v}}
\def\vg{\boldsymbol{g}}
\def\vn{\boldsymbol{n}}

\def\a{\alpha}
\def\b{\beta}

\def\vp{\varphi}
\def\1vp{\left(1-\vp\right)}
\def\vpa{{\varphi}_{\alpha}}
\def\v1pa{\left(1-{{\varphi_\alpha}}\right)}

\def\r{\rho}
\def\O{\Omega}

\def\T{\tau}
\def\G{\Gamma}

\def\g{\nabla}
\def\p{\partial}

\def\bS{\boldsymbol{\sigma}}
\def\bT{\boldsymbol{\tau}}

\def\dint{\displaystyle\int}
\def\dsum{\displaystyle\sum}

\def\dfrac{\displaystyle\frac}

\def\bd{\boldsymbol}

\def\wtd{\widetilde}

\def\mb{\mbox}

\title{Multilayer models for shallow two-phase debris flows
	with dilatancy effects}
\author{J. Garres-D{\'i}az \thanks{Dpto. Matem{\'a}ticas. Edificio Einstein - Universidad de C\'ordoba.
		Campus de Rabanales, 14014-C\'ordoba, Spain, (jgarres@uco.es)}\ , F. Bouchut \thanks{Universit\'e Paris-Est, Laboratoire d’Analyse et de Mathé\'ematiques Appliqu\'ees (UMR 8050), CNRS, UPEM,
		UPEC, F-77454, Marne-la-Vall\'ee, France (francois.bouchut@u-pem.fr)}, E.D. Fern\'andez-Nieto \thanks{Dpto. Matem\'atica Aplicada I. ETS Arquitectura - Universidad de Sevilla.
		Avda. Reina Mercedes S/N, 41012-Sevilla, Spain, (edofer@us.es, gnarbona@us.es)}\ ,\\ A. Mangeney
	\thanks{Institut de Physique du Globe de Paris, Seismology team, University Paris-Diderot, Sorbonne Paris Cit\'e, 75238, Paris, France, (mangeney@ipgp.fr)}\ \thanks{ANGE team, CEREMA, INRIA, Lab. J. Louis Lions, 75252, Paris, France}\ , G. Narbona-Reina $^\ddagger$}

\begin{document}
	\date{}
	\maketitle
	
	\begin{abstract}
		We present here a multilayer model for shallow grain-fluid mixtures with dilatancy effects. It can be seen as a generalization of the depth-averaged model presented in [Bouchut et al. {\it A two-phase two-layer model for fluidized granular flows with dilatancy effects}. J. Fluid Mech., 801:166-221, 2016], that includes dilatancy effects by considering a two-layer model, a mixture grain-fluid layer and an upper fluid layer, to allow the exchange of fluid between them. In the present work the approximation of the mixture layer is improved including normal variations of the velocities and concentrations of the two phases thanks to the multilayer approach.
		In the model presented here dilatancy effects induce in particular a non-hydrostatic pressure for both phases related to the excess pore fluid pressure. Contrary to the single-layer model, the computation of this excess pore pressure entrains a serious difficulty due to the multilayer approach. We identified here one of the main numerical difficulty of solving two-phase shallow debris flows models: the strongly non-linear behaviour and abrupt changes of the excess pore fluid pressure when starting from non-equilibrium conditions. We propose a simplified approach to approximate the excess pore fluid pressure in the simple case of uniform flows in the downslope direction and quantify the error made. Our method makes it possible to introduce two or three layers in the normal directions with a reasonable approximation. Analytical solutions for uniform grain-fluid flows over inclined planes, with and without side wall friction, are calculated and compared to the proposed model. 
		The presented model preserves the total solid granular mass as in \cite{bouchut:2016}. 
		In the numerical results, we observe that the proposed model with a two layer description of the mixture accurately represents the velocity measured at the surface of the mixture in the laboratory experiments. This is obviously poorly represented by the depth-averaged velocity in single-layer models while the other quantities (solid volume fraction, basal excess pore fluid pressure) are similar to those obtained with single-layer models. Our numerical results show a significant impact of the parameters involved in dilatancy law, in particular on the calculation of the time evolution of the excess pore fluid pressure.

	\end{abstract}

	\newpage
	\tableofcontents
	
	\newpage
	\section{Introduction}
	
	Many efforts have been devoted to the study of granular flows (aerial, sub-aerial, debris flows,...) in recent years. Gravity driven flows such as landslides, submarine avalanches, or rock avalanches are important natural hazards. One of the challenges of those studies is to predict the velocity and the runout distance in rapid landslides in order to provide new tools for prediction and prevention systems.
	The physical understanding of these flows, and their theoretical description is still a challenge from the geophysical, mathematical and numerical point of view. \\
	In last years, mathematical modelling has contributed to investigate granular flows. However, many questions remain open regarding the internal behavior of these flows, as the rheology and the fluid-grain interaction in fluidized flows.\\
	
	On one hand, rheological laws able to explain the complex behavior of granular flows have been widely investigated. Currently, the most accepted law is the so-called $\mu(I)$-rheology, introduced in \cite{jop:2006}, and other equivalent laws that consider a variable friction coefficient in the definition of the stress tensor (see Capart {\it et al.} \cite{capart:2015}).
	On the other hand, most of natural landslides involve a fluid (commonly water) mixed with the granular material. Interactions between the fluid and solid phases play a key role in the dynamics of these debris flows. This interaction basically depends on the pore fluid pressure, which determines the effective friction between the fluid and the grains. Many studies have been devoted to the study of dilatancy effects in granular materials. Compression/dilatation implies a decrease/increase of the fluid pressure that has a strong effects on the dynamics of fluidized flows (see e.g. \cite{bouchut:2016,pailha:2009}).\\
	
	Most of the models in the literature describing two-phase (grain-fluid) flows are based on the Jackson's model \cite{jackson:2000}, which considers the buoyancy force and the friction between the phases. This model has five unknowns: the solid (grain) volume fraction, the fluid and solid pressures, and solid and fluid velocities. However, only four equations are used: mass and momentum conservation equations for each phase. Therefore, the Jackson's model is underdetermined and a closure equation has to be added. It seems convenient to formulate this closure equation in term of the contraction/dilatation effect as discussed in \cite{bouchut:2016}.\\
	
	As it is well known, the computational cost of solving 3D models is huge. For this reason, the thin-layer (i. e. shallow) approximation have been commonly used to obtain depth-averaged models from the Jackson's model. Pitman and Le \cite{pitman:2005} and later Petanti {\it et al.} \cite{pelanti:2008} proposed a depth-averaged model where no closure relation was included but used an extra boundary condition instead. This approach makes it impossible to obtain a dissipative energy balance, leading therefore to a model that is physically meaningless. This is widely discussed in \cite{bouchut:2015}, where a model with a consistent energy balance is proposed based on the incompressibility of the solid phase. However, the granular phase is not incompressible and dilatancy effects have to be accounted for.
	
	Roux and Radjai introduced a dilatancy model in \cite{roux:1998}, namely the rate of the solid volume change is defined in terms of the shear rate of the granular material $\dot{\gamma}$ and a dilatation angle $\psi$ as $-\dot{\gamma}\tan\psi$. Actually, they write the evolution of the solid volume fraction $\vp$ as
	$$
	\p_t \vp + v\cdot\g\vp = -\vp\,\dot{\gamma}\,\tan\psi,
	$$
	where $v$ is the solid velocity. Pailha and Pouliquen \cite{pailha:2009} used this relation as a closure of the Jackson's model to obtain a depth-averaged model, but they also assume an extra boundary condition and therefore their system is overdetermined. When the previous equation is combined with the mass conservation equation, it leads to
	\begin{equation}
	\label{eq:intro}
	\g\cdot v = \dot{\gamma}\tan\psi,
	\end{equation}
	which is the closure relation considered by Bouchut {\it et al.} \cite{bouchut:2016}. A key point in that model is to consider a thin layer of fluid over the mixture layer. Then, the contraction/dilatation of the mixture is allowed through a fluid transference between the two layers. As in the Pailha and Pouliquen model \cite{pailha:2009}, the dilatancy effects appear in this model in particular through an excess pore fluid pressure term representing a deviation from the hydrostatic solid and fluid pressures.
	
	A different approach to two-phase models is to assume a single-phase flow (e.g. Iverson \cite{iverson:1997}, George and Iverson \cite{george:2011,george:2014}, Iverson and George \cite{iverson:2014}) where the mass and velocity of the mixture, which are defined as an average in terms of the solid volume fraction, are used instead of the masses and velocities of each phase. Therefore, the relative motion between the fluid and solid is not explicit in the model. In \cite{iverson:2014} they consider dilatancy effects through a modification of the dilatancy law of Roux and Radjai \cite{roux:1998} and took into account the compressibility of the granular material.    \\
	
	The previous depth-averaged models share an important limitation. The dilatancy effect is written in terms of the shear rate of the solid phase $\dot{\gamma}$, which strongly depends on the variations in the direction normal to the topography. Therefore, these 'single-layer' models give a poor approximation of this term. Moreover, contraction/dilatation produces a relative motion of the fluid and solid phases in the normal direction that these models hardly reproduce.
	Multilayer models were introduced as an intermediate step between depth-averaged and 3D Navier-Stokes models (e.g. \cite{audusse:2005,fernandezNieto:2014}). These models allow us to recover the normal structure of the flow, and consequently to get a better approximation of the terms that depend on the normal variations. Fern\'andez-Nieto {\it et al.} \cite{fernandezNieto:2016} introduced a multilayer model for dry granular flows with the $\mu(I)$ rheology. They showed that the rheological terms are better approximated thanks to this approach, making it possible to numerically reproduce relevant behaviour of dry granular flows. In particular, they recover the position and evolution of the flow/no-flow transition. Furthermore, their model were able to simulate naturally the change of velocity profiles from Bagnold to S-shaped. This was deeply investigated by Fern\'andez-Nieto {\it et al.} \cite{fernandezNieto:2018}, where side walls friction was added, changing significantly the flow dynamics.\\
	
	The goal of this work is to propose a multilayer extension of model \cite{bouchut:2016} for fluidized granular flows with dilatancy effects. To this aim, we start from the Jackson's model with the closure relation \eqref{eq:intro}, and the final model is obtained from an asymptotic analysis and the multilayer approach. Then, the velocity and solid volume fraction is no more constant along the normal direction. Actually, both are piecewise constant functions, with discontinuities at the internal interfaces of the multilayer domain.
	One of the key point of this work is to highlight the main numerical difficulty in such multilayer  (or 3D)  models for debris flows: to deal with the strongly non-linear and abrupt changes of the excess pore fluid pressure when starting from non-equilibrium conditions. We propose an approximation of this excess pore fluid pressure in simple uniform flows and compare our numerical results with previous depth-averaged (i. e. single-layer) models and with analytical solutions to quantify the errors made in the proposed numerical approximation and how much adding layers in the normal direction improves the comparison with laboratory experiments compared to single-layer models. 
	\\
	
	The paper is organized as follows: in section \ref{se:initialSystem} the governing equations and the rheology are presented. Section \ref{se:twophase_model} is devoted to the details of the derivation of the multilayer model, from the dimensional analysis of the Jackson's model to the final two-phase multilayer model. The numerical tests are presented in section \ref{se:numericalTest} and some conclusions are introduced in section \ref{se:conclusions}. 
	The explicit expression of the excess pore pressure is done in Appendix \ref{Apend_B}. Appendix \ref{Apend_C} collects all the details of the special case of the 2 layer model preserving the granular mass, named PGM-2 model. This model is the multilayer model with a  smaller computational cost and which we compare with depth-averaged models. We show that it reproduces accurately experimental data, mainly the velocity measured at the surface of the mixture, as can be seen in section \ref{se:numericalTest}. 
	
	\section{The initial system}\label{se:initialSystem}
\begin{figure}[ht]
		\begin{center}
			\includegraphics[width=0.5\textwidth]{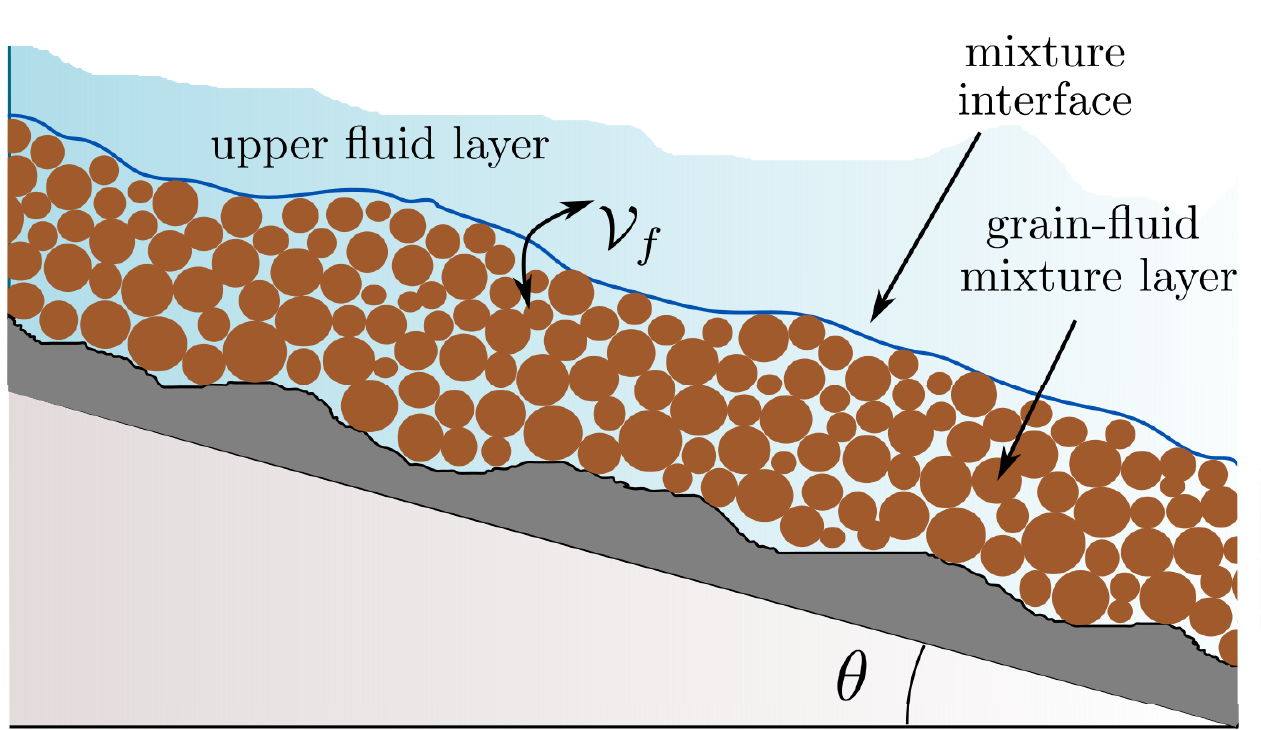}
		\end{center}
		\caption{\label{fig:domain} \it{Sketch of the domain. The term $\mathcal{V}_f$, which may be no-zero, represents the fluid transference at the mixture interface with the upper fluid domain. }}
	\end{figure}

	In this section we present the governing equations together with the closure relations and the rheology describing fluidized granular flows, as well as the boundary conditions of the system. As in \cite{bouchut:2016}, the domain consists of a grain-fluid mixture layer and a fluid layer on its top. The role of this upper layer is to allow the exchange of fluid $\mathcal{V}_f$ between the two layers (Figure \ref{fig:domain}) that is a key point on the dilatancy effect. This exchange is given through the kinematic boundary condition at the mixture interface.
	
	\subsection{Governing equations}\label{sec:gov}
	As mentioned before, the goal of this paper  is to derive a new multilayer model for fluidized granular flows that can be seen as an extension of the work developed in \cite{bouchut:2016}. We focus on the modelling of the mixture layer and the model for the single upper layer is adopted from \cite{bouchut:2016} since it does not change. 
		Thus, we focus on the detailed modelling of the mixture layer in the following.\\
		
		\noindent We consider a granular material and a fluid with constant densities $\r_s, \r_f$. The governing equations are given by  the Jackson's model for fluidized granular materials. The system consists of the concatenation of the mass and the momentum equations for the granular and fluid phases. The equations for the solid phase are considered only in the mixture domain and the fluid equations are considered in both domains. We denote $\vu_f$ the velocity of the fluid in the top layer and $\vv,\vu\in\mathbb{R}^2$ the solid and fluid velocities respectively in the mixture layer.\\
		Thus, for the upper layer the mass and momentum equations read:
		\begin{equation*}\label{eq:onlyfluid}
		\left\{
		\begin{array}{l}
		\nabla\cdot \vu_f=0,\\
		\rho_f (\partial_t \vu_f+(u_f\cdot\nabla) \vu_f)=-\nabla \cdot \bS_f
		+\rho_f \vg,
		\end{array}
		\right.
		\end{equation*}
		whereas in the mixture domain, the Jackson's complete system reads:
		\begin{equation}
		\label{eq:jackson}
		\left\{
		\begin{array}{l}
		\r_s\big( \p_t\vp + \g\cdot\left(\vp\vv\right)\big)\ = \ 0,\\[2mm]
		\r_f\big( \p_t\1vp + \g\cdot\left(\1vp\vu\right)\big)\ = \ 0,\\
		\\
		\r_s\vp\big(\p_{t}\vv+\left(\vv\cdot\g\right)\vv\big) =  \g\cdot\bS_s\ -\vp\g p_{f_m} + f + \r_s\vp\vg,\\[2mm]
		\r_f\1vp\big(\p_{t}\vu+\left(\vu\cdot\g\right)\vu\big) =  \g\cdot\bS_{f_m}\ +\vp\g p_{f_m} - f + \r_f\1vp\vg,
		\end{array}
		\right.
		\end{equation}
		where $\bd{g}$ is the gravity force and $0<\vp<1$ is the solid volume fraction. The total stress tensors are
		\begin{equation*}
		\label{eq:sigma}
		\bS_f=-p_f \mathcal{I}+\bd{\tau}_f,\qquad\bS_{s}= -p_{s}\bd{\mathcal{I}} + \bd{\tau}_{s},\qquad \bS_{f_m}= -p_{f_m}\bd{\mathcal{I}} + \bd{\tau}_{f_m}, 
		\end{equation*}
		with $p\in\mathbb{R}$ the pressure for the solid and fluid phases in the mixture (subscript ${\it s}$ and $f_m$ respectively), and for the fluid in the upper layer (subscript ${\it f}$). $\bd{\mathcal{I}}$ is the 2D identity tensor and the deviatoric stress tensors are given by
		$$\bd{\tau}_f=2\eta_f D(\vu_f),\qquad \bT_{s} =2\eta_{s} D(\vv),\qquad \bT_{f_m} =2\eta_{f_m} D(\vu),$$
		where $\eta_f\in\mathbb{R}$ is the constant viscosity of the fluid phase, and the viscosity of the solid phase $\eta_s\in\mathbb{R}$ is defined by the rheological law for the granular phase. The strain-rate tensors are defined as usual by
		$$
		D(\vu_f)=\dfrac12(\nabla \vu_f +(\nabla \vu_f)'),\qquad D(\vv)=\frac12 (\nabla \vv +(\nabla \vv)'),\qquad D(\vu)=\frac 1 2 (\nabla \vu +(\nabla \vu)').
		$$

	In system \eqref{eq:jackson}, the terms $\vp\g p_{f_m}$ and $f$ are the buoyancy and the drag forces between the phases. We consider here that
	\begin{equation*}
	\label{eq:drag}
	f = \b\left(\vu-\vv\right),
	\end{equation*}
	where $\b$ is the drag coefficient. Following precedent works, this coefficient is defined as
	\begin{equation}
	\label{eq:beta}
	\b = \1vp^2\dfrac{\eta_f}{\kappa},\quad\text{with}\quad \kappa = \dfrac{d_s^2\left(1-\vp\right)^3}{150\vp^2},
	\end{equation}
	where $\kappa$ is the hydraulic permeability of the granular mass and $d_s$ the grain diameter \cite{pailha:2009, iverson:2014, bouchut:2016}. The Jackson's model is underdetermined and a closure equation must be added. In this work we assume the closure introduced in \cite{bouchut:2016} including the dilatancy effects
	\begin{equation}
	\label{eq:dilatancy}
	\g\cdot \vv = \Phi,
	\end{equation}
	where the dilatancy function $\Phi$ is defined as in \cite{pailha:2009} and it is detailed in the subsection \ref{se:muI}.\\ 
	
	\subsubsection{Boundary conditions}
In this work we consider analogous boundary condition to \cite{bouchut:2016} that we summarize here, we refer the reader to this reference for more details.

		\begin{itemize}
			\item At the bottom we consider the non penetration conditions
			\begin{equation*}
			\vu\cdot n=0,\qquad \vv\cdot n=0\qquad \textrm{at the bottom,}
			\label{eq:bcbot}
			\end{equation*}
			where $n$ is the space unit normal pointing out of the domain.
			This is completed with friction conditions for solid and fluid phases. At first, a Coulomb friction law,
			\begin{equation*}
			(\bS_s n)_\tau = -\tan\mu_{\rm eff} \,\frac{\vv}{|\vv|}(\bS_s n)\cdot n\qquad
			\textrm{at the bottom},\label{eq:coulomb}
			\end{equation*}
			where $\mu_{\rm eff}$ is the effective friction coefficient given in terms of the rheology introduced in the next subsection and the subscript $\tau$ denotes the tangential projection.
			For the fluid phase we consider a Navier friction condition,
			\begin{equation*}
			(\bS_{f_m} \,n)_\tau=-k_b \vu\qquad \textrm{at the bottom,}
			\label{eq:noslip}
			\end{equation*}
			for some coefficient $k_b\geq 0$.
			
			\item At the free surface we assume no tension for the fluid and the kinematic condition
			\begin{equation*}
			\bS_{f} N_X=0;\qquad 	N_t+\vu_f\cdot N_X=0\qquad \textrm{at the free surface,}
			\label{eq:notension}
			\end{equation*}
			where $N=(N_t,N_X)$ is a time-space normal to the free surface.
			\item At the interface, we consider the kinematic condition for the solid phase
			\begin{equation*}
			\tilde N_t+\vv\cdot \tilde N_X=0 \qquad \textrm{at the interface,}
			\label{eq:kinem_s}
			\end{equation*}
			where $\tilde{N}=(\tilde{N}_t,\tilde{N}_X)$ is a time-space downward normal to the interface.
			The fluid exchange between the mixture and the upper fluid layer is given through the Rankine-Hugoniot condition:
			\begin{equation*}\label{interf_mass}
			\tilde N_t+\vu_f \cdot \tilde N_X
			=(1-\varphi^*)( \tilde N_t+\vu\cdot \tilde N_X)\equiv \mathcal{V}_f\qquad
			\textrm{at the interface},
			\end{equation*}
			where $\varphi^*$ is the value of the solid volume fraction at the interface (the limit is taken
			from the mixture side), since $\varphi$ is discontinuous at the interface.
			The term $\mathcal{V}_f$ defines the fluid mass that is transferred from the mixture
			to the fluid-only layer.
			
			The conservation of the total momentum gives,
			\begin{equation*}\label{interf_mom}
			\rho_f \mathcal{V}_f (\vu-\vu_f)+(\bS_s+\bS_{f_m})\tilde N_X
			=\bS_f \tilde N_X\qquad\textrm{at the interface}.
			\end{equation*}
			The energy balance through the interface yields the stress transfer condition
			\begin{equation*}
			\bS_s\tilde N_X=\Biggl(\frac{\rho_f}{2}\biggl((\vu-\vu_f)\cdot\frac{\tilde N_X}
			{|\tilde N_X|}\biggr)^2
			+\biggl((\bS_{f_m} \tilde N_X)\cdot \frac{\tilde N_X}
			{|\tilde N_X|^2}-p_{f_m}\biggr)\frac{\varphi^*}{1-\varphi^*}\Biggr) \tilde N_X.
			\label{energy_inter}
			\end{equation*}
			Finally a Navier fluid friction condition is considered at this level
			\begin{equation*}\label{tanf}
			\Bigl( \frac{\bS_{f_m}+\bS_f}{2} \tilde N_X \Bigr)_\tau
			=-k_i (\vu_f-\vu)_\tau \qquad \textrm{at the interface},
			\end{equation*}
			where $k_i\geq 0$ is a friction coefficient.
		\end{itemize}

	\subsection{Rheology for fluidized granular flows}\label{se:muI}
	A rheology accounting for the behavior of immersed granular flows must be considered. For dry granular flows the $\mu(I)$-rheology \cite{jop:2006} is usually considered where the viscosity of the granular material is given by $\frac{\mu(I) p_s}{\|D(\vv)\|}$, with $\mu(I)$ being the friction coefficient and $I$ the inertial number. Dilatancy affects the friction law and can be introduced as follows (see \cite{pailha:2009,fernandezNieto:2016}):
	
	\begin{equation*}
	\label{eq:viscosity_prev}
	\eta_s = \dfrac{\left(\mu(I)+\tan\psi\right)}{p_s{\|D(\vv)\|}},
	\end{equation*}
	where $\|D\| = \sqrt{0.5\;D:D}$ and $\psi$ is the dilatation angle as is defined by equation (\ref{eq:phieq}) (see \cite{roux:1998, pailha:2009, bouchut:2016}). We introduce $\mu_{\rm eff}=\mu(I)+\tan\psi$ the effective friction coefficient, appearing also in the Coulomb friction law.\\
	The friction coefficient is
	\begin{equation}
	\label{eq:muI_1}
	\mu(I) = \mu_{s} + \dfrac{\mu_2-\mu_s}{I_{0}+I}I ,
	\end{equation}
	which depends on the inertial number $I$, where $I_{0}, \mu_{2}>\mu_s$ are constant values depending on the material. The main difference between the dry and the fluidized case is the definition of the inertial number $I$ (see \cite{andreotti:2013}). In the case of immersed granular flows in the viscous regime, this dimensionless number is defined as
	\begin{equation} \label{eq:def_I}
	I = \dfrac{2\,\eta_{f}\,\|D(\vv)\|}{p_s},
	\end{equation}
	in contrast to $I_{dry} =\frac{2d \|D(\vv)\|}{\sqrt{p_s/\rho_s}}$ that is the inertial number for the dry case with $d$ being the particle diameter.\\
	Following this idea, we can also consider an alternative simplified rheology where the friction coefficient is defined as
	\begin{equation}
	\label{eq:muI_2}
	\mu(I) = \mu_{s} + K_1 I ,
	\end{equation} where $K_1$ is a constant value. In this case, the viscosity coefficient can be written as follows,
	\begin{equation*}
	\label{eq:viscosity_prev2}
	\eta_s = \dfrac{\left(\mu_s+\tan\psi\right)p_s}{\|D(\vv)\|} + 2 K_1\eta_f .
	\end{equation*}

	Note that the viscosity coefficient is not defined in the case $\|D(\vv)\| = 0$. As in previous works for dry granular flows \cite{fernandezNieto:2016,fernandezNieto:2018} we consider a regularization method (e. g. \cite{lagree:2011,lusso:2017a}, since the results are good enough (although zero velocity is not obtained in static solutions, but very small velocities) and they are cheaper computationally than using a duality method (e. g. \cite{ionescu:2015, martin:2017})). Then, the regularized viscosity coefficient is
	\begin{equation}\label{eq:regularisation}
	\eta_s = \dfrac{\left(\mu(I)+\tan\psi\right)\,p_s}{\sqrt{\|D(\vv)\|^2 + \delta^2}},
	\end{equation}
	with $\delta>0$ the regularization parameter. \\
	
	Finally, the dilatancy function is adopted from \cite{pailha:2009} and it is given by:
	\begin{equation}
	\label{eq:dil_funcion}
	\Phi = \dot{\gamma} \tan \psi
	\end{equation} with $\dot{\gamma} = 2\|D(\vv)\|$ the shear rate, $\psi$ the dilatation angle and
	\begin{equation}
	\label{eq:phieq}
	\tan \psi = K\left(\vp-\vp_c^{eq}\right),\qquad
	\vp_c^{eq} = \vp_c^{stat} - K_2 I,
	\end{equation}
	which allows us to rewrite \eqref{eq:dilatancy} as  \begin{equation*}
	\g\cdot v = K\left(\vp \,-\, \vp_c^{stat} \,+\, K_2 I\right),
	\end{equation*}
	where $K$ is the dilatation constant and $K_2$ is a constant value, $\vp_{c}^{stat}$ is a constant volume fraction corresponding to a static equilibrium, and $\varphi^{eq}_c$ the critical-state equilibrium compacity.
	This value determines if the current state of the granular medium is dilatation, contraction or equilibrium. Assuming a deformation occurs, $\dot{\gamma} > 0$, the solid dilates if $\vp > \vp_c^{eq}$ and contracts if $\vp < \vp_c^{eq}$. In the dilatation case, the fluid is sucked into the fluid-solid mixture and the pore pressure decreases. In the other case the fluid is expelled from the mixture and the pore pressure increases. If $\vp = \vp_c^{eq}$ then there is no dilatation nor contraction.

	\subsection{Local coordinates}\label{sec:bc}

	We write the system for the mixture in tilted coordinates. Let $\wtd{b}(x)$ be an inclined fixed plane of constant angle $\theta$ with respect to the horizontal axis, we define the coordinates $(x,z)\in\O \times \mathbb{R}^+ \subset \R^{2}$. The $x$ (respectively $z$) axis is measured along the inclined plane (respectively the normal direction) as depicted in Figure \ref{fig:Multilayers}. In this reference frame the gravity force is written as
	$$\bd{g} = (-g\;\sin\theta,-g\;\cos\theta)'.$$
	We also set $b(x)$ an arbitrary bottom topography and the layer of the fluidized material over it with thickness $h(t,x)$, which are measured in the normal direction to the inclined plane $\wtd{b}(x)$. The fluid and granular velocities are $\vu = \left(u,u^z\right)$ and $\vv = \left(v,v^z\right)$ where $u,v$ denotes the downslope velocities of the fluid and granular phase, and $u^z,v^z$ the normal components of $\vu,\vv$. Finally, we set $\smash{\g = (\p_{x}, \p_{z})}$, the usual differential operator in the space variables.\\
	
	We denote the components of the total stress tensor for the fluid phase as:
	$$
	\bT_{f_m} = \left( \begin{matrix}
	\T^{xx}_{f_m} & \T^{xz}_{f_m}\\
	\T^{xz}_{f_m} &  \T^{zz}_{f_m}
	\end{matrix}\right)
	\quad \mbox{ and }\quad
	D(\vu)=\frac 1 2 \left( \begin{matrix}
	2\p_x u & \p_{z}u+\p_{x}u^z\\
	\\
	\p_{z}u+\p_{x}u^z &  2\p_{z}u^z
	\end{matrix}\right),
	$$
	\noindent and analogously defined for the solid phase. With these definitions and embedding the mass equations into the momentum equations, system \eqref{eq:jackson} is written as
	\begin{equation}\label{eq:jackson_3D}
	\left\{
	\begin{array}{l}
	\r_s\big(\p_t \vp + \p_{x} \left(\vp v\right)+\p_{z}\left(\vp v^z\right)\big)= 0,\\[2mm]
	\r_f\big(\p_t \1vp + \p_{x} \left(\1vp u\right)+\p_{z}\left(\1vp u^z\right)\big)= 0,\\
	\\
	\r_s\big(\p_{t}\left(\vp v\right)+\p_{x}\left(\vp v^2\right)+\p_{z}\left(\vp v v^z\right) \big)+\p_{x}p_s = \p_{x}\T_s^{xx}+\p_{z}\T_s^{xz}-\vp\p_x p_{f_m}\\[2mm]
	\qquad+\, \b\left(u-v\right) - \r_s \vp g\sin\theta,\\[2mm]
	\r_f\big(\p_{t}\left(\1vp u\right)+\p_{x}\left(\1vp u^2\right)+\p_{z}\left(\1vp u u^z\right) \big)+\p_{x}p_{f_m} = \p_{x}\T_{f_m}^{xx}+\p_{z}\T_{f_m}^{xz} \\[2mm]
	\qquad +\vp\p_x p_{f_m}- \,\b\left(u-v\right)-\, \r_f \1vp g\sin\theta,\\
	\\
	\r_s\big(\p_{t}\left(\vp v^z\right)+\p_{x}\left(\vp v v^z\right)+\p_{z}\left(\vp (v^z)^2\right) \big)+\p_{z}p_s = \p_{x}\T_s^{xz}+\p_{z}\T_s^{zz}-\vp\p_z p_{f_m}\\[2mm]
	\qquad +\, \b\left(u^z-v^z\right) - \r_s \vp g\cos\theta,\\[2mm]
	\r_f\big(\p_{t}\left(\1vp u^z\right)+\p_{x}\left(\1vp u u^z\right)+\p_{z}\left(\1vp (u^z)^2\right) \big)+\p_{z}p_{f_m} = \p_{x}\T_{f_m}^{xz}+\p_{z}\T_{f_m}^{zz} \\[2mm]
	\qquad +\vp\p_z p_{f_m}-\, \b\left(u^z-v^z\right) - \r_f \1vp g\sin\theta ,
	\end{array}
	\right.
	\end{equation}
	and the dilatancy closure equation is
	\begin{equation*}
	\label{eq:dil_3D}
	\p_x v + \p_z v^z = \Phi.
	\end{equation*}
	The boundary conditions are detailed directly in section \ref{se:twophase_model} after the asymptotic analysis.
	
	\section{Two-phase multilayer models with dilatancy}\label{se:twophase_model}
	In this section we deduce the multilayer model for fluidized granular flows focusing on the points that are different from previous multilayer models (e.g. \cite{fernandezNieto:2016}). The model is obtained from a dimensional analysis and considering a multilayer approach of the domain in the vertical direction. We focus on the approximation of the deviatoric tensor at the internal interfaces, as well as the definition of the vertical velocity and the stress for both, the solid and the fluid phases.
	
	A difficulty of this model is that the solid pressure is not purely hydrostatic: an excess pore fluid pressure appears in the equations as a consequence of the contraction/dilatation of the granular material (see \eqref{eq:pres_solid}-\eqref{eq:press_fluid_e_exp}).
	This term also appears in the vertical velocities of both phases \eqref{eq:vert_vel_sol}. Finally, the model and the explicit expressions for the mass transference terms are presented.
	
	%
	
	\subsection{Dimensional analysis}\label{se:dimensional_analysis}
	
	We perform here a dimensional analysis of the system \eqref{eq:jackson_3D}. We consider a shallow domain by assuming that the dimensionless parameter $\varepsilon = H/L$ is small, where $H$ and $L$ are the characteristic height and length of the flowing mass, respectively. We define the dimensionless variables, denoted with the tilde symbol ($\tilde{.}$), as follows:
	\begin{equation*}
	\label{nondim_var}
	\begin{array}{c}
	(x,z,t) = (L\wtd{x},H\wtd{z},(L/U)\wtd{t}),
	\quad
	h = H\wtd{h},
	\quad
	\r_{s,f} = \r_{0}\wtd{\r_{s,f}},
	\quad\vp = \vp_0\wtd{\vp}\\
	\\
	(u,u^z) =  (U\wtd{u},\varepsilon U\wtd{u^z}),
	\quad
	(v,v^z) =  (U\wtd{v},\varepsilon U\wtd{v^z}),\\
	\\
	p_{s,f} = \r_{0}U^{2}\wtd{p_{s,f}},
	\quad
	\eta_{s,f}= \varepsilon\r_{0}UH\wtd{\eta_{s,f}}, 
	\quad 
	\kappa_i = U\wtd{\kappa_i}, \quad k_b = U\wtd{k_b}
	\\
	\\
	\left(\T_{s,f}^{xx},\T_{s,f}^{xz},\T_{s,f}^{zz}\right) = \vp_0\r_{0}U^2\left(\varepsilon^2\wtd{\T_{s,f}^{xx}},\varepsilon\wtd{\T_{s,f}^{xz}},\varepsilon^2\wtd{\T_{s,f}^{zz}}\right),\quad \mathcal{V}_{f} = \varepsilon U \wtd{\mathcal{V}_{f}}.
	\end{array}
	\end{equation*}
	Note that since
	$$
	D(\vu) = \dfrac{U}{H}\ \frac12 \left( \begin{matrix}
	2\varepsilon^2 \p_{\wtd{x}} \wtd{u}  & \varepsilon\p_{\wtd{z}}\wtd{u}+\varepsilon^3\p_{\wtd{x}}\wtd{u^z}\\
	\\
	\varepsilon\p_{\wtd{z}}\wtd{u}+\varepsilon^3\p_{\wtd{x}}\wtd{u^z} &  2\varepsilon^2\p_{\wtd{z}}\wtd{u^z}
	\end{matrix}\right),
	$$
	we obtain that
	$$
	\begin{array}{lll}
	\wtd{\T^{xx}_f}= \wtd{\eta} \p_{\wtd{x}} \wtd{u}, & \quad
	\wtd{\T^{xz}_f}= \frac{\wtd{\eta}}{2} \left( \p_{\wtd{z}}\wtd{u}+\varepsilon^2\p_{\wtd{x}}\wtd{u^z} \right),  &
	\quad
	\wtd{\T^{zz}_f}= \wtd{\eta} \p_{\wtd{z}}\wtd{u^z},
	\end{array}
	$$
	and $D(\vv)$ and the components of $\T_s$ are analogously written. In this work we have considered that the characteristic velocity for the solid phase is equal to the one of the fluid, and we also take the characteristic solid volume fraction is $\vp_0=1$. We also define the Froude number
	\[Fr = \frac{U}{\sqrt{gH\cos\theta }},\]
	and the nondimensional drag coefficient
	\begin{equation}
	\label{eq:beta_friccion}
	\b = \dfrac{\r_0 U}{L}\,\varepsilon^k\,\wtd{\b},\qquad\mbox{where}\quad k=\left\{\begin{array}{cc}
	-1 & \mbox{strong friction};\\
	0 & \mbox{moderate friction},
	\end{array}\right.
	\end{equation}
	and the dilatancy function
	$$
	\Phi = \dfrac{U}{H}\wtd{\Phi}.
	$$

	Thus, the system \eqref{eq:jackson_3D} can be rewritten using dimensionless variables as (tildes have been dropped for simplicity):
	\begin{subequations}
		\label{eq:nondim_jackson_3D}
		\begin{align}[left = \empheqlbrace\,]
		&\r_s\big(\p_t \vp + \p_{x} \left(\vp v\right)+\p_{z}\left(\vp v^z\right)\big)= 0,\label{eq:nondim_jackson_3D_a}\\[2mm]
		&\r_f\big(\p_t \1vp + \p_{x} \left(\1vp u\right)+\p_{z}\left(\1vp u^z\right)\big)= 0,\label{eq:nondim_jackson_3D_b}\\
		&\nonumber\\
		&\r_s\big(\p_{t}\left(\vp v\right)+\p_{x}\left(\vp v^2\right)+\p_{z}\left(\vp v v^z\right) \big)+\p_{x}p_s = \varepsilon^2\p_{x}\T_s^{xx}+\p_{z}\T_s^{xz}\nonumber\\[2mm]
		&\qquad -\,\vp\p_x p_{f_m} + \varepsilon^k\b\left(u-v\right) - \dfrac{\r_s \vp}{\varepsilon Fr^2} \tan\theta,\label{eq:nondim_jackson_3D_c}\\[2mm]
		&\r_f\big(\p_{t}\left(\1vp u\right)+\p_{x}\left(\1vp u^2\right)+\p_{z}\left(\1vp u u^z\right) \big)+\p_{x}p_{f_m} = \varepsilon^2\p_{x}\T_{f_m}^{xx}+\p_{z}\T_{f_m}^{xz} \nonumber\\[2mm]
		&\qquad +\,\vp\p_x p_{f_m}- \varepsilon^k\b\left(u-v\right) - \dfrac{\r_f \1vp}{\varepsilon Fr^2} \tan\theta,\label{eq:nondim_jackson_3D_d}\\
		\nonumber\\
		&\varepsilon^2\r_s\big(\p_{t}\left(\vp v^z\right)+\p_{x}\left(\vp vv^z\right)+\p_{z}\left(\vp (v^z)^2\right) \big)+\p_{z}p_s = \varepsilon^2\p_{x}\T_s^{xz}+\varepsilon^2\p_{z}\T_s^{zz}\nonumber\\[2mm]
		&\qquad -\,\vp\p_z p_{f_m}+ \varepsilon^{k+2}\b\left(u^z-v^z\right) - \dfrac{\r_s \vp}{Fr^2},\label{eq:nondim_jackson_3D_e}\\[2mm]
		&\varepsilon^2\r_f\big(\p_{t}\left(\1vp u^z\right)+\p_{x}\left(\1vp u u^z\right)+\p_{z}\left(\1vp (u^z)^2\right) \big)+\p_{z}p_{f_m} = \varepsilon^2\p_{x}\T_{f_m}^{xz} \nonumber\\[2mm]
		&\qquad +\varepsilon^2\p_{z}\T_{f_m}^{zz}+\,\vp\p_z p_{f_m}- \varepsilon^{k+2}\b\left(u^z-v^z\right) - \dfrac{\r_f \1vp}{Fr^2}\label{eq:nondim_jackson_3D_f}
		\end{align}
		and
		\begin{equation}
		\label{eq:nondim_dil_3D}
		\p_x v + \p_z v^z = \dfrac{1}{\varepsilon}\Phi.
		\end{equation}
	\end{subequations}

		Before applying the multilayer approach it is suitable to write the system in matrix notation. To this aim we collapse all terms coming from the stress tensors under the following notation for the fluid phase (analogously for the solid phase) $\bS_{\varepsilon,{f_m}} = -p_{f_m}\mathcal{E} + \varepsilon\bT_{\varepsilon,{f_m}}$ where the subindex $\varepsilon$ marks the dependence of these terms on $\varepsilon$, as in \cite{fernandezNieto:2016}, with 
	
	$$
	\bT_{\varepsilon,{f_m}} = \eta D_{\varepsilon,{f_m}}(\vu), \quad D_{\varepsilon,{f_m}}(\vu):= \dfrac{1}{2}\left( \begin{matrix}
	2\varepsilon^2 \p_x u & \p_{z}u+\varepsilon^2\p_{x}u^z\\
	\\
	\p_{z}u+\varepsilon^2\p_{x}u^z &  2\;\p_{z}u^z
	\end{matrix}\right), \quad\mbox{ and }\quad
	\mathcal{E} = \left(\begin{matrix}
	\varepsilon & 0\\
	\\
	0 & 1/\varepsilon
	\end{matrix}\right).
	$$
	Then, defining $\bd{F}= \left(\dfrac{\tan\;\theta}{Fr^{2}},\;\dfrac{1}{\varepsilon Fr^{2}}\right)'$, the system \eqref{eq:nondim_jackson_3D} is finally written as

\begin{equation}
	\label{eq:jackson_matricial}
	\left\{
	\begin{array}{l}
	\r_s\big( \p_t\vp + \g\cdot\left(\vp\vv\right)\big)\ = \ 0,\\[2mm]
	\r_f\big( \p_t\1vp + \g\cdot\left(\1vp\vu\right)\big)\ = \ 0,\\
	\\
	\r_s\big(\p_{t}\left(\vp\vv\right)+\g\cdot\left(\vp\vv\otimes\vv\right)\big)-\dfrac{1}{\varepsilon}\g\cdot\bS_{\varepsilon,s}  + \dfrac{1}{\varepsilon}\vp\g\cdot\left(p_{f_m}\mathcal{E}\right) \\[2mm]
	\qquad =\,  \varepsilon^k\b\left(\vu-\vv\right) - \dfrac{1}{\varepsilon}\r_s\vp \bd{F},\\[2mm]
	\r_f\big(\p_{t}\left(\1vp\vu\right)+\g\cdot\left(\1vp\vu\otimes\vu\right)\big)-\dfrac{1}{\varepsilon}\g\cdot\bS_{\varepsilon,{f_m}}  - \dfrac{1}{\varepsilon}\vp\g\cdot\left(p_{f_m}\mathcal{E}\right) \\[2mm]
	\qquad =\,  -\varepsilon^k\b\left(\vu-\vv\right) - \dfrac{1}{\varepsilon}\r_f\1vp \bd{F},\\
	\\
	\g\cdot \vv = \dfrac{1}{\varepsilon}\Phi.
	\end{array}
	\right.
	\end{equation}
	
	Next we specify the asymptotic expressions of the boundary conditions described in section \ref{sec:gov}. Taking into account the asymptotic analysis up to second order performed in \cite{bouchut:2016} (see its Appendixes A and B for the details), these conditions read
	\begin{equation}
		\label{eq:boundCond_surf}
		\begin{array}{clc}
		\left(1-\vp^*\right)\left(\p_{t}h + u_{|_{z=b+h}}\,\p_{x}\left(b+h\right) - u^z_{|_{z=b+h}}\right) = -\mathcal{V}_f; 
		\\[4mm]
		\p_{t}h + v_{|_{z=b+h}}\,\p_{x}\left(b+h\right) - v^z_{|_{z=b+h}}= 0;\\
		\\
		p_{{f_m}|_{z=b+h}} = \r_f g\cos\theta h_f,\qquad p_{s|_{z=b+h}} = 0,
		\end{array}
		\end{equation}
	
	The energy balance gives
	$$
	\left(\dfrac{\eta_f}{2}\p_{z}u\right)_{|_{z=b+h}} =  \dfrac{1}{\varepsilon}\left(\left(\dfrac{\varepsilon}{2} \r_f\mathcal{V}_f-\kappa_i\right)\left(u_f-u\right) \right)_{|_{z=b+h}}
	$$
	where $u_f$ is the first component of the velocity of the water in the upper layer.
	
	At the bottom $z=b$ the no penetration and the friction conditions read
	\begin{equation*}
	\label{eq:boundCond_bot}
	\begin{array}{c}
	u_{|_{z=b}}\ \p_{x} b = u^z_{|_{z=b}};\qquad \left(\eta_f\,\p_{z}u\right)_{|_{z=b}} = \dfrac{1}{\varepsilon}\,k_b \,u_{|_{z=b}},\\
	\\
	v_{|_{z=b}}\ \p_{x} b = v^z_{|_{z=b}}; \qquad \left(\dfrac{\eta_s}{2}\p_{z}v\right)_{|_{z=b}} = \dfrac{1}{\varepsilon}\left(\left(\mu(I)+\tan\psi\right)\,p_s\,\dfrac{v}{\abs{v}} \right)_{|_{z=b}}.
	\end{array}
	\end{equation*}
	Moreover we consider that there is no transference of mass at the bottom, neither solid nor fluid.

	\subsection{A multilayer approach}\label{se:multilayer_approach}
\begin{figure}
		\begin{center}
			\includegraphics[width=1\textwidth]{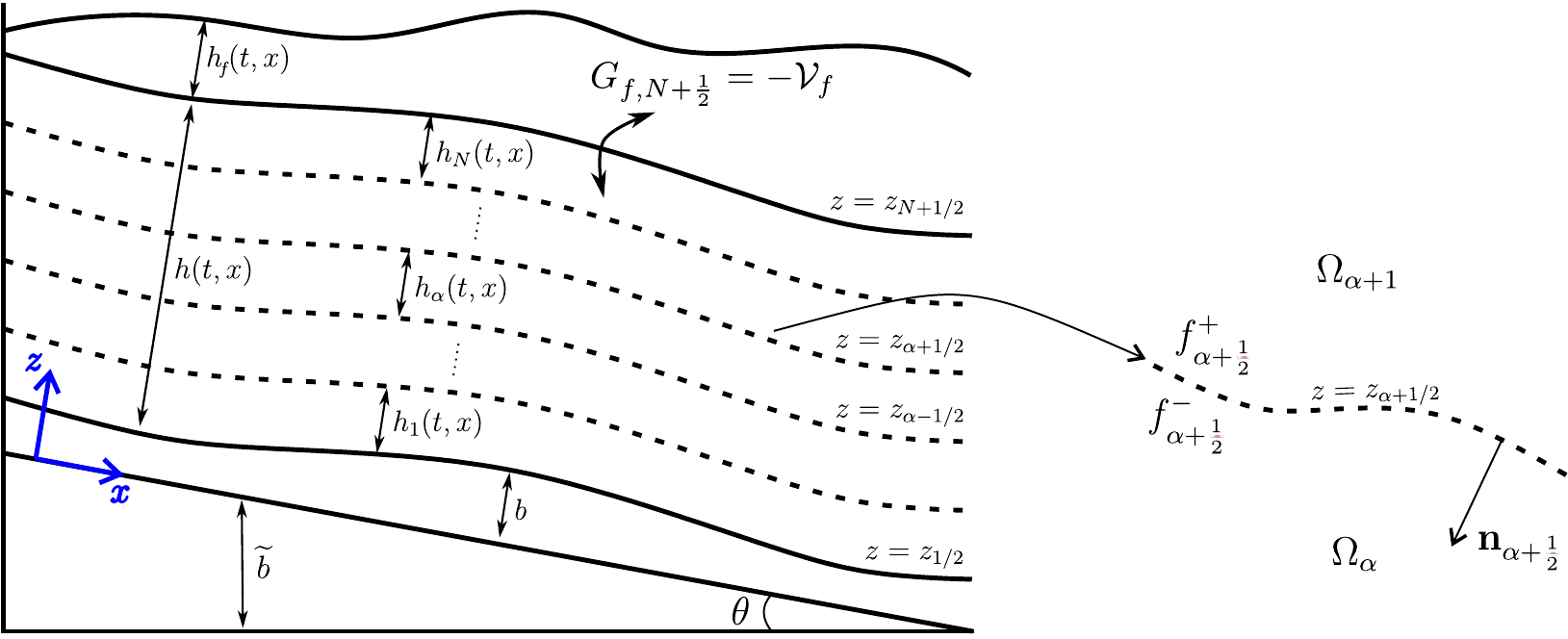}
		\end{center}
		\caption{\label{fig:Multilayers} \it{Sketch of domain with the particular multilayer division for the mixture grain-fluid layer and its notation.}}
	\end{figure}

	We remind the notation of the multilayer domain in order to apply this approach to system \eqref{eq:jackson_matricial}.
	The mixture domain is denoted by $\Omega_{F}(t)$ and
	\[I_{F}(t) = \Big\{x\in \mathbb{R} ; (x,z) \in \O_F(t)\Big\},\]
	is its projection on the reference inclined plane, for any time $t>0$. The idea of this approach is to divide the domain along the direction normal to the plane in $N\in\mathbb{N}\setminus\{0\}$ shallow layers with thickness $h_\a(t,x)$, such that the total height is $h=\sum_{\a=1}^{N}h_\a$ (see Figure \ref{fig:Multilayers}). In practice, the normal partition is previously fixed by using the positive coefficient $l_\a$ satisfying
	$$
	h_{\a} = l_{\a}h \quad\text{ for } \a = 1,...,N; \quad\quad
	\dsum_{\a=1}^N l_{\a} = 1.
	$$

	The $N+1$ interfaces separating the vertical layers are denoted by
	$\G_{\a+\frac12}(t)$, and are described by $z = z_{\a+\frac12}(t,x)$ for $\a = 0,1,..,N$, $x\in I_{F}(t)$. These interfaces, which are assumed to be smooth enough, shut the subdomain $\O_{\a}(t)$ up, i.e.,
	$$\O_{\a}(t) =  \Big\{(x,z); \; x\in I_{F}(t) \mb{ and } z_{\a-\frac{1}{2}}< z <z_{\a+\frac{1}{2}}\Big\}.$$
	
	A key point of this approach is the approximation of the variables at the interfaces. Then, for an arbitrary function $\smash f$ and for $\smash\a=0,1,...,N$, we set
	$$ f_{\a+\frac{1}{2}}^{-} := ( f_{|_{\O_{\a}(t)}})_{|_{\G_{\a+\frac{1}{2}}(t)}} \mb{\;\;and\;\;}  f_{\a+\frac{1}{2}}^{+} := (f_{|_{\O_{\a+1}(t)}})_{|_{\G_{\a+\frac{1}{2}}(t)}},$$
	where if the function $\smash f$ is continuous, then
	$$ f_{\a+\frac{1}{2}} := f_{|_{\G_{\a+\frac{1}{2}}(t)}} = f_{\a+\frac{1}{2}}^+ = f_{\a+\frac{1}{2}}^-.$$
	Finally, the normal vectors at the interfaces are defined. For a time $t\geq 0$,
	\begin{equation*}
	\label{eq:unit_norm}
	\vn_{T,\a+\frac12}=\dfrac{\left(\p_t z_{\a+\frac{1}{2}}, \p\!_{x}z_{\a+\frac{1}{2}}, -1\right)^{'}}{\sqrt{1+\left(\p\!_{x}z_{\a+\frac{1}{2}}\right)^2 + \left(\p_t z_{\a+\frac{1}{2}}\right)^2}} \hspace{0,3cm}\mbox{and}\hspace{0,3cm} \vn_{\a+\frac12}=\dfrac{\left(\p\!_{x}z_{\a+\frac{1}{2}}, -1\right)^{'}}{\sqrt{1+\left(\p\!_{x}z_{\a+\frac{1}{2}}\right)^2}}
	\end{equation*}
	denote the space-time unit normal vector and the space unit normal vector to the interface $\smash \Gamma_{\a+\frac12}(t)$ outward  to the layer $\smash \Omega_{\a+1}(t)$  for $\smash\a=0,...,N$ (see detail in Figure \ref{fig:Multilayers}).
	\subsubsection{Weak solution with discontinuities}\label{se:weak_sol}
	In order to apply the multilayer approach, we look for a particular weak solution $(\vu,\vv,p_{f_m},p_s,\r_f,\r_s,\vp)$ of \eqref{eq:jackson_matricial} (see \cite{fernandezNieto:2014,fernandezNieto:2016}) satisfying that:
	\begin{itemize}
		\item[(i)] $(\vu,\vv,p_{f_m},p_s,\r_f,\r_s,\vp)$ is a standard weak solution of \eqref{eq:jackson_matricial} in each layer $\Omega_\alpha(t)$,
		\item[(ii)] $(\vu,\vv,p_{f_m},p_s,\r_f,\r_s,\vp)$ satisfies the normal flux jump condition for mass and momentum equations for each phase, at the interfaces $\Gamma_{\alpha+\frac 1 2 }(t)$, namely:
		\begin{equation}
		\label{eq:jumpCond_mass}
		\begin{array}{c}
		\left[\left(\rho_s\vp\,;\rho_s\vp\vv\right)\right]_{\alpha+\frac 1 2 }\ \vn_{T,\a+\frac12}=0,\\
		\\
		\left[\left(\rho_f\1vp\,;\rho_f\1vp\vu\right)\right]_{\alpha+\frac 1 2 }\ \vn_{T,\a+\frac12}=0,
		\end{array}
		\end{equation}
		\begin{equation}
		\label{eq:jumpCond_mom}
		\begin{array}{c}
		[(\rho_s\vp\vv;\rho_s\vp\vv\otimes\vv-\dfrac{1}{\varepsilon}\bS_{\varepsilon,s})\big]_{\alpha+\frac 1 2 }\ \vn_{T,\a+\frac12}=0,\\
		\\
		\big[(\rho_f\1vp\vu;\rho_f\1vp\vu\otimes\vu-\dfrac{1}{\varepsilon}\bS_{\varepsilon,{f_m}})\big]_{\alpha+\frac 1 2 }\ \vn_{T,\a+\frac12}=0,\\
		\end{array}
		\end{equation}
		
		\noindent where $[(a;b)]_{\alpha+\frac12}$ denotes the jump of $(a;b)$ across the interface $\Gamma_{\alpha+\frac12}(t)$.
	\end{itemize}
	In the previous normal flux jump condition for the momentum equation, we have used that the fluid pressure $p_{f_m}$ is continuous and the gravitational force and friction between the phases are source terms, therefore there is no jump in the bouyancy term and all the terms at the interfaces.
	
	Thus, there is no significant differences between the jump condition in \cite{fernandezNieto:2016} and our case. A particular family of test function is considered where the horizontal component of the velocity does not depend on $z$, and the vertical one is linear on $z$. Then, the velocities of the granular and the fluid phase in each layer are
	$$ \vu_{|_{\O_{\a}(t)}} := \vu_{\a} := (u_{\a}, u^z_{\a})^{'},\quad\vv_{|_{\O_{\a}(t)}} := \vv_{\a} := (v_{\a}, v^z_{\a})^{'},
	$$
	where $u_{\a}$,$v_{\a}$ and $u^z_{\a},v^z_{\a}$ are the horizontal and normal velocities, respectively, on layer $\alpha$. Then,
	\begin{equation*}
	\label{eq:strucvp}
	\partial_z u_{\a}=0,\ \p_z v_\a = 0;  \quad\mbox{and}\quad\partial_z u^z_{\a}=d_{f,\a}(t,x),\ \partial_z v^z_{\a}=d_{s,\a}(t,x)
	\end{equation*}
	for some smooth functions $d_{f,\a}(t,x),d_{s,\a}(t,x)$. We also consider that the solid volume fraction is constant in each layer $$\vp_{|_{\O_{\a}(t)}} := \vp_{\a} = \vp_{\a}(t,x).$$
	
	From the jump condition for the mass equations \eqref{eq:jumpCond_mass} we can easily obtain the definition of the solid mass flux ($G_{s,\a+\frac12}$) and the fluid mass flux ($G_{f,\a+\frac12}$):
	\begin{equation*}
	G_{s,\a+\frac12} := G_{s,\a+\frac12}^+ = G_{s,\a+\frac12}^- \quad\mbox{and}\quad G_{f,\a+\frac12} := G_{f,\a+\frac12}^+ = G_{f,\a+\frac12}^-
	\end{equation*}
	with
	\begin{subequations}
		\label{eq:G}
		\begin{equation}\label{eq:G_s}
		G_{s,\a+\frac12}^\pm = \vp_{\a+\frac12}^\pm\left(\p_{t}z_{\a+\frac{1}{2}} + v_{\aum}^\pm\ \partial_{x}z_{\a+\frac{1}{2}} - (v^z_{\a+\frac{1}{2}})^{\pm}\right),
		\end{equation}
		\begin{equation}\label{eq:G_f}
		G_{f,\a+\frac12}^\pm = \1vp_{\a+\frac12}^\pm\left(\p_{t}z_{\a+\frac{1}{2}} + u_{\aum}^\pm\ \partial_{x}z_{\a+\frac{1}{2}} - (u^z_{\a+\frac{1}{2}})^{\pm}\right).
		\end{equation}
	\end{subequations}
	Note that condition \eqref{eq:boundCond_surf} gives the mass transference at the top of the mixture layer $z=z_{N+1/2}(t,x)$, i.e. the exchange with the upper fluid domain,
	\begin{equation}\label{eq:gn12nuf}
	G_{f,N+1/2} = -\mathcal{V}_{f} \qquad\mbox{and}\qquad G_{s,N+1/2} = 0.
	\end{equation}
	Now, after some straightforward calculations we get the approximation of the deviatoric tensor at the interface $\G_{\a+\frac12}$ using \eqref{eq:jumpCond_mom} as in \cite{fernandezNieto:2016}:
	\begin{equation*}
	\label{eq:aprox_T_a12_s}
	\begin{array}{ll}
	\bT_{\varepsilon,s,\a+\frac{1}{2}}^{\pm}\ \vn_{\a+\frac{1}{2}}= & \displaystyle \widetilde{\bT}_{\varepsilon,s,\a+\frac{1}{2}} \ \vn_{\a+\frac{1}{2}}
	\displaystyle \pm \frac{1}{2} \dfrac{\r_s G_{s,\a+\frac{1}{2}}}{\sqrt{1+\abs{\p\!_{x}z_{\a+\frac{1}{2}}}^2 }}\ \left[\vv\right]_{|_{\G_{s,\a+\frac{1}{2}}}},
	\end{array}
	\end{equation*}
	where $\widetilde{\bT}_{\varepsilon,s,\a+\frac{1}{2}}$ is an approximation of $\big(\eta_s D_{\varepsilon}(\vv_{\a})\big)_{|\Gamma_{\a+\frac{1}{2}}}$, defined by
	\begin{equation*} \label{eq:D_gorro}
	\widetilde{\bT}_{\varepsilon,s,\a+\frac{1}{2}} = \eta_{s,\aum} \widetilde{D}_{\varepsilon,s,\a +\frac12}=\dfrac12\eta_{s,\aum}
	\left(
	\begin{array}{ccc}
	\displaystyle 2 \varepsilon^2 \p_x \left( \frac{v_{\a+\frac{1}{2}}^+ + v_{\a+\frac{1}{2}}^-}{2} \right) &\quad &
	\widetilde{D}_{\varepsilon,s,\a+\frac12}^{xz}
	\\
	&&\\
	\left( \widetilde{D}_{\varepsilon,s,\a+\frac{1}{2}}^{xz} \right)' & \quad & 2 \, \QsV_{,\a+\frac{1}{2}} \\
	\end{array}
	\right),
	\end{equation*}
	with
	$$
	\widetilde{D}_{\varepsilon,s,\a+\frac{1}{2}}^{xz}=
	\varepsilon^2 \p\!_{x} \left( \dfrac{(v^z_{\a+\frac{1}{2}})^+ + (v^z_{a+\frac{1}{2}})^-}{2} \right)  + \QsH_{,\a+\frac{1}{2}}.
	$$
	As in previous works, $\Qs=(\QsH,\QsV)$ is an approximation of the derivatives in $z$ of the velocity
	\begin{equation*} \label{eq:def_q}
	\Qs-\partial_z \vv=0, \qquad \textrm{at the interfaces.} 
	\end{equation*}
	Analogously, we can define for the fluid phase $\widetilde{\bT}_{\varepsilon,{f_m},\a+\frac{1}{2}}$, $\widetilde{D}_{\varepsilon,{f_m},\a +\frac12}$ and $\Q_{,\a+\frac{1}{2}}$
	and get the approximation of the deviatoric tensor of the fluid phase:
	\begin{equation*}
	\label{eq:aprox_T_a12_f}
	\begin{array}{ll}
	\bT_{\varepsilon,{f_m},\a+\frac{1}{2}}^{\pm}\ \vn_{\a+\frac{1}{2}}= & \displaystyle \widetilde{\bT}_{\varepsilon,{f_m},\a+\frac{1}{2}} \ \vn_{\a+\frac{1}{2}}
	\displaystyle \pm \frac{1}{2} \dfrac{\r_f G_{f,\a+\frac{1}{2}}}{\sqrt{1+\abs{\p\!_{x}z_{\a+\frac{1}{2}}}^2 }}\ \left[\vu\right]_{|_{\G_{f,\a+\frac{1}{2}}}}.
	\end{array}
	\end{equation*}
	
	\bigskip
	
	\noindent Finally, we need to approximate $\eta_{s,\a+\frac12}$, the viscosity coefficient \eqref{eq:regularisation}, up to first order in $\varepsilon$. We consider
	\begin{equation*} \label{eq:approx_norm_Du_o1}
	\|D(\vv)\|_{{\a+\frac{1}{2}}} \approx \dfrac12\abs{\QsH_{,\a+\frac{1}{2}} },
	\end{equation*}
	then, the viscosity coefficient is
	\begin{equation} \label{eq:approx_eta_interfaz}
	\eta_{s,\a+\frac{1}{2}} = \dfrac{\left(\mu(I_{\a+\frac{1}{2}})+\left(\tan\psi\right)_{\a+\frac12}\right) p_{s,\a+\frac{1}{2}}}{\sqrt{\abs{ \QsH_{,\a+\frac{1}{2}} }^2 / 4 + \delta^2}}\, ,
	\end{equation}
	with
	\begin{equation*}\label{velalfa}
	\QsH_{,\a+\frac{1}{2}}= \frac{v_{\a+1}-v_{\a}}{h_{\a+\frac{1}{2}}}, \quad \quad \mbox{for} \quad \alpha=1, \dots,N-1, \quad
	\end{equation*}
	and $h_{\a+\frac{1}{2}}$ the distance between the midpoints of layers $\a$ and $\a+1$. For the particular case $\a=0$ we take
	\begin{equation}\label{velb}
	\QsH_{,\frac{1}{2}}=\frac{\lambda v_1}{h_1},
	\end{equation}
	with $\lambda$ depending on the friction condition at the bottom, $\lambda=1$ (friction), $\lambda=2$ (no slip). $p_{s,\a+\frac{1}{2}}$ is the solid pressure at the interface $z_{\a+\frac12}$, whose expression will be given later, and
	\begin{equation} \label{eq:def_I_interfaz}
	I_{\a+\frac{1}{2}} = \frac{\eta_f \abs{ \QsH_{,\a+\frac{1}{2}} }}{p_{s,\a+\frac{1}{2}}},\,  \quad \mbox{for} \quad \alpha=0, \dots, N-1.
	\end{equation}
	Finally, thanks to \eqref{eq:boundCond_surf} we trivially have $\eta_{s,N+1/2}=0$.

	\subsubsection*{Vertical velocities}
	The procedure to recover the vertical velocities is the same as in \cite{fernandezNieto:2014}. In order to compute the vertical velocity for the solid phase, we use both the solid mass transference term and the dilatancy closure equation. Firstly, since there is not transference of solid mass at the bottom level,  we have $G_{s,1/2} = 0$, i.e.,  we get
	\begin{subequations}
		\label{eq:vert_vel_sol}
		\begin{equation*}
		\label{eq:vert_vel_sol_a}
		(v^z_{\frac12})^+ = v_1 \partial_x b+\partial_t b.
		\end{equation*}
		Now, integrating the dilatancy equation \eqref{eq:nondim_dil_3D}  between $z_{\a-\frac12}$ and $z\in\left(z_{\a-\frac12},z_{\a+\frac12}\right)$ we obtain that
		\begin{equation*}
		\label{eq:vert_vel_sol_b}
		v^z_{\a}(z) \;=\; (v^z_{\a-\frac{1}{2}})^+ \;-\; (z-z_{\a-\frac{1}{2}})\left(\p\!_{x}v_{\a} - \dfrac{1}{\varepsilon}\Phi_\a\right),
		\end{equation*}
		and from the solid mass transference term \eqref{eq:G_s} we get
		\begin{equation*}
		\label{eq:vert_vel_sol_c}
		(v^z_{\a+\frac{1}{2}})^{+} \;=\; \dfrac{1}{\vp_{\a+1}}\Big( \left(\vp_{\a+1} - \vpa\right)\p_t z_{\a+\frac12} + (\vp_{\a+1} u_{\a+1} - \vpa u_{\a})\;\p\!_{x}z_{\a+\frac{1}{2}} \;+\; \vpa (v^z_{\a+\frac{1}{2}})^{-}\Big),
		\end{equation*}
		with
		\begin{equation*}
		\label{eq:vert_vel_sol_d}
		(v^z_{\a+\frac{1}{2}})^{-} =(v^z_{\a-\frac{1}{2}})^{+} - h_{\a}\left(\p\!_{x}v_{\a} - \dfrac{1}{\varepsilon}\Phi_\a\right).
		\end{equation*}
	\end{subequations}
	
	Computing the vertical velocity of the fluid phase is possible by using the incompressibility of the mixture
	\begin{equation}
	\label{eq:incomp_mixt}
	\g\cdot\left( \vp\vv + \1vp\vu\right) = 0.
	\end{equation}
	Analogously to the procedure above, we can obtain the vertical velocity for the fluid phase using \eqref{eq:incomp_mixt}, \eqref{eq:G_f} and \eqref{eq:vert_vel_fl_b}, resulting:
	\begin{subequations}
		\label{eq:vert_vel_fl}
		\begin{equation}
		\label{eq:vert_vel_fl_a}
		(u^z_{\frac12})^+ = u_1 \partial_x b+\partial_t b,
		\end{equation}
		\begin{equation}
		\label{eq:vert_vel_fl_b}
		u^z_{\a}(z) \;=\; (u^z_{\a-\frac{1}{2}})^+ \,-\, \dfrac{\left(z-z_{\a-\frac12}\right)}{\v1pa} \bigg(  \p\!_{x} \left(\vpa v_{\a}+\v1pa u_{\a}\right) - \vpa\p_x v_\a + \dfrac{1}{\varepsilon}\vpa\Phi_\a\bigg) ,
		\end{equation}
		\begin{equation}
		\label{eq:vert_vel_fl_c}
		\begin{array}{ll}
		(u^z_{\a+\frac{1}{2}})^{+} = &\dfrac{1}{\left(1-\vp_{\a+1}\right)}\Big( \left(\vp_{\a} - \vp_{\a+1}\right)\p_t z_{\a+\frac12} \\
		& +\, \big(\left(1-\vp_{\a+1}\right) u_{\a+1} - \v1pa u_{\a}\big)\p\!_{x}z_{\a+\frac{1}{2}}+\; \v1pa (u^z_{\a+\frac{1}{2}})^{-}\Big),
		\end{array}
		\end{equation}
		and
		\begin{equation}
		\label{eq:vert_vel_fl_d}
		(u^z_{\a+\frac{1}{2}})^{-} =(u^z_{\a-\frac{1}{2}})^{+} - \dfrac{h_{\a}}{\v1pa}\bigg(\p\!_{x} \left(\vpa v_{\a}+\v1pa u_{\a}\right) - \vpa\p_x v_\a + \dfrac{1}{\varepsilon}\vpa\Phi_\a\bigg).
		\end{equation}
	\end{subequations}

	\subsubsection*{Explicit expression for fluid and solid pressures: $p_{f_m},p_s$}
	From the nondimensional vertical momentum equation \eqref{eq:nondim_jackson_3D_f} we obtain up to first order that
	\begin{equation}
	\label{eq:nodim_pres_fluid}
	\p_z p_{f_m} = \vp\p_z p_{f_m} - \varepsilon^{k+2} \b (u^z-v^z) - \dfrac{\r_f\1vp}{Fr^2}.
	\end{equation}
	By integrating this expression from $z$ to $b+h$ (thanks to the continuity of the dynamic fluid pressure), we obtain
	\begin{subequations}\label{eq:pres_fluid}
		\begin{equation}
		\label{eq:pres_fluid_a}
		p_{{f_m},\a}(z) = \dfrac{\r_f}{Fr^2}\left(b+h+h_f-z\right) \,+\, p_{f,\a}^e,
		\end{equation}
		where
		\begin{equation}
		\label{eq:pres_fluid_e}
		p_{f,\a}^e = p_{f,\a+\frac12}^e  \,+\, \varepsilon^{k+2} \dint_{z}^{z_{\a+\frac12}} \dfrac{\b}{\1vp} \left(u^z-v^z\right)\,dz',
		\end{equation}
		with
		\begin{equation}
		\label{eq:pres_fluid_e_a12}
		p_{f,\a+\frac12}^e = p_{f,\a+1}^{e}\left(z=z_{\a+\frac12}\right) = \varepsilon^{k+2}\dsum_{\gamma = \a+1}^N \dint_{z_{\gamma-\frac12}}^{z_{\gamma+\frac12}} \dfrac{\b}{\1vp} \left(u^z-v^z\right)\,dz'.
		\end{equation}
	\end{subequations}
	
	Once we have the fluid pressure, we can obtain an explicit expression for the solid one. From the vertical momentum equation \eqref{eq:nondim_jackson_3D_e} for the solid phase we have
	\begin{equation*}
	\label{eq:nodim_pres_solid}
	\p_z p_s = -\vp\p_z p_{f_m} + \varepsilon^{k+2} \b (u^z-v^z) - \dfrac{\r_s\vp}{Fr^2}.
	\end{equation*}
	Using \eqref{eq:nodim_pres_fluid} in the previous equation and integrating from $z$ to $z_{\a+\frac12}$ we get
	\begin{equation*}
	\label{eq:pres_solid_prev}
	p_{s,\a}(z) = p_{s,\a+\frac12} + \dfrac{\vpa\left(\r_s-\r_f\right)}{Fr^2}\left(z_{\a+\frac12}-z\right) \,-\,  \varepsilon^{k+2} \dint_{z}^{z_{\a+\frac12}} \dfrac{\b}{\1vp} \left(u^z-v^z\right)\,dz',
	\end{equation*}
	where
	\begin{equation}
	\label{eq:pres_solid_inter}
	p_{s,\a+\frac12} = \dfrac{\left(\r_s-\r_f\right)}{Fr^2}\dsum_{\gamma=\a+1}^N h_\gamma\vp_{\gamma} \,-\, p_{f,\a+\frac12}^e.
	\end{equation}
	We can write
	\begin{equation}
	\label{eq:pres_solid}
	p_{s,\a}(z) = \dfrac{\left(\r_s-\r_f\right)}{Fr^2}\dsum_{\gamma = \a+1}^N \vp_\gamma h_\gamma \,+\,\dfrac{\vpa\left(\r_s-\r_f\right)}{Fr^2}\left(z_{\a+\frac12}-z\right)  \,-\, p_{f,\a}^e, \end{equation}
	where $p_{f,\a}^e$ is given by \eqref{eq:pres_fluid_e}. The computation of the excess pore pressure is subtle. It is carried out using the dilatancy equation, the mass transference terms for the solid and the fluid, and the fact that we obtain the incompressibility of the mixture with velocity $\left(\vp \vv + \1vp\vu\right)$ from the mass equations in \eqref{eq:jackson_matricial}. We also assume that we are in the case where the friction between the solid and the fluid phase is strong ($k=-1$ in \eqref{eq:beta_friccion}). This leads to (details in appendix \ref{Apend_B})
	\begin{subequations}
		\label{eq:press_fluid_e_exp}
		\begin{equation}
		p_{f,\a}^e(z) = p_{f,\a+\frac12}^e - \b_\a\dfrac{h_\a^2 - \left(z-z_{\a-\frac12}\right)^2}{2\v1pa^2}\Phi_\a \,-\, \dfrac{\b_\a}{\vpa\v1pa^2}\left(z_{\a+\frac12}-z\right)\mathlarger{\dsum}_{\gamma=1}^{\a-1} \vp_{\gamma}h_{\gamma}\Phi_{\gamma} ,
		\end{equation}
		with
		\begin{equation}\label{eq:pres_solid_exc_inter}
		p_{f,\a+\frac12}^e = p_{f,\a+1}^{e}\left(z=z_{\a+\frac12}\right) = \mathlarger{\mathlarger{\dsum}}_{\xi=\a+1}^N\dfrac{-\b_\xi\,h_{\xi}}{\vp_\xi\left(1-\vp_\xi\right)^2}\,\left(\mathlarger{\dsum}_{\gamma = 1}^{\xi-1}\vp_{\gamma}h_\gamma\Phi_\gamma \,+\, \dfrac{\vp_{\xi}h_\xi\Phi_\xi}{2}\right).
		\end{equation}
	\end{subequations}

	\subsection{Final model}\label{se:FinalModel}
	We derive the final model  for the mixture by looking for a particular weak solution of system \eqref{eq:jackson_matricial} with the procedure introduced in \cite{fernandezNieto:2014}, which was applied to the case of a variable viscosity in \cite{fernandezNieto:2016}.   As we mentioned before, for the upper layer we adopt the model derived in \cite{bouchut:2016}.\\ 
	Finally, the resulting system must be written in dimensional variables. As conclusion, the final model reads 
	, for $\smash\a = 1,...,N$,
	
	\begin{subequations}
		\label{eq:FinalModel}
		\begin{align}[left = \empheqlbrace\,]
		&\partial_t h_f+\partial_x(h_f u_f)=-G_{f,N+\frac12},\label{eq:massf}\\
		&\rho_f\bigg(\partial_t \left(h_fu_f\right)+ \p_x\left(h_fu_f^2 \right) + g\cos\theta h_f\partial_x(b+\wtd{b}+h+h_f) \bigg) \nonumber\\
		&\quad =  -\dfrac12 \r_f G_{f,N+\frac12}(u_f+u) - k_i(u_f-u),\label{eq:momf}\\[4mm]
		&l_{\a}\vpa \bigg( \p_{t} h + \p\!_{x}(hv_{\a}) \bigg) = G_{s,\a+\frac{1}{2}} - G_{s,\a-\frac{1}{2}} + \vpa l_\a h\Phi_\a, \label{eq:FinalModel_a}\\
		&l_{\a} \bigg( \p_{t}\left( h\vpa\right) + \p\!_{x}(h\vpa v_{\a}) \bigg) = G_{s,\a+\frac{1}{2}} - G_{s,\a-\frac{1}{2}}, \label{eq:FinalModel_b}\\
		&l_{\a} \bigg( \p_{t}\left( h\v1pa\right) + \p\!_{x}(h\v1pa u_{\a}) \bigg) = G_{f,\a+\frac{1}{2}} - G_{f,\a-\frac{1}{2}}, \label{eq:FinalModel_c}\\[4mm]
		&l_{\a} \bigg( \r_s\p_{t}\left(h\vpa v_{\a}\right) \;+\; \r_s\p\!_{x}\left(h\vpa v_{\a}^2\right) \, +  g\cos\theta h\vpa\,\p_{x} \left(\r_s\left( b+\tilde b+h\right) + \r_f h_f\right)	\nonumber\\	
		&\quad  + \,g \cos\theta h\left(\r_s-\r_f\right)\left(\dsum_{\gamma=\a+1}^N h_\gamma\p_x\vp_{\gamma} + \dfrac{h_\a}{2}\p_x\vpa\right)\nonumber\\
		&\quad + g \cos\theta h\left(\r_s-\r_f\right)\dsum_{\gamma=\a+1}^N \left(\vp_\gamma-\vpa\right) \p_x h_\gamma - h\v1pa\overline{\p_xp_{f,\a}^e}\bigg) \, = l_\a h\overline{\b_\a}\left(u_\a - v_\a\right) \nonumber\\
		&\quad +\, K_{s,\a-\frac{1}{2}} - K_{s,\a+\frac{1}{2}}\; +\;\dfrac{1}{2} \r_s G_{s,\a+\frac{1}{2}}\left(v_{\a+1} + v_{\a}\right) \;-\; \dfrac{1}{2} \r_s G_{s,\a-\frac{1}{2}}\left(v_{\a} + v_{\a-1}\right),\label{eq:FinalModel_d} \\[4mm]
		&l_{\a} \bigg( \r_f\p_{t}\left(h\v1pa u_{\a}\right) \;+\; \r_f\p\!_{x}\left(h\v1pa u_{\a}^2\right)   \nonumber\\
		&\quad+ \r_f g\cos\theta h\v1pa\,\p_{x} \left(b+\tilde b+h+h_f\right) + \,  h\v1pa\overline{\p_xp_{f,\a}^e}\bigg) \, = -\,l_\a h\overline{\b_\a}\left(u_\a - v_\a\right)\nonumber\\
		&\quad +\, K_{f,\a-\frac{1}{2}} - K_{f,\a+\frac{1}{2}}\; +\;\dfrac{1}{2} \r_f G_{f,\a+\frac{1}{2}}\left(u_{\a+1} + u_{\a}\right) \;-\; \dfrac{1}{2} \r_f G_{f,\a-\frac{1}{2}}\left(u_{\a} + u_{\a-1}\right), \label{eq:FinalModel_e}
		\end{align}
		\\
		where
		\begin{equation}
		\label{eq:b_pe_gorro}
		\overline{\b_\a} = \dfrac{1}{h_\a}\dint_{z_{\a-\frac12}}^{z_{\a-\frac12}}\b_\a\,dz\qquad \mbox{and}\quad \overline{\p_x p_{f,\a}^e} = \dfrac{1}{h_\a}\dint_{z_{\a-\frac12}}^{z_{\a-\frac12}}\p_x p_{f,\a}^e\,dz,
		\end{equation}
		where $\b_\a$ and $p_{f,\a}^e$ are given by \eqref{eq:beta} and \eqref{eq:press_fluid_e_exp}, respectively. The mass transference terms $G_{s,\a+\frac{1}{2}},G_{f,\a+\frac{1}{2}}$ are given by \eqref{eq:G}, and
		\begin{equation}\label{eq:FinalModel_f}
		\begin{array}{lll}
		K_{s,\a+\frac{1}{2}} = - \dfrac12\eta_{s,\a+\frac{1}{2}}(\QsH_{,\a+\frac{1}{2}})\, \QsH_{,\a+\frac{1}{2}},  & &
		K_{s,\frac{1}{2}} = -\r_s g\,\cos\theta\,h\,\left(\mu\big(I_{\frac12}\big)+\left(\tan\psi\right)_{1/2}\right)\dfrac{v_{1}}{\abs{v_{1}}},\\
		K_{f,\a+\frac{1}{2}} = - \eta_{f}\, \QH_{,\a+\frac{1}{2}}, &
		& K_{f,\frac{1}{2}} = -k_b\,u_1,
		\end{array}
		\end{equation}
	\end{subequations}
	for $\QsH_{,\a+\frac{1}{2}}, \ \eta_{s,\a+\frac{1}{2}}$ defined in (\ref{eq:approx_eta_interfaz})-(\ref{eq:def_I_interfaz}).
	
	For the specific case of the bottom and mixture interface, we introduce the following definitions:
		$$
		u_0=0;\qquad v_0=(1-\lambda)v_1; \qquad u_{N+1}=u_f;\qquad v_{N+1}=0;
		$$
		notice that the definition of $v_0$ comes from \eqref{velb}. Moreover, the frictions between the mixture and the upper fluid layer are defined by
		$$
		K_{s,N+\frac12} = 0, \qquad \mbox{and}\qquad K_{f,N+\frac12} = k_i(u_f-u).
		$$
		At the mixture interface we also remind that the transference terms are given in \eqref{eq:gn12nuf}.
	
	Note that although the previous system has $5N$ equations, only $5N-1$ are linearly independent, therefore we have a system with $5N-1$ equations and unknowns, which are:$$( h, \{\vpa, h u_{\a},h v_\a\}_{\a=1,\dots,N}, \{G_{f,\a+\frac{1}{2}},G_{s,\a+\frac{1}{2}} \}_{\a=1, \dots,N-1} ).$$
	However, an explicit expression can be found  for the mass transference terms, as detailed below.
	
	\subsubsection{Explicit mass transference terms}
	In order to obtain explicit expressions for the solid mass transference term, we start by combining the averaged mass equations for the solid and fluid phase \eqref{eq:FinalModel_b}-\eqref{eq:FinalModel_c}, giving
	
	\begin{equation}
	\label{eq:eqmasa_aux}
	l_{\a} \bigg( \p_{t}h + \p\!_{x}\Big(h\big(\vpa v_{\a} + \v1pa u_\a\big)\Big) \bigg) = \left(G_{s,\a+\frac{1}{2}} + G_{f,\a+\frac{1}{2}}\right) - \left(G_{s,\a-\frac{1}{2}}+G_{f,\a-\frac{1}{2}}\right).
	\end{equation}
	Next, denoting
	$$
	\wtd{U_\a} =\vpa v_{\a} + \v1pa u_\a,
	$$
	and summing up the previous equation from $\a=1,\dots,N$ we obtain
	\begin{equation}
	\label{eq:eqmasa_h}
	\p_{t}h + \p\!_{x}\bigg(\mathlarger{\mathlarger{\dsum}}_{\a=1}^{N}l_\a h\,\wtd{U_\a} \bigg) = G_{s,N+\frac12}+G_{f,N+\frac12}.
	\end{equation}
	Here we have used that there is no mass transference at the bottom, i.e. $G_{s,\frac12} = G_{f,\frac12} = 0$. The transference terms $G_{s,N+\frac12}$ and $G_{f,N+\frac12}$ are defined in \eqref{eq:gn12nuf}.
	Using previous equation in the averaged dilatancy equation \eqref{eq:FinalModel_a}, it leads to
	$$
	l_\a\,\vpa\Bigg(-\p\!_{x}\bigg(\mathlarger{\mathlarger{\dsum}}_{\gamma=1}^{N}l_\gamma h\,\wtd{U_\gamma}\bigg) + \p_x\left(hv_\a\right) - h\Phi_\a \Bigg) + l_\a\vpa\left(G_{s,N+\frac12}+G_{f,N+\frac12}\right) = G_{s,\a+\frac{1}{2}} -G_{s,\a-\frac{1}{2}},
	$$
	and summing up the previous equation from $\a=1,\dots,N$, it gives
	\begin{subequations}\label{eq:GsGf}
		\begin{equation}
		\label{eq:Gs}
		G_{s,\a+\frac{1}{2}} = \mathlarger{\mathlarger{\dsum}}_{\b=1}^\a  l_\b\,\vp_\b \Bigg(\p\!_{x}\bigg(hv_\b-\mathlarger{\dsum}_{\gamma=1}^{N}l_\gamma h\,\wtd{U_\gamma} \bigg) - h\Phi_\b \Bigg) +  \left(G_{s,N+\frac12}+G_{f,N+\frac12}\right)\dsum_{\b=1}^\a l_\b\vp_{\b}.
		\end{equation}
		
		Once we have an explicit expression for the solid mass transference term, we can use it in order to obtain the fluid mass transference term. We embbed \eqref{eq:eqmasa_h} in \eqref{eq:eqmasa_aux}, and sum it up from $\a=1,\dots,N$, it gives
		\begin{equation}
		\label{eq:Gf}
		G_{f,\a+\frac{1}{2}} = \mathlarger{\mathlarger{\dsum}}_{\b=1}^\a  l_\b \p\!_{x}\bigg(h\,\wtd{U_\b}-\mathlarger{\dsum}_{\gamma=1}^{N}l_\gamma h\,\wtd{U_\gamma} \bigg) \,-\,G_{s,\a+\frac{1}{2}} +\left(G_{s,N+\frac12}+G_{f,N+\frac12}\right)\dsum_{\b=1}^\a l_\b.
		\end{equation}
	\end{subequations}
	
	\bigskip
	
	Taking into account those expressions for the mass transference terms, system \eqref{eq:FinalModel} can be rewritten as a system with $3N+1$ equations and unknowns. The unknowns are now the total height ($h$), the solid volume fraction ($\{\varphi_{\alpha} \}_{\alpha=1}^N$) and the solid and fluid velocities ($\{v_{\alpha}\}_{\alpha=1}^N$, $\{u_{\alpha}\}_{\alpha=1}^N$). Thus, the equations of the final system are (\ref{eq:FinalModel_b}), (\ref{eq:FinalModel_d}), (\ref{eq:FinalModel_e}) and (\ref{eq:eqmasa_h}).

	Note that we need a closure relation for the fluid and solid mass transference at the free surface introduced in \eqref{eq:gn12nuf}. Moreover, some important properties of the model, as the mass conservation, depend on how these terms are defined. This is discussed in subsection \ref{se:closure_GN12}.
	
	Even in simple configurations, the numerical approximation of our model involves important difficulties, namely when approximating the pressure. For this reason, in the following we study the particular case of immersed uniform flows, for which numerical tests will be presented in section \ref{se:numericalTest}.
	
	\subsection{Particular configuration: immersed uniform flows}	
\begin{figure}[!ht]
		\begin{center}
			\includegraphics[width=0.5\textwidth]{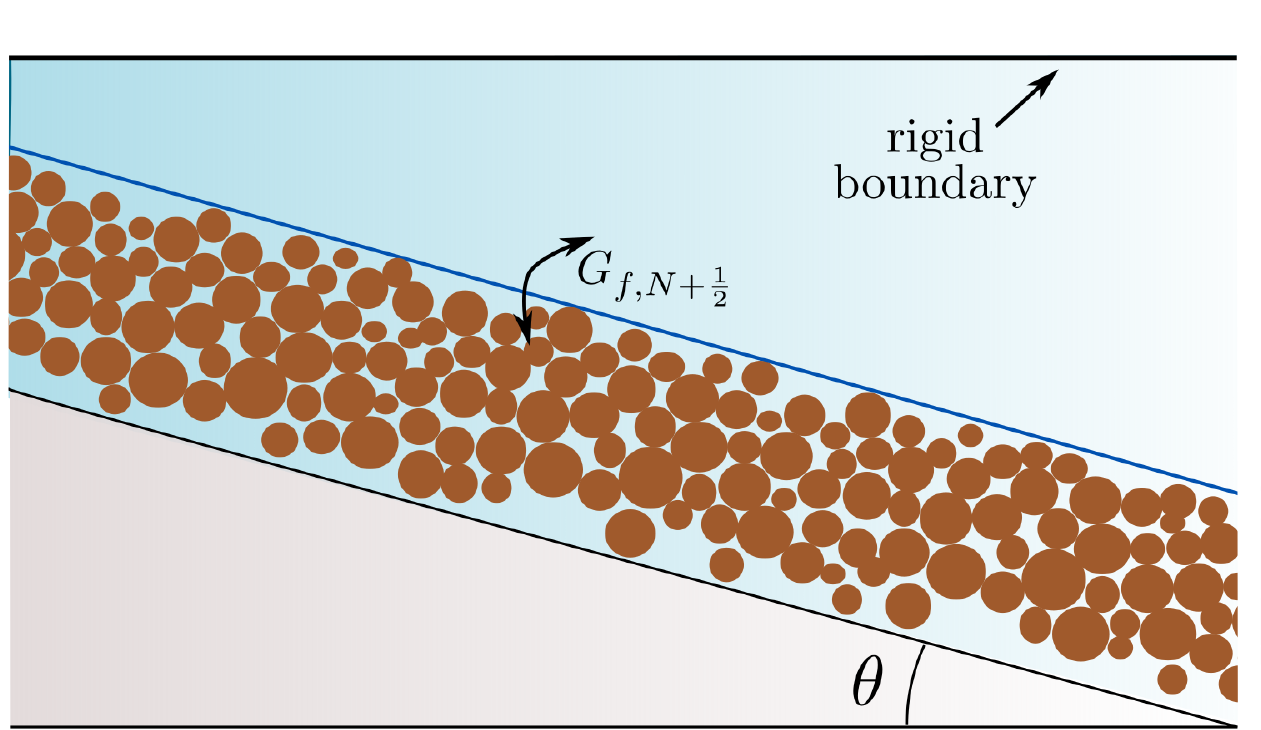}
		\end{center}
		\caption{\label{fig:domain_immersed} \it{Sketch of the domain in the immersed configuration and flat free surface.}}
	\end{figure}

	We consider the analogous case to \cite{bouchut:2016}, i.e. an immersed flow with rigid top boundary at the surface (see Figure \ref{fig:domain_immersed}), in order to compare with their results. To this aim, we rewrite the derivative of the fluid depth taking into account that we have a rigid lid boundary ($b=0$). This implies that
	$$
	\p_x\left(x\tan\theta + h + h_f\right)=0.
	$$
	Since we consider the case of a uniform flow we also have that $\p_x h = 0$. Then, from the previous equation  we deduce that $\p_x h_f = -\tan\theta$.\\  
	Taking these hypotheses into account in \eqref{eq:FinalModel_a}-\eqref{eq:FinalModel_f} with $u_f=0$, $k_i=0$, we obtain the following simplified system
	\begin{subequations}
		\label{eq:FinalModel_uniform}
		\begin{align}[left = \empheqlbrace\,]
		&\p_t h = G_{s,N+\frac12} + G_{f,N+\frac12}, \label{eq:FinalModel_uniform_a}\\[4mm]
		&\p_{t}\vpa = -\,\vpa\Phi_\a, \label{eq:FinalModel_uniform_b}\\[4mm]
		&l_{\a}\,\r_s\,\p_{t}\left(h\vpa v_{\a}\right) = l_\a\,h\,\overline{\b_\a}\left(u_\a - v_{\a}\right)\,-\,\left(\r_s-\r_f\right)g\sin\theta l_\a h \vpa  \nonumber\\[2mm]
		&\quad+ K_{s,\a-\frac{1}{2}} - K_{s,\a+\frac{1}{2}}\; +\;\dfrac12\r_s G_{s,\a+\frac{1}{2}}\left(v_{\a+1} + v_{\a}\right) \;-\; \dfrac12\r_s G_{s,\a-\frac{1}{2}}\left(v_{\a} + v_{\a-1}\right),\label{eq:FinalModel_uniform_c} \\[4mm]
		&l_{\a}\,\r_f\,\p_{t}\left(h\left(1-\vpa\right) u_{\a}\right) = \,-\,l_\a\,h\,\overline{\b_\a}\left(u_\a - v_{\a}\right)\nonumber\\[2mm]
		&\quad+ K_{f,\a-\frac{1}{2}} - K_{f,\a+\frac{1}{2}}\; +\;\dfrac12\r_f G_{f,\a+\frac{1}{2}}\left(u_{\a+1} + u_{\a}\right) \;-\; \dfrac12\r_f G_{f,\a-\frac{1}{2}}\left(u_{\a} + u_{\a-1}\right),\label{eq:FinalModel_uniform_d}
		\end{align}
	\end{subequations}
	Note that \eqref{eq:FinalModel_uniform_b} is deduced from \eqref{eq:FinalModel_a} and \eqref{eq:FinalModel_b}. We also consider the following definitions: for the drag coefficient (see \eqref{eq:beta})
	$$
	\overline{\b_\a} = \dfrac{150\vpa^2}{d_s^2\left(1-\vpa\right)}\eta_f,
	$$
	and from \eqref{eq:dil_funcion}-\eqref{eq:phieq} the dilatancy function is
	\begin{equation*}
	\Phi_\a = \dot{\gamma}_\a K\left(\vp_\a-\vp_{c,\a}^{eq}\right),\quad\text{with}\quad \vp^{eq}_{c,\a} = \vp_c^{stat} - K_2\,I_\a,\quad\mbox{with}\quad
	\dot{\gamma}_\a = \dfrac{\abs{v_\a-v_{\a-1}}}{h_\a}.
	\end{equation*}
	Note that in the definition of  $\dot{\gamma}_\a$ we set a backward approximation because this is the one that gives us better results, namely at the bottom where the friction boundary condition must be taken into account.
	
		The system \eqref{eq:FinalModel_uniform} is written in terms of the mass transference at the mixture interface, appearing in the first equation \eqref{eq:FinalModel_uniform_a}. An important property of this system that is shared with the system proposed in \cite{bouchut:2016} is that it preserves the solid mass in the mixture domain under the definitions given in \eqref{eq:gn12nuf}, that is, $G_{s,N+\frac12} = 0$ and $G_{f,N+\frac12}$. 
		In this case we get	
		\begin{equation*}
		\label{eq:h_gN12}
		\p_t h = G_{f,N+\frac12},
		\end{equation*}
		hence, thanks to \eqref{eq:eqmasa_h} the total solid mass is preserved:
		$$
		\p_t\left(\dsum_{\b=1}^N h_\b\vp_{\b}\right) = 0.
		$$
		
		In order to give an expression for the mass transference terms in the multilayer domain, we first use that $G_{s,N+\frac12} = 0$ in \eqref{eq:Gs} to obtain the fluid mass transference at the free surface, so finally we have
		\begin{equation}\label{eq:transmasPGM}
		G_{s,N+\frac12} = 0,\qquad G_{f,N+\frac12} = \dfrac{\dsum_{\gamma = 1}^{N}l_\gamma\vp_{\gamma}h\Phi_{\gamma}}{\dsum_{\gamma = 1}^{N}l_\gamma\vp_{\gamma}}.
		\end{equation}
		With this definition, the solid and fluid mass transference terms are written as
		\begin{equation}\label{eq:transmasPGM2}
		G_{s,\a+\frac12} = \dsum_{\b = 1}^{\a}l_\b\vp_{\b}\left(G_{f,N+\frac12}-h\Phi_{\b}\right) \quad\mbox{and}\quad G_{f,\a+\frac12} = G_{f,N+\frac12}\dsum_{\b=1}^\a l_\b - G_{s,\a+\frac12},
		\end{equation} for $\a=1,\dots,N-1$.
\medskip

		From now on we refer to model \eqref{eq:FinalModel_uniform}, \eqref{eq:transmasPGM}, \eqref{eq:transmasPGM2} as PGM model ({\it Preserving Granular Mass}).
				Actually, this model is the multilayer extension of the model proposed by Bouchut {\it et al.} \cite{bouchut:2016}, where the thickness of the mixture ($h_m$ in their notation) is not preserved. In fact, their case is the single-layer case of the proposed model. In that case, as authors assume there is no solid mass transference between these layers, therefore $G_{s,3/2} = 0$ and from \eqref{eq:Gs} the fluid mass transference at the free surface remains
		$$
		G_{f,3/2} = h_1\Phi_1 + \p_x\big( h_1 \left(1-\vp_1\right) \left(u_1-v_1\right) \big),
		$$
		which matches with $-\mathcal{V}_f$ as defined in \cite{bouchut:2016}, where the minus sign comes from the fact that they consider the upward normal vectors while we work here with downward normal vectors.

	\bigskip
	
	This section is organized as follows:  in subsection \ref{se:closure_GN12} we propose a multilayer extension preserving the mixture height (as the Pailha and Pouliquen model \cite{pailha:2009}). An analytical solution is deduced in subsection \ref{se:analytical_uniform}, by neglecting  the friction between the phases. In subsection \ref{se:pared} we deduce an analytical solution that includes the effect of the side walls friction for confined flows, which plays an important role in the case of granular flows, namely in laboratory experiments.  The numerical approximation of the resulting system (\ref{eq:FinalModel_uniform}) is a difficult task even in this simple configuration of a uniform flow. The main difficulty is the approximation of the excess pore fluid pressure term $p^e_{f,\a}$, which is necessary to define $I_{\alpha}$. Its numerical approximation is a key point of the numerical scheme, which is detailed in subsection \ref{se:approx_pressure}.

		\subsubsection{A multilayer model preserving the mixture height}\label{se:closure_GN12}
		As discussed in \cite{bouchut:2016} one of the main difference between their model and the one of Pailha and Pouliquen is that this last one preserves the total height of the mixture and not the total mass. This implies different steady state solutions for each model. In order to perform a similar comparison in the numerical tests we show a multilayer model that preserves the height instead of the granular mass in the mixture. \\\\
		
		In order to preserve the total height of the mixture layer it is required that the solid mass leaving the mixture domain equals the fluid mass entering the domain, i.e., $G_{s,N+\frac12} = -G_{f,N+\frac12}$ in the model \eqref{eq:FinalModel_uniform}.
				 Note that in this case, the model in the upper layer should also take into account the granular phase.
				 				 
		From \eqref{eq:FinalModel_uniform_a} we obtain that
		$$\p_t h = 0,$$
		and \eqref{eq:FinalModel_b} and \eqref{eq:Gs}, under the uniform flow assumption, lead to
		\begin{equation}\label{eq:transmasPH}
		\p_t\left(\dsum_{\b=1}^N h_\b\vp_{\b}\right) = G_{s,N+\frac12},\quad\mbox{with}\quad G_{s,N+\frac12} = -\dsum_{\b=1}^N l_\b h \vp_\b\Phi_{\b}.
		\end{equation}
Therefore, the solid mass of the considered mixture may be not preserved at the domain. The solid and fluid mass transference terms \eqref{eq:GsGf} are
		\begin{equation}\label{eq:transmasPH2}
				G_{s,\a+\frac{1}{2}} = -\dsum_{\b=1}^\a  l_\b\,h\,\vp_\b \Phi_\b\qquad\mbox{and}\qquad G_{f,\a+\frac{1}{2}} = -\,G_{s,\a+\frac{1}{2}}.
		\end{equation} for $\a=1,\dots,N$.
		From now on we refer to model \eqref{eq:FinalModel_uniform}, \eqref{eq:transmasPH}, \eqref{eq:transmasPH2} as the PH model ({\it Preserving Height}).

	\subsubsection{Analytical solution}\label{se:analytical_uniform}
	The analytical solution for a uniform immersed flow can be obtained with an analogous procedure to the case of a granular flow, i.e, Bagnold flow (see e.g. \cite{lagree:2011}).
	
	Firstly, since we focus on steady uniform flows, where $G_{s,N+\frac12} =G_{f,N+\frac12} = 0$, once the final height is computed (which will be different in each model), the analytical solution will be the same for both models analysed in previous section.
	
	Starting from system \eqref{eq:jackson_3D} in the immersed configuration ($\p_x h_f = -\tan\theta$),
	and assuming a steady uniform flow ($\tan\psi = 0$), we obtain
	\begin{equation}
	\label{eq:analitica_ini}
	\begin{array}{c}
	\p_z \T_{s}^{xz} = \left(\r_s-\r_f\right) \vp(z)\,g\sin\theta,\\[2mm]
	\p_z p_{s} = -\left(\r_s-\r_f\right) \vp(z)\,g\cos\theta.
	\end{array}
	\end{equation}
	where we have not considered the friction between the phases, that is, $\beta(\vu-\vv)$. The influence of this term will be studied in the numerical tests (see section \ref{se:nosidewalls}). Moreover we consider
	$$
	\T_s(z) = \mu(I(z))\,p_s(z),\quad \mu(I(z)) = \mu_s + K_1\,I(z), \quad\mbox{and}\quad\vp(z) = \vp_c^{eq}(z) = \vp_c^{stat} - K_2\,I(z).
	$$
	Integrating from $z>b$ to the free surface $b+h$ we obtain
	\begin{equation}
	\label{eq:analitica_ini_3}
	\begin{array}{c}
	p_s(z) = \left(\rho_s-\rho_f\right)g\cos\theta \dint_{z}^{b+h} \vp(s) ds,
	\end{array}
	\end{equation}
	\begin{equation*}
	\label{eq:analitica_ini_4}
	\T_{s}^{xz} (z)   = -\left(\r_s-\r_f\right)\,g\sin\theta \dint_{z}^{b+h} \vp(s) ds  .
	\end{equation*}
	Using that
	$$\T_s^{xz}(z) \,=\, \mu(I(z))\,p_s(z),$$
	we get
	$$-\left(\r_s-\r_f\right)\,g\sin\theta \dint_{z}^{b+h} \vp(s) ds \,=\, \mu(I(z))\left(\rho_s-\rho_f\right)g\cos\theta  \dint_{z}^{b+h} \vp(s) ds,$$
	so
	$$( \mu(I(z))+\tan\theta )\dint_{z}^{b+h} \vp(s) ds \,=0.$$
	This leads to a constant friction coefficient $\mu(I) = -\tan\theta$. Thus, we obtain that the inertial number $I(\mu_s,\, \Delta \mu=\mu_2-\mu_s,\tan\theta,K_1)$ is constant
	$$
	I = \dfrac{-\tan\theta - \mu_s}{\Delta \mu + \mu_s + \tan\theta} I_0,
	\quad\mbox{or}\quad I = \dfrac{-\tan\theta - \mu_s}{K_1},
	$$
	depending of the considered rheology \eqref{eq:muI_1} or \eqref{eq:muI_2}, and therefore the solid volume fraction
	$$
	\vp = \vp_{c}^{eq} = \vp_c^{stat} - K_2 I,
	$$
	is also constant. In addition, using the definition of the inertial number \eqref{eq:def_I} we get an expression for $\p_z v$. Integrating from $z=b$ to $0<z<b+h$ and using the no-slip condition at the bottom, the velocity profile reads:
	$$
	v(z) = \dfrac{\left(\r_s-\r_f\right)\,I\,\vp\cos\theta }{2\eta_f}\bigg(\left(b+h\right)^2-\left(b+h-z\right)^2\bigg).
	$$
	
	
	\subsubsection{Side walls friction for confined flows: analytical solution}\label{se:pared}
	To model side wall friction the following additional viscous term is considered at the momentum solid phase (see \cite{fernandezNieto:2018} and \cite{martin:2017}),
	\begin{equation}
	\label{eq:fric_lateral}
	-\dfrac{2}{W} \mu_w p_s(z) \dfrac{v}{|v|},
	\end{equation}
	where $W$ is the width of the domain and $\mu_w$ the friction coefficient at the lateral walls.

	In order to obtain an analytical solution in this case we start again from \eqref{eq:analitica_ini}, where the side walls friction term is added
	\begin{equation*}
	\label{eq:analitica_ini_2}
	\begin{array}{rcl}
	\p_z \T_{s}^{xz} (z) - \dfrac{2}{W} \mu_w p_s(z) &=& \left(\r_s-\r_f\right) \vp(z)\,g\sin\theta,\\[2mm]
	\p_z p_{s}(z) &=& -\left(\r_s-\r_s\right) \vp(z)\,g\cos\theta.
	\end{array}
	\end{equation*}
	where we have also assumed that $sign(\p_z v) = sign(v) = 1$.
	Following the same procedure as in the previous section we get
	\begin{equation}
	\label{eq:analitica_ini_5}
	\T_{s}^{xz} (z) + \dfrac{2}{W} \mu_w \left(\rho_s-\rho_f\right)g\cos\theta \dint_{z}^{b+h} \left(\dint_s^{b+h}\vp(\xi) d\xi\right) ds  = -\left(\r_s-\r_f\right)\,g\sin\theta \dint_{z}^{b+h} \vp(s) ds  .
	\end{equation}
	and writing \eqref{eq:analitica_ini_5} in terms of $\vp(z)$, we obtain
	\begin{equation}\label{eq:analitica_int}
	\begin{array}{rcl}
	\left(\mu_s + K_1 \dfrac{\vp_c^{stat} - \vp(z)}{K_2}\right) \dint_{z}^{b+h} \vp(s) ds &+&  \dfrac{2}{W} \mu_w \dint_{z}^{b+h} \left(\dint_s^{b+h}\vp(\xi) d\xi\right) ds \\[6mm]
	&=& -\tan\theta \dint_z^{b+h}\vp(s) ds.
	\end{array}
	\end{equation}
	Now, deriving the previous equation, it leads to the integro-differential equation
	$$
	\vp'(z) \dfrac{K_1}{K_2} \dint_z^{b+h}\vp(s) ds \,+\, \left(\mu_s + K_1 \dfrac{\vp_c^{stat} - \vp(z)}{K_2}\right)\vp(z) \,+\, \dfrac{2}{W} \mu_w \dint_{z}^{b+h} \vp(s) ds\,=\,-\tan\theta\vp(z).
	$$
	Considering the variable
	$$
	F(z) = \dint_{z}^{b+h} \vp(s) ds,
	$$
	we obtain an initial value problem with two differential equations
	$$
	\left(
	\begin{array}{c}
	F(z)  \\[2mm]
	\vp(z)
	\end{array}
	\right)' = \left(
	\begin{array}{c}
	-\vp(z)  \\[2mm]
	-\dfrac{K_2}{K_1 F(z)}\bigg(\left(\mu_s + K_1 \dfrac{\vp_c^{stat} - \vp(z)}{K_2}\right)\vp(z) + \dfrac{2}{W} \mu_w F(z) + \tan\theta \vp(z)\bigg)
	\end{array}
	\right),
	$$
	where
	$$
	\begin{array}{l}
	F(\b+h) = \epsilon, \quad \epsilon \ll 1, \\[2mm]
	\vp(b+h) = \vp_c^{stat} + \dfrac{K_1}{K_2}\left(\tan\theta + \mu_s\right).
	\end{array}$$
	This system is solved using a numerical ODE solver, obtaining the concentration $\vp(z)$ and $F(z)$. The pressure is computed using \eqref{eq:analitica_ini_3} and the inertial number is $I(z) = \left(\vp_c^{stat} - \vp(z)\right)/K_2$. Using the definition of $I$ we find the analytical solid velocity solving the differential equation
	$$
	v'(z) = \dfrac{p_s(z)I(z)}{\eta_f}, \quad\mbox{with}\quad v(0) = 0.
	$$
	\begin{remark}\label{re:phi_constant}
		Note that if we take $(\mu_w = 0)$ in \eqref{eq:analitica_int} we recover the previous case without side walls friction (subsection \ref{se:analytical_uniform}). In particular, it leads to $\mu(I) = -\tan\theta$, which implies that both, the inertial number $I$ and the solid volume fraction $\vp$, are constant.
	\end{remark}
	\begin{remark}\label{re:phi_pared}
		In the case of uniform flows for dry granular flows and a hydrostatic pressure it is equivalent to consider the additional viscous term (\ref{eq:fric_lateral}) or the following modification of the friction coefficient (see \cite{jop:2005})
		\begin{equation*}\label{eq:muI_mod}
		\overline{\mu(I)} = \mu(I) + \mu_w\dfrac{b+h -z}{W}.
		\end{equation*}
		We show in what follows that for the case considered in this paper, where we deal with a two-phase flow, this modification of the friction coefficient, under the same hypothesis, is not equivalent but it is a second order approximation of it.

		It is enough to use the trapezoidal rule to  approximate the double integral in \eqref{eq:analitica_int}. We obtain, up to second order,
		$$
		\left(\mu(I) + \mu_w\dfrac{\left(b+h-z\right)}{W} + \tan\theta\right) \dint_{z}^{b+h} \vp(s) ds =0.
		$$
		Note that as a consequence we can deduce an explicit expression that is a second order approximation of the analytical solution. We obtain the following approximation,
		\begin{equation*}
		\begin{array}{rl}
		\vp = \vp_{c}^{eq} =& \vp_c^{stat} + C_2\big(\left(b+h-z\right) - \left(b+h-h^*\right)\big),\\[4mm]	
		\dfrac{v(z)}{C_1} =& \dfrac{C_2}{4} \big(\left(b+h-z\right)^4 - \left(b+h-h^*\right)^4\big) \\[3mm]
		-& \dfrac{2C_2\left(b+h-h^*\right)-\vp_c^{stat}}{3}\big(\left(b+h-z\right)^3 - \left(b+h-h^*\right)^3\big)\\[3mm]
		-& \dfrac{\vp_c^{eq}-C_2\left(b+h-h^*\right)}{2}\left(b+h-h^*\right)\big(\left(b+h-z\right)^2 - \left(b+h-h^*\right)^2\big),
		\end{array}
		\end{equation*}
		for $z>h^*$, where
		$$h^* = (b+h) - \dfrac{\tan\theta-\mu_s}{\mu_w}W,\qquad
		C_1 = \dfrac{\mu_w\left(\rho_s-\rho_f\right)g\cos\theta}{K_1\eta_f W}, \quad\mbox{and}\quad C_2 = \dfrac{K_2\mu_w}{K_1 W}.
		$$
		Note that in this case, $h^*$ is the position of the static/flowing interface.
	\end{remark}

	\subsubsection{Numerical approximation: computation of the linearized solid pressure} \label{se:approx_pressure}
	We approximate the system of ODEs for uniform flows \eqref{eq:FinalModel_uniform} (for the two cases described in section \ref{se:closure_GN12}) using a standard first order semi-implicit scheme to go from a time $t^n$ to $t^{n+1} = t_n + \Delta t$ as follows:
	\begin{itemize}
		\item The first equation is trivially discretized as $h^{n+1} = h^n + \Delta t G^n_{f,N+\frac12}$.
		\item For the equation \eqref{eq:FinalModel_uniform_b} for the solid volume fraction, we consider an explicit approximation
		$$
		\vpa^{n+1} = \vpa^n - \Delta t \vpa^n \Phi_a^n.
		$$
		We can also use an implicit discretization or the exact solution of \eqref{eq:FinalModel_uniform_b}. We have chosen the explicit one for the sake of simplicity, since similar results are obtained with all of them.
		
		\item For the momentum equations \eqref{eq:FinalModel_uniform_c} and \eqref{eq:FinalModel_uniform_d} we use a semi-implicit scheme in order to avoid the restriction on the time step that the viscous terms usually involve:
		$$
		\begin{array}{lll}
		\r_s l_\a h^{n+1}\vpa^{n+1} v_\a^{n+1} &=& \r_s l_\a h^{n}\vpa^{n} v_\a^{n} + l_\a h^n \overline{\b_\a}^n\left(u_\a^n-v_\a^n\right) - \left(\r_s-\r_f\right)\, l_\a h^n\,\vpa^n\,g\sin\theta \\[3mm]
		&+& \r_s G_{s,\a+\frac12}^n v_{\a+\frac12}^n - \r_s G_{s,\a-\frac12}^n v_{\a-\frac12}^n\\[3mm]
		&+& \dfrac12\eta_{s,\a+\frac12}^n\dfrac{v_{\a+1}^{n+1}-v_{\a}^{n+1}}{l_{\a+\frac12}h^n} - \dfrac12\eta_{s,\a-\frac12}^n\dfrac{v_{\a}^{n+1}-v_{\a-1}^{n+1}}{l_{\a-\frac12}h^n} ,
		\end{array}
		$$
		$$
		\begin{array}{lll}
		\r_f l_\a h^{n+1}\v1pa^{n+1} u_\a^{n+1} &=& \r_f l_\a h^n\v1pa^n u_\a^n - l_\a h^n \overline{\b_\a}^n\left(u_\a^n-v_\a^n\right)\\[3mm]
		&+& \r_f G_{f,\a+\frac12}^n u_{\a+\frac12}^n - \r_f G_{s,\a-\frac12}^n u_{\a-\frac12}^n\\[3mm]
		&+& \eta_{f}\dfrac{u_{\a+1}^{n+1}-u_{\a}^{n+1}}{l_{\a+\frac12}h^n} - \eta_{f}\dfrac{u_{\a}^{n+1}-u_{\a-1}^{n+1}}{l_{\a-\frac12}h^n} .
		\end{array}
		$$
		For $\a=1$ and $\a=N$ the definitions of $K_{\a+\frac12}$ must be taken into account. The new values are obtained solving a tridiagonal linear system with $N$ equations and unknowns.
		
		If the side walls friction \eqref{eq:fric_lateral} is considered, the term
		\begin{equation*}
		\label{eq:fric_lateral_multi}
		-\dfrac{2}{W} \mu_w h^n_\a p^n_{s,\a} \dfrac{v^{n+1}_\a}{\sqrt{|v^n_\a|^2 + \delta^2}}
		\end{equation*}
		must be added to the right hand side of the momentum equation for the solid phase, where $p_{s,\a} = \left(p_{s,\a+\frac12}+p_{s,\a-\frac12}\right)/2$ is the pressure in the midpoint of layer $\a$.
		
	\end{itemize}
	
	\noindent Once we have $h^{n+1},\vpa^{n+1}, \left(h\vpa v_\a\right)^{n+1},\left(h\v1pa v_\a\right)^{n+1}$ we find $u_\a^{n+1},v_\a^{n+1}$.
	The difficulty of the scheme comes when computing the solid pressure at the interfaces $p_{s,\a-\frac12}^n$ at time $t^n$ (see \eqref{eq:pres_solid_inter} and \eqref{eq:pres_solid_exc_inter}):
	\begin{subequations}\label{eq:eq_pressure}
		\begin{equation}
		p_{s,\a-\frac12}^n = \left(\r_s-\r_f\right)g\cos\theta\mathlarger{\mathlarger{\dsum}}_{\gamma=\a}^N h^n_\gamma\vp^n_{\gamma} \,+\, \mathlarger{\mathlarger{\dsum}}_{\xi=\a}^N\dfrac{\overline{\b}^n_\xi\,h^n_{\xi}}{\vp^n_\xi\left(1-\vp^n_\xi\right)^2}\,\left(\mathlarger{\dsum}_{\gamma = 1}^{\xi-1}\vp_{\gamma}^n h^n_\gamma\Phi^n_\gamma \,+\, \dfrac{\vp^n_{\xi}h^n_\xi\Phi^n_\xi}{2}\right),
		\end{equation}
		depending on the dilatation function
		\begin{equation}
		\label{eq:phi_n}
		\Phi_\b^n = \dot{\gamma}^n_\b K\left(\vp_\b^n- \vp_c^{stat} + K_2\, I_\b^n\right),\qquad\mbox{for}\quad\b = 1,\dots,N.
		\end{equation}
		In previous equation the inertial number must be approximated in the middle of the layer. For the sake of simplicity in the computation of $I_\b^n$ in \eqref{eq:phi_n}, 
		we approximate $p_{s,\a}^n$ by the value in the interface below. Thus we take
		\begin{equation}
		I_\b^n = \dfrac{\eta_f\dot{\gamma}^n_\b}{p_{s,\b-\frac12}^n},
		\end{equation}
	\end{subequations}
	Note that this approximation follows the same idea of what it is made in single-layer models, see for example \cite{pailha:2009}. In this paper the basal pressure is used to define the dilatancy relation arguing that the dilatancy at the bottom gives the right order of magnitude of the dilatancy inside the layer. Here, this hypothesis is adopted for each layer $\beta$, thus being $p_{s,\b-\frac12}^n$.

	Therefore, we obtain a nonlinear system with $N$ equations and unknowns (the solid pressures), where the main difficulty is the fact that the system is fully coupled, i.e., each equation depends on all the rest, and this is due to the excess pore pressure. 
	
	The first attempt that we carried out was to linearise the problem embedding into \eqref{eq:phi_n} the pressure at the previous time step $p_{s,\b-\frac12}^{n-1}$. However, due to the abrupt time variations of the excess fluid pressure, namely at short times, this method is inadequate.  We propose to consider a numerical method that is based on an approximation of the function $f(x) = 1/x$ by the firth-order term of its Taylor polynomial. Thus, in \eqref{eq:phi_n} we write
	$$
	\dfrac{1}{p_{s,\b-\frac12}^n}\, = \, \dfrac{2}{p_{s,\b-\frac12}^{n-1}} \,-\, \dfrac{1}{\left(p_{s,\b-\frac12}^{n-1}\right)^2}\,p_{s,\b-\frac12}^{n}+ {\mathcal O} (p_{s,\b-\frac12}^{n}-p_{s,\b-\frac12}^{n-1} )^2,\qquad\mbox{for}\quad\b = 1,\dots,N.
	$$
	Thanks to this approximation, and defining the coefficients
	$$
	A_\a^n = \dfrac{150\vpa^n h_\a^n \eta_f}{d_s^2\left(1-\vpa^n\right)^3},\quad B_\a^n = \vpa^n h_\a^n \dot{\gamma}^n_\a K\left(\vpa^n-\vp_c^{stat}\right),   \quad C^n_\a = 2\dfrac{\vpa^n h^n_\a \left(\dot{\gamma}_\a^n\right)^2 K K_2\eta_f}{p_{s,\a-\frac12}^{n-1}},\quad D_\a^n = \dfrac{C_\a^n}{2p_{s,\a-\frac12}^{n-1}},
	$$
	the previous nonlinear system becomes, up to second order, the following $N\times N$ linear system
	$$
	\left(
	\begin{array}{cccc}
	a_{11}^n & a_{12}^n & \cdots & a_{1N}^n\\
	a_{21}^n & a_{22}^n & \cdots & a_{2N}^n\\
	\cdots & \cdots & \cdots & \cdots\\
	a_{N1}^n & a_{N2}^n & \cdots & a_{NN}^n\\
	\end{array}
	\right)
	\left(
	\begin{array}{c}
	p^n_{s,\frac12} \\
	p^n_{s,\frac32} \\
	\cdots \\
	p^n_{s,N-\frac12}\\
	\end{array}
	\right) =
	\left(
	\begin{array}{c}
	b_1^n \\
	b_2^n \\
	\cdots \\
	b_N^n\\
	\end{array}
	\right),
	$$
	where
	$$
	\begin{array}{lcr}
	a_{ij} = D_j \left(\dsum_{k=j+i}^N A_k +\dfrac{A_j}{2}\right),& \mbox{if}& i<j,\\
	a_{jj} = 1\,+\,D_j \left(\dsum_{k=j+i}^N A_k +\dfrac{A_j}{2}\right),&  & \\
	a_{ij} = D_j \,\dsum_{k=i}^N A_k, & \mbox{if}& i>j,\\
	\end{array}
	$$
	and
	$$
	b_j = \left(\r_s-\r_f\right)g\cos\theta\mathlarger{\mathlarger{\dsum}}_{\xi=j}^N h_\xi \vp_\xi + \mathlarger{\mathlarger{\dsum}}_{\xi=j}^N \left(A_\xi\dsum_{\gamma=1}^{\xi-1}\left(B_\gamma+C_\gamma\right) + \dfrac{A_\xi}{2}\left(B_\xi+C_\xi\right) \right),
	$$
	that can be solve using a direct method.

\section{Numerical tests for uniform flows}\label{se:numericalTest}

In this section we present some numerical results in order to validate our model. In this work we only deal with uniform immersed flows and consider dilatancy effects by starting with initially loose or initially dense configurations. 
With the purpose of  comparing the results of the proposed model with model in \cite{bouchut:2016}, we consider the rheology given by \eqref{eq:muI_2}, i.e., $\mu(I) = \mu_s + K_1\,I$, in the immersed configuration. 

First, we compare the results obtained for the proposed model with the analytical solution with and without side walls friction. For sake of simplicity, we only use in this case the model preserving the total height (PH model). The reason is that it is more simple since the final height is known and therefore we can directly compute the velocity and the solid volume fraction profiles.\\
In a second step, we show that the method proposed to approximate the pressure in the previous section \ref{se:approx_pressure} makes it possible to recover the appropriate root of the non-linear system defined by \eqref{eq:eq_pressure}, for the particular cases $N=1,2$. However, the error made when approximating the pressure increases with the number of layers, so we decide to use the PGM model with only 2 layers to compare with experimental data (see section \ref{sec:choice}). Then, we briefly analyse the influence of the dilatation constant $K$ (see \eqref{eq:phieq}) and compare the results of the 2 layer PGM model with previous depth-averaged single-layer models in the literature, namely the models introduced in \cite{bouchut:2016,pailha:2009} and with laboratory experiments in \cite{pailha:2009}. For this comparison we use the model preserving the total solid mass because this is the only physically relevant model.

\bigskip

For the tests presented in this section we set the same values for the parameters than in \cite{bouchut:2016}, i.e., the ones proposed in \cite{pailha:2009} for low and high fluid viscosity. For each case a dense and a loose initial configuration is simulated.  The physical and rheological parameters that are common for these tests are detailed in Table \ref{tabla_bifasico}. The specific parameters in the high or low viscosity cases and for the loose or dense initial conditions are shown in Table \ref{tabla_bifasico2}. In all the presented tests, we consider no-friction with the bottom for the fluid phase, $k_b = 0$, and a no-slip condition (see \eqref{velb}) for the solid one. At the mixture interface, no friction between the fluid in the mixture layer and the fluid in the upper layer is also assumed, i.e., $k_i=0$.
	
\begin{table}[!htb]
	\begin{center}
		\begin{tabular}{c|c|c|c|c|c|c}
			$\r_s$ (Kg/m$^3$) & $d_s$ ($\mu$m)  & $\mu_s$ & $\vp_c^{stat}$ & $K$ & $K_1$ & $K_2$\\
			\hline			
			2500 & 160 & 0.415 & 0.582 & 4.09 & 90.5 & 25\\
		\end{tabular}
		\caption{\it{Common physical and rheological parameter for the two-phase simulations.}}
		\label{tabla_bifasico}
	\end{center}
\end{table}

\begin{table}[!htb]
	\begin{center}
		\begin{tabular}{c|c|c|c|c|c|c}
			$\eta_f$ case & $\r_f$ (Kg/m$^3$) & $\eta_f$ (Pa/s)  & $\theta$ ($^\circ$) & $h_0$ (mm) & $\vp_0$ (loose) & $\vp_0$ (dense)\\
			\hline			
			High & 1041 & 96$\times 10^{-3}$ & -25 & 4.9 & 0.562 & 0.588 \\
			Low & 1026 & 9.8$\times 10^{-3}$ & -28 & 6.1 & 0.576 & 0.592  \\
		\end{tabular}
		\caption{\it{Parameter for the two-phase simulations depending on the high/low fluid viscosity and loose/dense initial configuration.}}
		\label{tabla_bifasico2}
	\end{center}
\end{table}

\subsection{Comparison with analytical solution}

In this test we focus on steady uniform flows, where $G_{f,N+\frac12} =G_{s,N+\frac12} = 0$. Therefore, once the final height is computed (which will be different from the proposed PGM model and from the PH model described in subsection \ref{se:closure_GN12}), the analytical solution is the same for both models PH model and PGM model. For the sake of simplicity, we only consider in the comparison with the analytical solution the PH model, which is simpler than the PGM model since the height is prescribed by the initial condition.


We simulate a uniform flow initially at rest ($u_\a = v_\a = 0$, for $\a=1,\dots,N$), with height $h_0$, and initial solid volume fraction $\vp_{\a} = \vp_0$ for $\a=1,\dots,N$, until a steady state is reached. 

In the low viscosity case we use $50$ vertical layers and a time step $\Delta t_L=10^{-5}$ while 20 layers are used in the high viscosity case and $\Delta t_H=10^{-6}$. Note that $\Delta t_H = \Delta t_L/10$, i.e., we need a smaller time step for the high viscosity case, and therefore 20 layers are used in that case because the time needed to complete the simulation is huge. But although some differences can be found in the transient regime, the steady solution is the same in both cases ($20$ or more layers).

\subsubsection{No side walls friction} \label{se:nosidewalls}
\begin{figure}[!h]
 	\begin{center}
 		\includegraphics[width=0.49\textwidth]{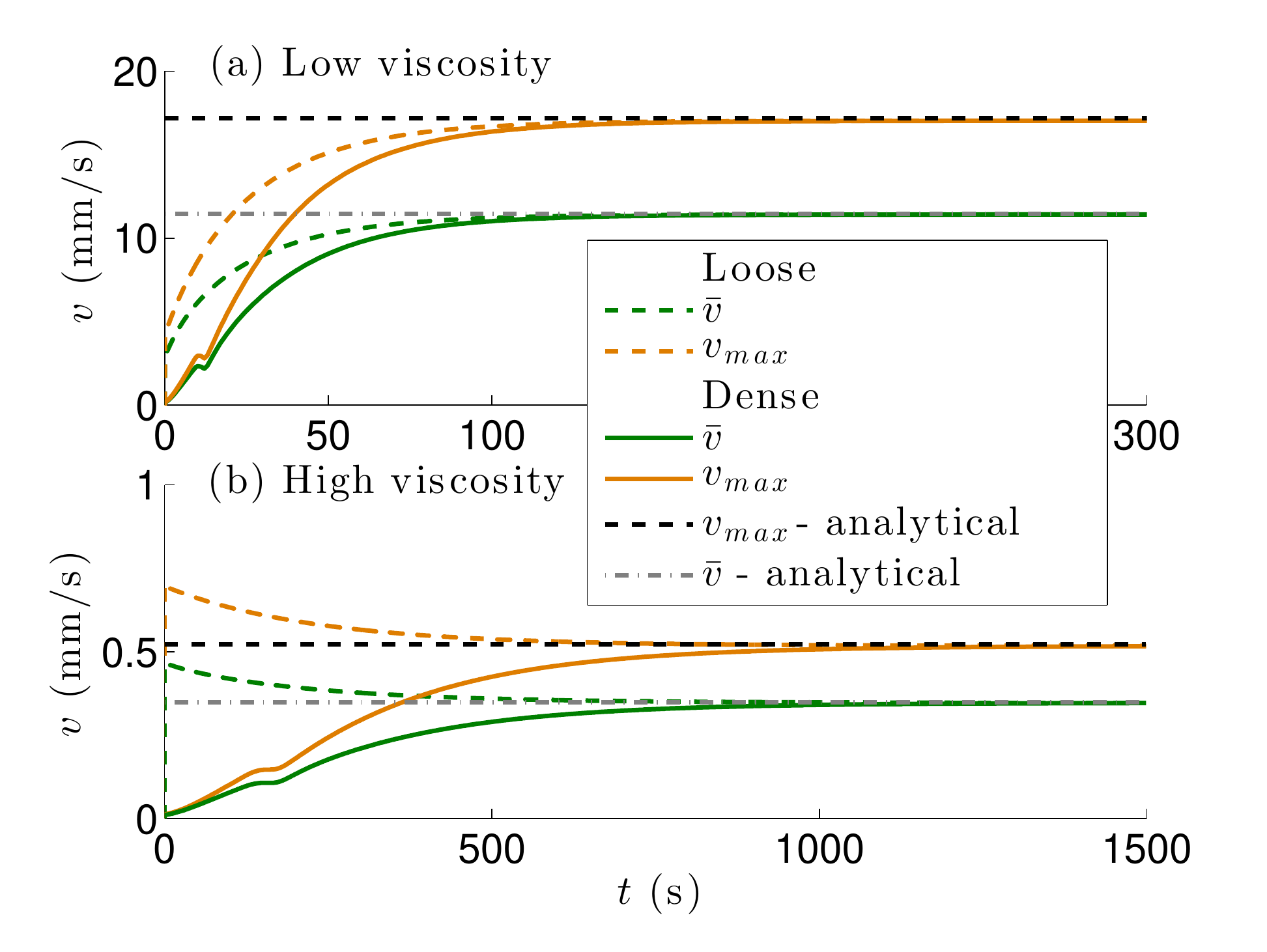}

 	\end{center}
 	\caption{\label{fig:analitica_max_med} \it{Time evolution of the maximum ($v_{max}$, brown lines) and averaged ($\bar{v}$, green lines) velocities in the (a) low and (b) high viscosity case. Dashed (resp. solid) lines are the solution starting from a loose (resp. dense) configuration, and the dash-dotted (resp. dashed) grey line is the analytical averaged (resp. maximum) velocity.}}
 \end{figure}

\begin{figure}[!t]
  	\begin{center}
  		\includegraphics[width=0.9\textwidth]{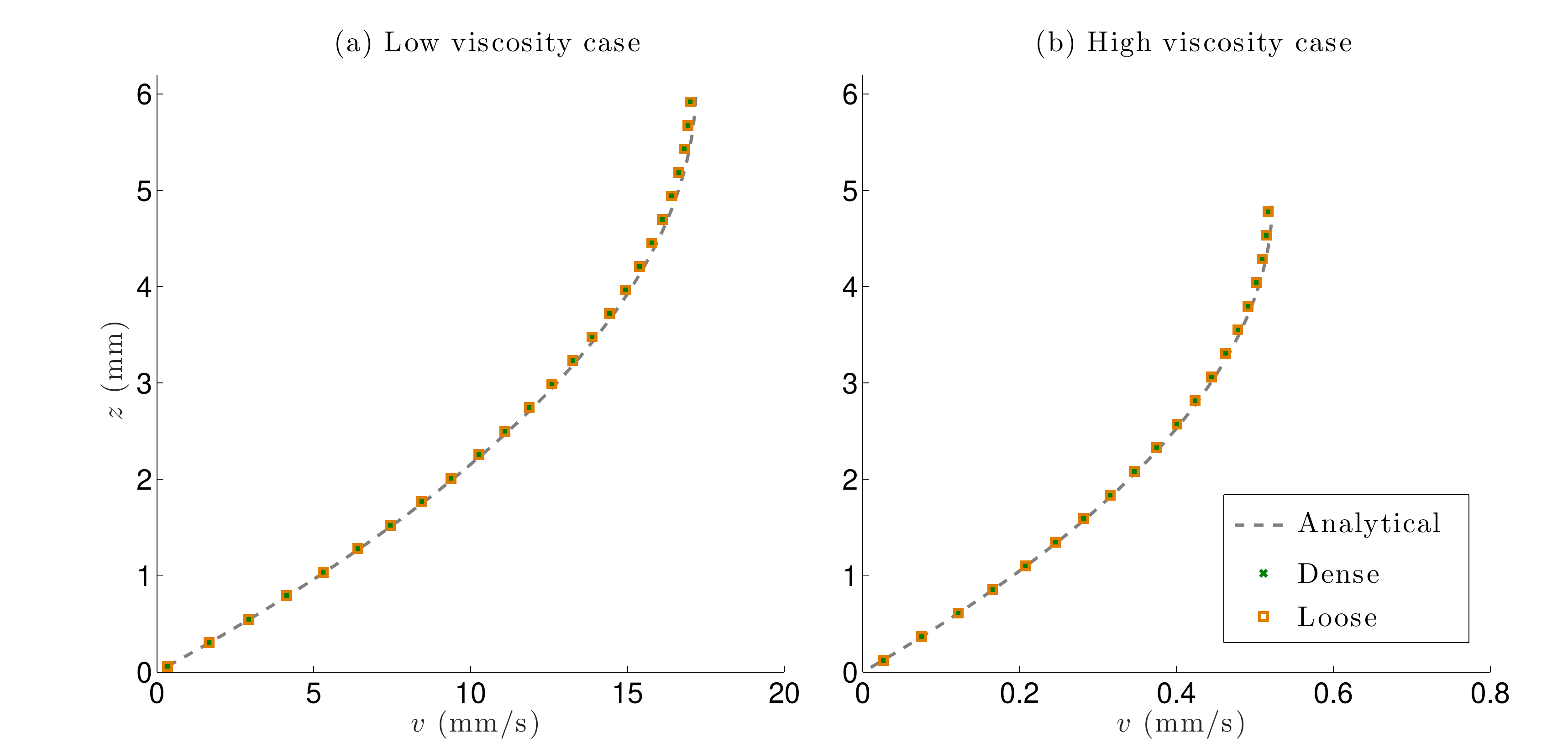}

  	\end{center}
  	\caption{\label{fig:analitica_perfil} \it{Normal profile of downslope velocity  in the (a) low and (b) high viscosity case. Green crosses (resp. brown squares) are the solution in the dense (resp. loose) initial configuration, and the dashed grey line is the analytical profile of velocity.}}
  \end{figure}

Firstly, we compare our results with the analytical solution for a flow without side walls friction. In Figure \ref{fig:analitica_max_med} we show that the model approximate properly, with loose and dense initial configuration, the analytical maximum velocity (the one on the top of the flow) and the averaged velocity $\bar{v} = \sum_{\a=1}^{N}l_\a v_\a$.
 We can see that the convergence in time is much slower in the high viscosity case than in the low viscosity case. This fact, together with the use of a smaller time step justify the use of $20$ vertical layers instead of $50$ as in the low viscosity case. The main advantage of multilayer models is that the vertical structure of the flow can be recovered. We show these profiles and the analytical profile of velocity in Figure \ref{fig:analitica_perfil}, where the steady states reached for the loose/dense initial configuration are the same. In these figures we only show the velocity of the solid phase because we obtain that both, the solid and the fluid velocities, are equal with a relative error of order $10^{-4}$.

\begin{figure}[!t]
	\begin{center}
		\includegraphics[width=0.49\textwidth]{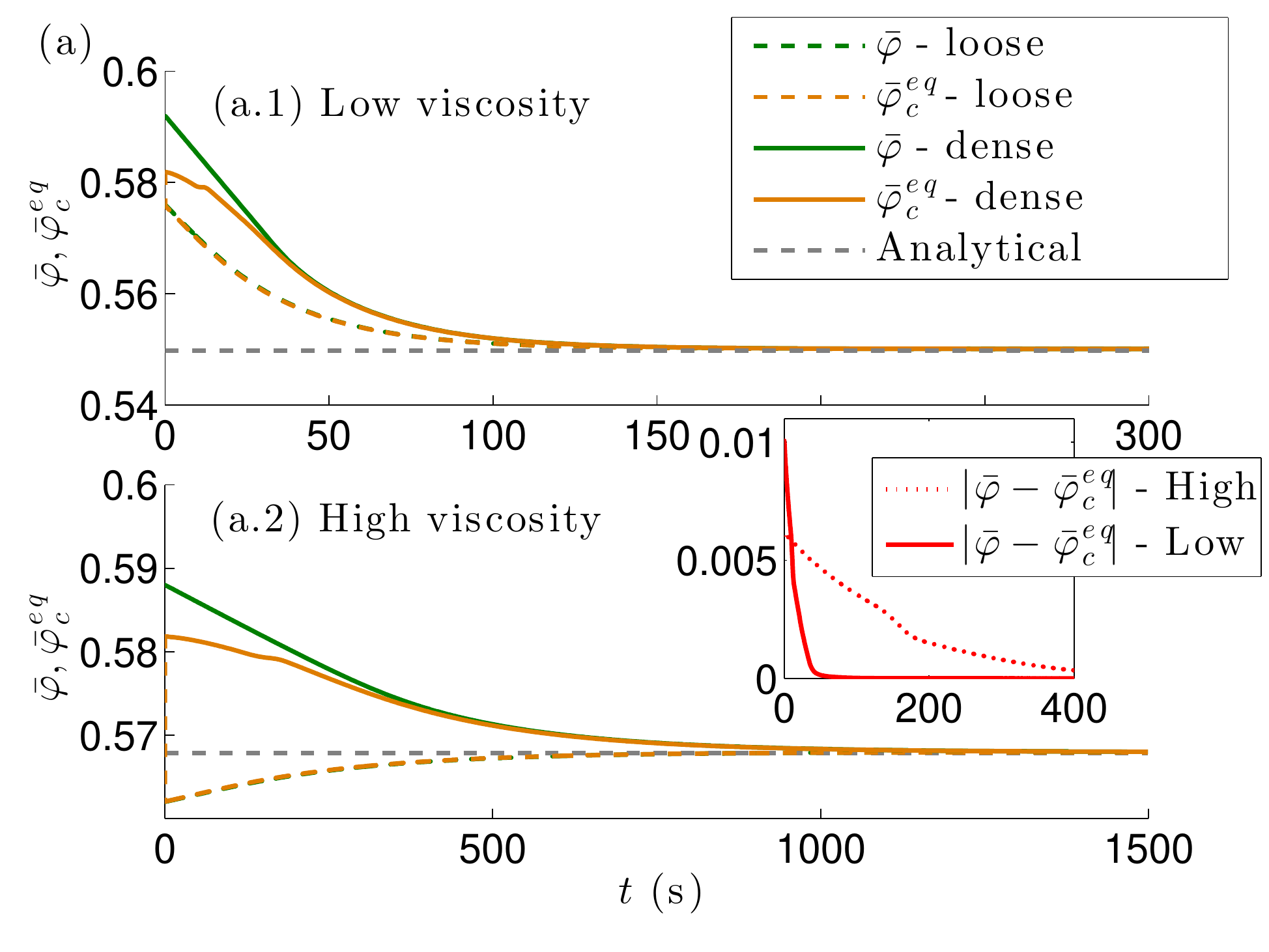}
		\includegraphics[width=0.49\textwidth]{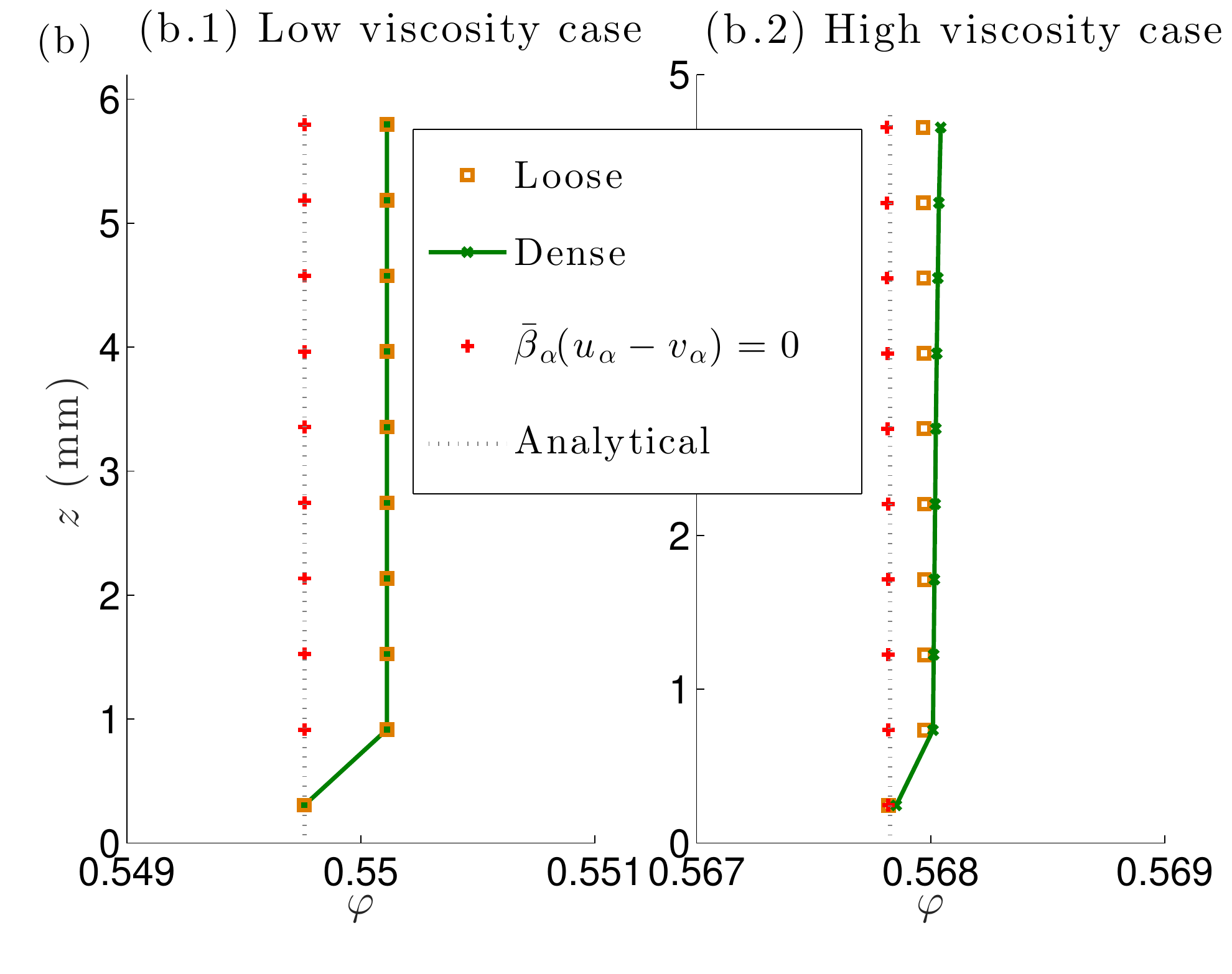}
	\end{center}
	\caption{\label{fig:analitica_concentracion} \it{(a) Time evolution of the averaged solid volume fraction (green lines) and equilibrium concentration (brown lines). Dashed (resp. solid) lines are the solution starting from a loose (resp. dense) configuration, and the dashed grey line is the analytical averaged solid volume fraction. The inner figure in (a.2) is the difference $\abs{\bar{\vp} - \bar{\vp}_c^{eq}}$ in the dense initial configuration.\newline
	(b) Normal profiles of solid volume fraction in the (b.1) low and (b.2) high viscosity case. The dash-dotted grey line is the analytical normal profile of the solid volume fraction. Red-crossed are the simulation removing the term $\overline{\b_\a}\left(u_\a-v_\a\right)$ from the momentum equation. }}
		
	\end{figure}
In Figure \ref{fig:analitica_concentracion}(a) we show the convergence in time of the averaged solid volume fraction ($\bar{\vp} = \sum_{\a=1}^{N}l_\a\vp_\a$) and the equilibrium concentration ($\bar{\vp}_c^{eq} = \sum_{\a=1}^{N}l_\a\vp_{c,\a}^{eq}$) towards its steady state.

\begin{figure}[!t]
	\begin{center}
		\includegraphics[width=0.49\textwidth]{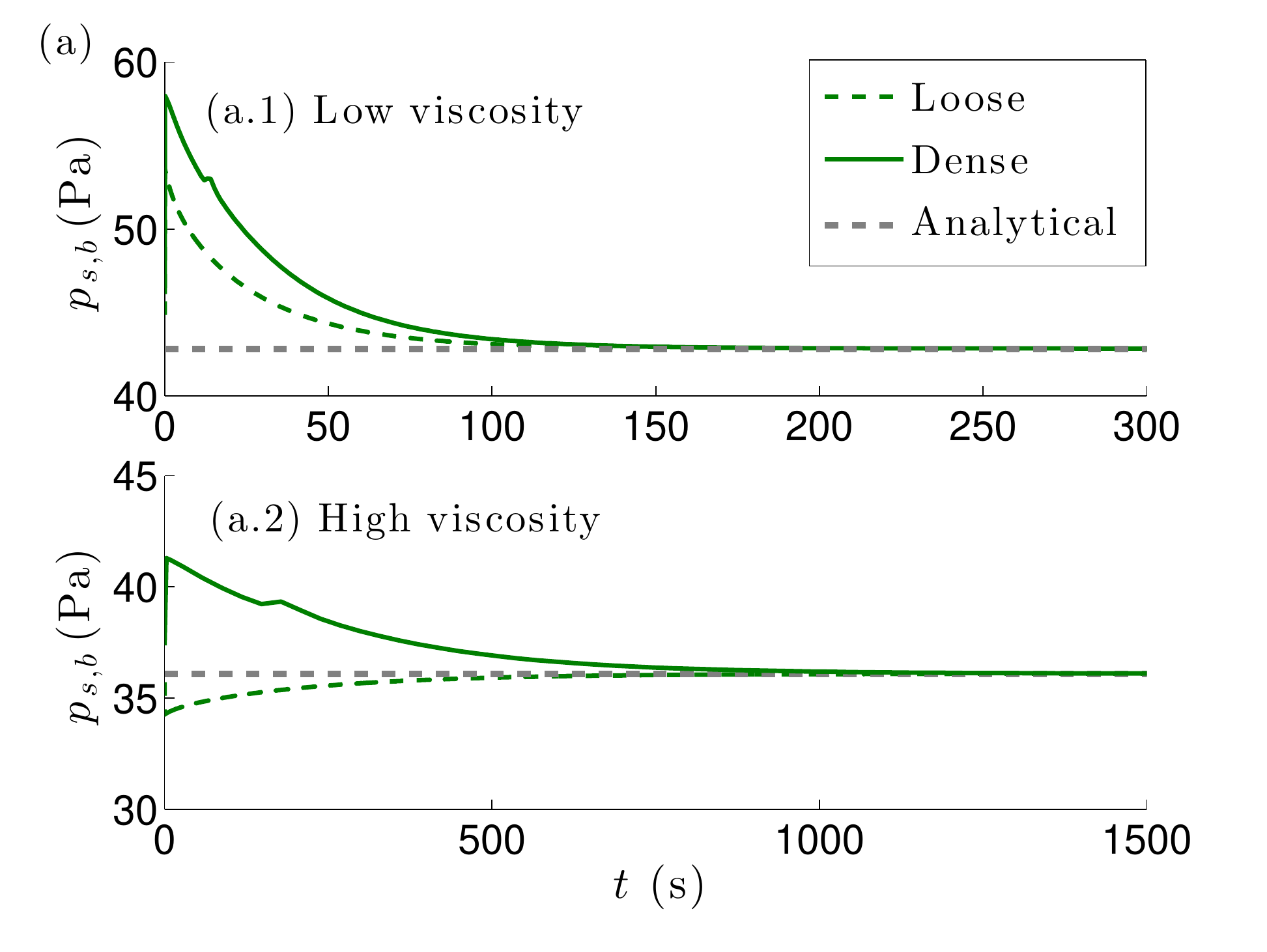}
		\includegraphics[width=0.49\textwidth]{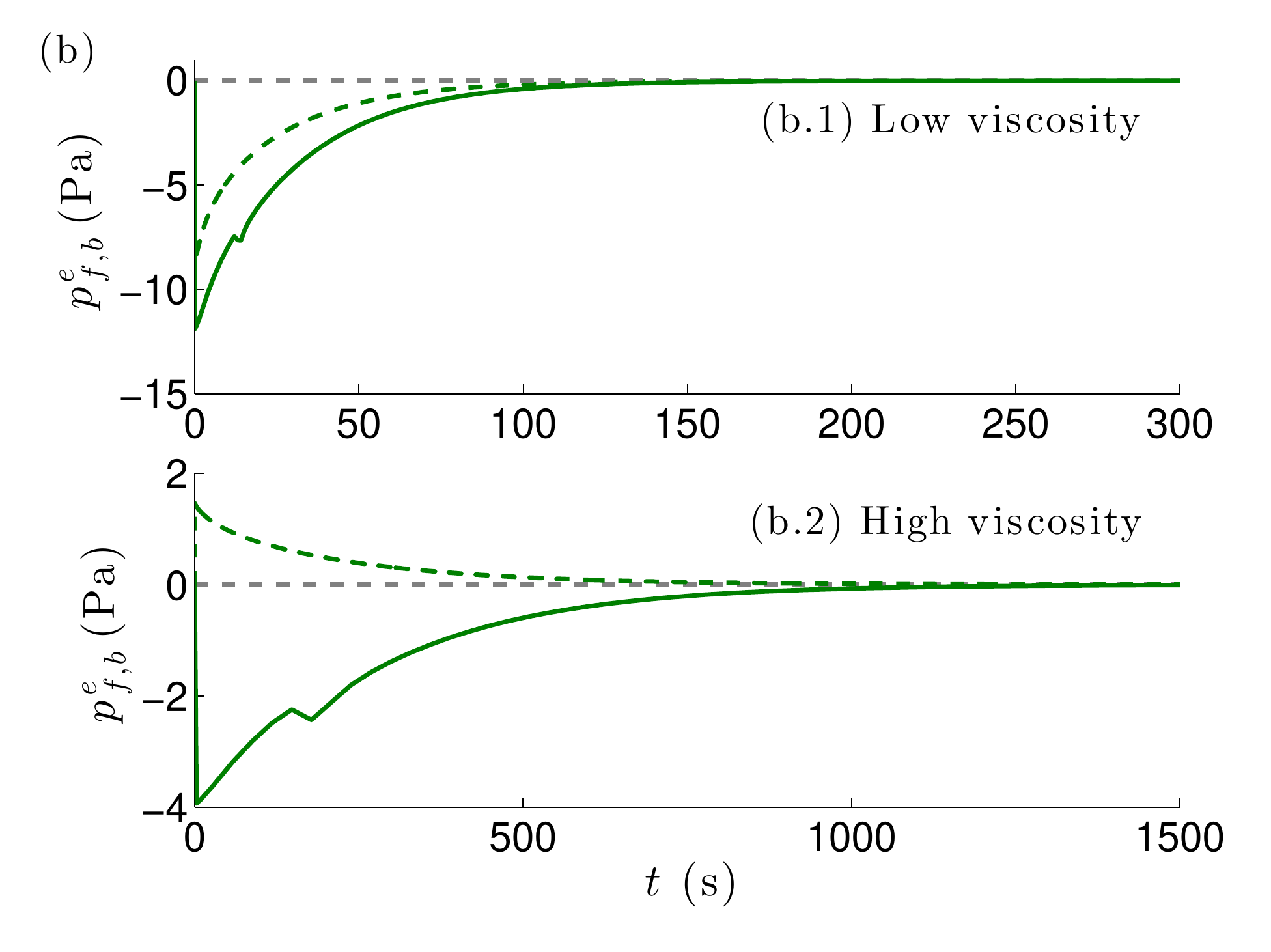}
	\end{center}
	\caption{\label{fig:analitica_presion} \it{Time evolution of (a) the pressure and (b) the  excess pore pressure at the bottom in the (x.1) low  and (x.2) high viscosity case. Dashed (resp. solid) lines are the solution starting from a loose (resp. dense) configuration, and the dashed grey lines are the analytical steady states.}}
\end{figure}
\begin{figure}[!ht]
		\begin{center}
			\includegraphics[width=0.49\textwidth]{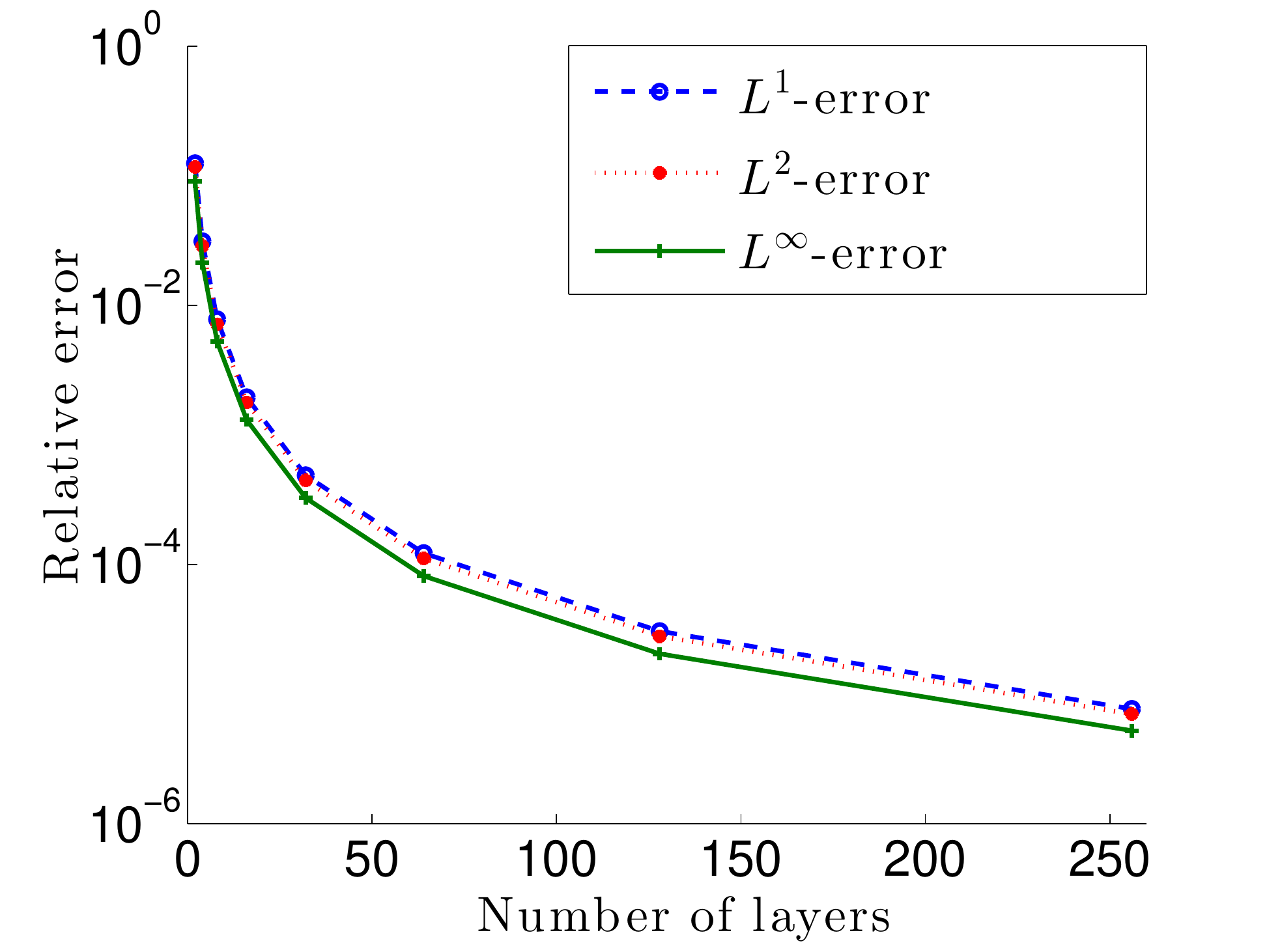}
		\end{center}
		\caption{\label{fig:error_noW} \it{Relative errors of the  velocity for the analytical solution without side walls friction and removing the friction term $\overline{\b_\a}\left(u_\a-v_\a\right)$.}}
	\end{figure}

In the inset in Figure \ref{fig:analitica_concentracion}(a.2) we also show that the solid volume fraction go to the equilibrium concentration quickly (we don't show the loose case because the dynamics towards equilibrium is even quicker in that case). Figure \ref{fig:analitica_concentracion}(b) shows the normal profile of solid volume fraction. In this case with no lateral friction, when the friction between the phases $\b_\a(u_\a-v_\a)$ is not considered in system \eqref{eq:FinalModel_uniform}, the analytical profile of the solid volume fraction is constant. Notice that this option represents an infinite friction between the phases, and therefore $u=v$. This is properly reproduced by our model except for the first layer (the closest to the bottom). Moreover, when increasing the number of vertical layers, we obtain the same behaviour. This loss of accuracy in Figure \ref{fig:analitica_concentracion}(b) is due to the friction between the solid and fluid phases. Actually, we have checked that if the term $\bar{\b}_\a \left(u_\a-v_\a\right)$ is removed from the momentum equations then the constant profile of solid volume fraction is obtained, (red-crossed solution in \ref{fig:analitica_concentracion}(b)). In addition, a convergence test in the number of vertical layers has been performed in this case. We see that the method is second order accurate for the velocity (see Table \ref{tabla1} and Figure \ref{fig:error_noW}). For the solid volume fraction, we always obtain an estimated error of $4.63\times10^{-8}$, which confirms the fact that the friction between the solid and the fluid phases is the responsible for not obtaining the constant profile in Figure \ref{fig:analitica_concentracion}(b).

\begin{table}[!ht]
	\begin{center}
		\begin{tabular}{ccccccc}
			\hline
			$N (\Delta z = 1/N)$ & $L^1$ - Error & $L^1$ - Order & $L^2$ - Error & $L^2$ - Order & $L^{\infty}$ - Error & $L^{\infty}$ - Order\\
			2 & 1.25$\times$10$^{-1}$ & --    & 1.17$\times$10$^{-1}$ & --      & 9.09$\times$10$^{-2}$ & --\\
			4 & 3.12$\times$10$^{-2}$ & 2.00 & 2.87$\times$10$^{-2}$ & 2.02 & 2.12$\times$10$^{-2}$ & 2.09\\
			8 & 7.81$\times$10$^{-3}$ & 2.00 & 7.14$\times$10$^{-3}$ & 2.00 & 5.23$\times$10$^{-3}$ & 2.02\\
			16 & 1.95$\times$10$^{-3}$ & 2.00 & 1.78$\times$10$^{-3}$ & 2.00 & 1.30$\times$10$^{-3}$ & 2.00\\
			32 & 4.88$\times$10$^{-4}$ & 1.99 & 4.45$\times$10$^{-4}$ & 2.00 & 3.25$\times$10$^{-4}$ & 2.00\\
			64 & 1.22$\times$10$^{-4}$ & 1.99 & 1.11$\times$10$^{-4}$ & 1.99 & 8.15$\times$10$^{-5}$ & 1.99\\
			128 & 3.05$\times$10$^{-5}$ & 1.99 & 2.79$\times$10$^{-5}$ & 1.99 & 2.04$\times$10$^{-5}$ & 1.99\\
			256 & 7.67$\times$10$^{-6}$ & 1.99 & 7.00$\times$10$^{-6}$ & 1.99 & 5.21$\times$10$^{-6}$ & 1.97\\
			\hline
		\end{tabular}
		\caption{\footnotesize  \it{Order of the error for the velocity of the analytical solution without side walls friction and removing the friction term $\overline{\b_\a}\left(u_\a-v_\a\right)$.}}
		\label{tabla1}
	\end{center}
\end{table}

\begin{figure}[!t]
	\begin{center}
		\includegraphics[width=0.8\textwidth]{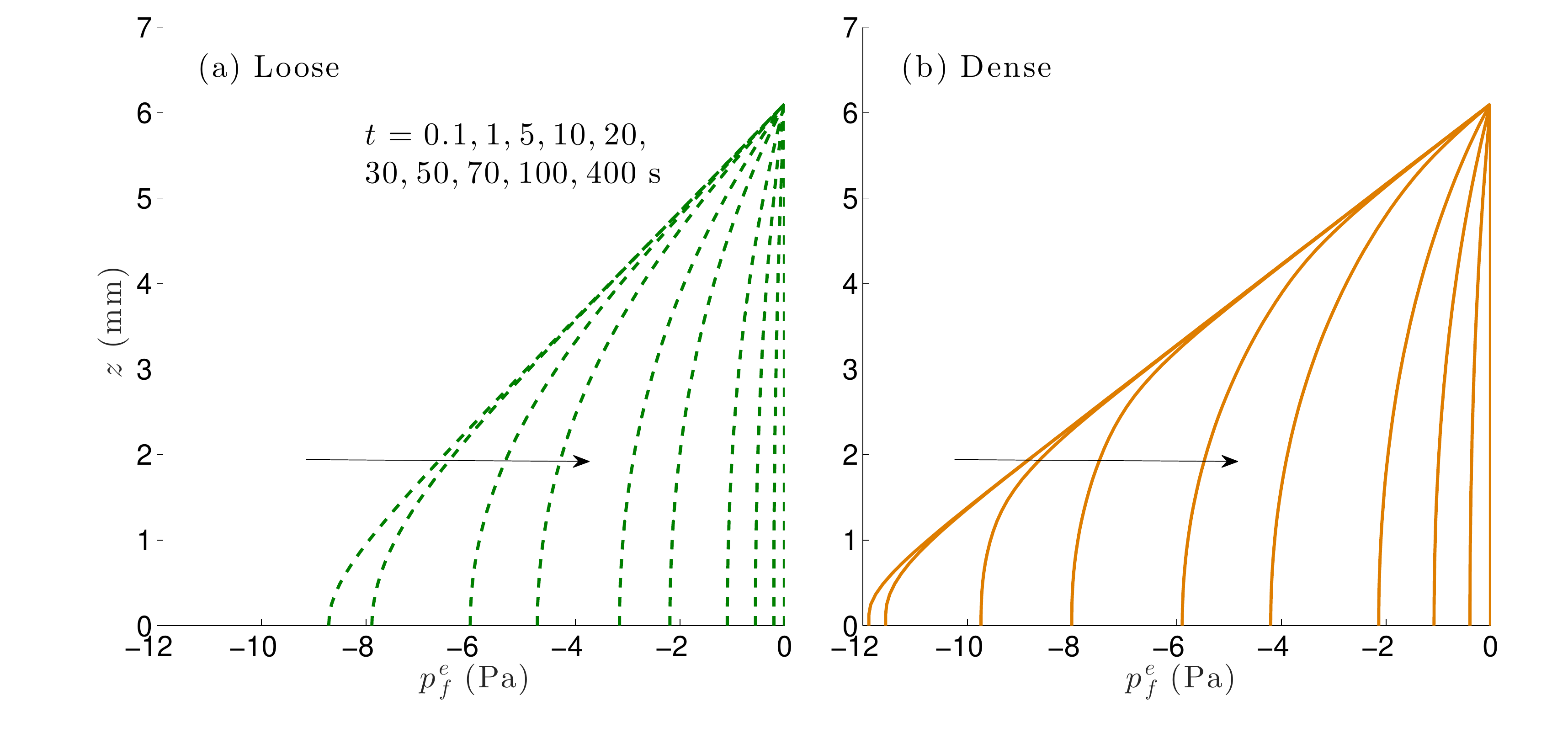}
	\end{center}
	\caption{\label{fig:analitica_presion_exc_low} \it{Normal profiles of  excess pore pressure for the low viscosity case, in the (a) loose and (b) dense configurations, at times $t=0.1, 1, 5, 10, 20, 30, 50, 70, 100, 400$ s.}}
\end{figure}
\begin{figure}[!t]
	\begin{center}
		\includegraphics[width=0.8\textwidth]{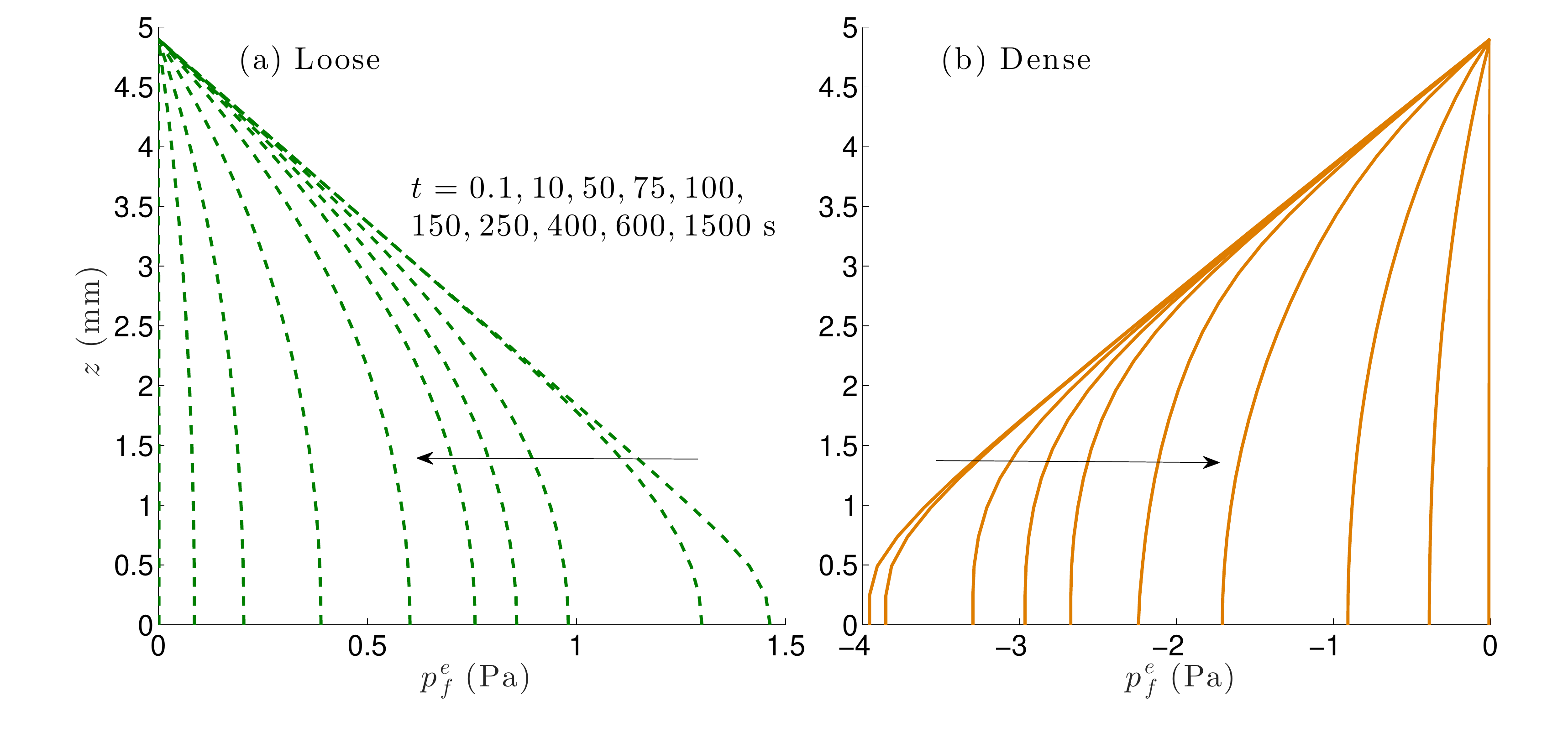}
	\end{center}
	\caption{\label{fig:analitica_presion_exc_high} \it{Normal profiles of  excess pore pressure for the high viscosity case, in the (a) loose and (b) dense configurations, at times $t=0.1,10,50,75,100,150,250,400,600,1500$ s.}}
\end{figure}

Figure \ref{fig:analitica_presion} shows the pressure and the excess pore pressure at the bottom. We see that the numerical solutions go to the analytical one in all the configurations, and that there is no  excess pore pressure in the steady state.

Figures \ref{fig:analitica_presion_exc_low} and \ref{fig:analitica_presion_exc_high}  show the normal profiles of  excess pore pressure at different times. We see that the profiles are approximately linear at short times, when the  excess pore pressure is larger, except near the bottom. 

\subsubsection{Side walls friction}
\begin{figure}[!htb]
	\begin{center}
		\hspace{-1.05cm}\includegraphics[width=0.53\textwidth]{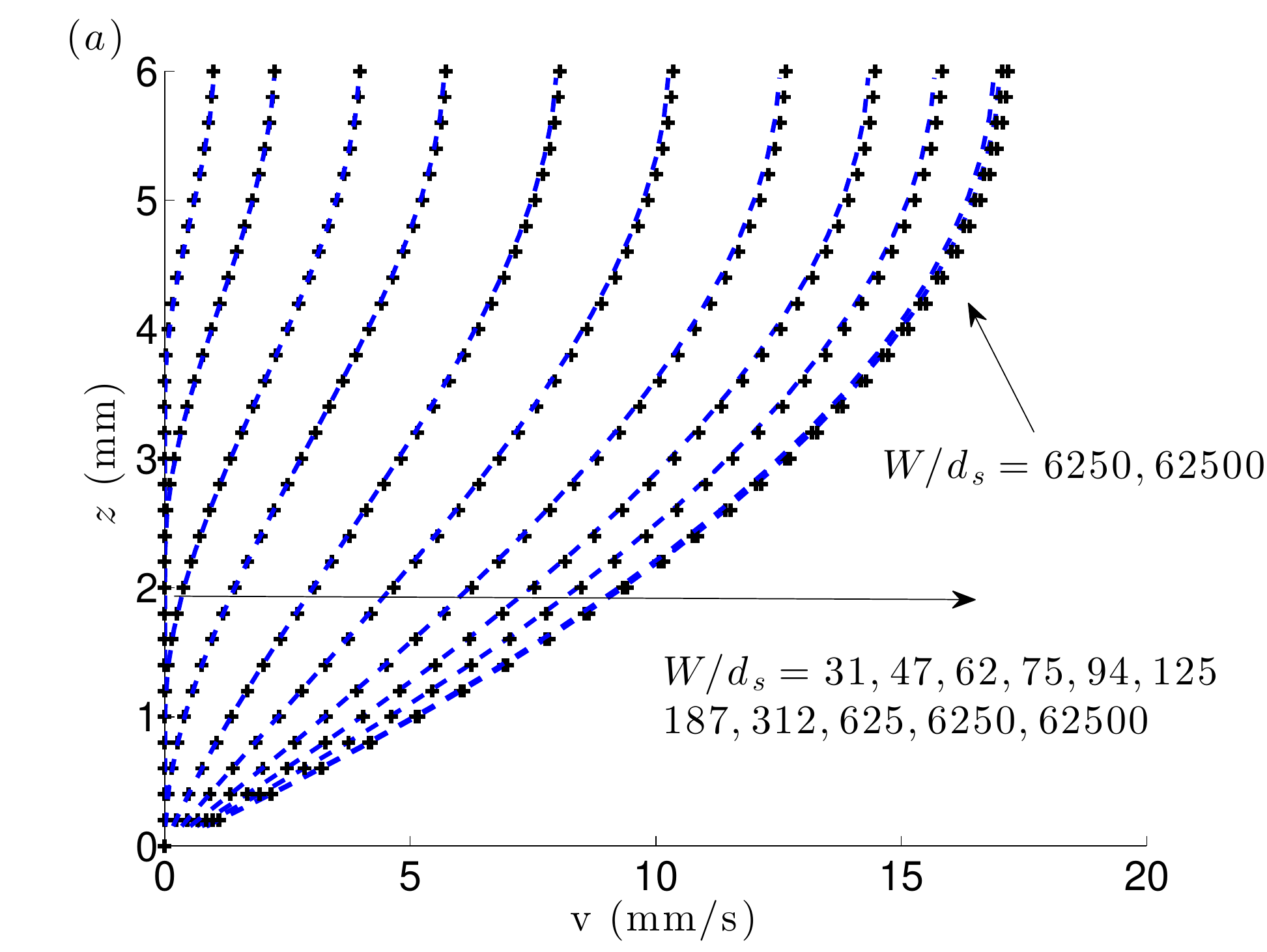}
		\includegraphics[width=0.53\textwidth]{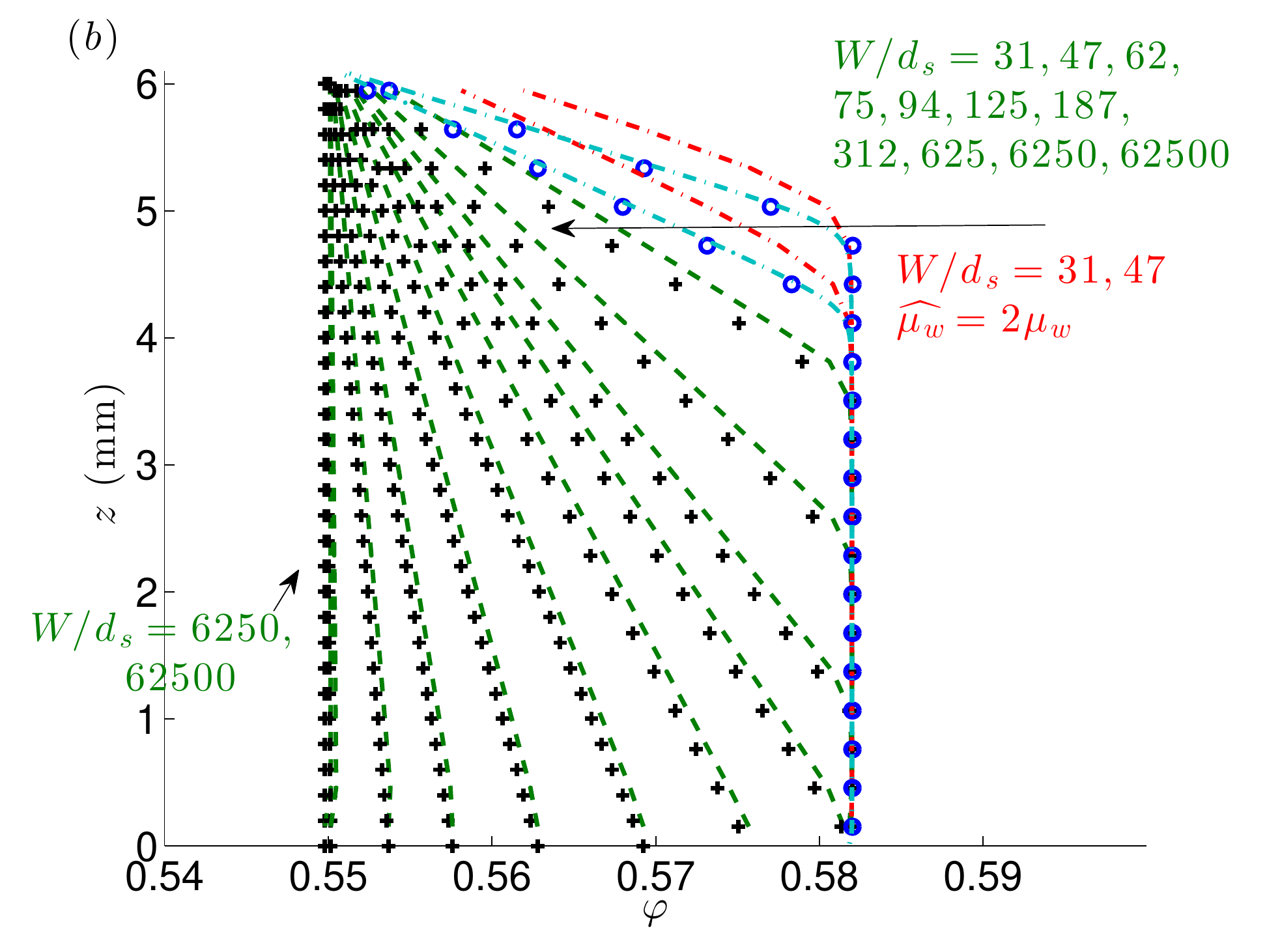}
	\end{center}
	\caption{\label{fig:analitica_vel_pared} \it{Normal profiles of velocity (a) and solid volume fraction (b) for different widths $W$. Dashed lines correspond to the simulations and black crosses to the analytical solutions. Blue symbols and red dash-dotted lines in figure  (b) are the simulations for widths $W= 31d_s, 47d_s$ and $\wh{\mu_w} = 2\mu_{w}$, additionally cyan dash-dotted lines are the simulations for this case with $160$ vertical layers.}}
\end{figure}

In subsection \ref{se:pared} the analytical solution is obtained in the case of a flow that is confined in a channel whose width is $W$. In that case, the side walls friction has an important effect on the dynamics of the flow. Mainly, the normal profile can become S-shaped instead of Bagnold type (see \cite{fernandezNieto:2018}) leading to the appearance of a flow/no-flow transition.

We consider a flow with the same physical and rheological properties as in previous subsection in the low viscosity regime and loose initial condition, and we add the side walls friction term with the friction coefficient $\mu_w = \tan\left(13.1^\circ\right)$. In this case we use $20$ layers in the multilayer system since the computational cost, until reaching the stationary solution, is very high.

Figure \ref{fig:analitica_vel_pared} shows the normal profiles of velocity and solid volume fraction in the presence of lateral walls, for several channel widths $W=0.005 , 0.0075, 0.01, 0.012, 0.015, 0.02, 0.03, 0.05, 0.1, 1, 10$ m. These values correspond to  $\a$ layers of granular particles, $W=\a d_s$, with $\a = 31, 47,\dots,62500$, approximately. 
	For the velocity profiles, the dynamics is similar to the case of a dry granular flow (see \cite{fernandezNieto:2018}). In the case of the solid volume fraction, a constant profile is obtained for large widths, similar to the case of no lateral wall friction.

However, looking at the analytical solution (see remark \ref{re:phi_pared}) we see that the analytical profile of the volume fraction is always linear, starting from the equilibrium volume fraction $\vp^{eq}_{c}$ at the top of the mixture layer. Actually, the linear counterpart is $K_2\mu_w z/(K_1W)$. Therefore, when $W$ is large, this contribution is nearly zero and we obtain an (almost) constant profile.
As observed on Figure \ref{fig:analitica_vel_pared}(b), wall friction makes the solid volume fraction vary linearly with a slope increasing when the channel width $W$ decreases. An interesting remark is that when a flow/no-flow transition appears in the velocity profile, the solid volume fraction becomes a constant, whose value is $\vp^{stat}_c = 0.582$, from the flow/no-flow position to the bottom representing a region of rigid material (zero velocity). It decreases from the flow/no-flow transition up to the upper surface of the mixture. In Figure \ref{fig:analitica_vel_pared} we also see the influence of the friction parameter $\mu_w$. For widths $W=31d_s,47d_s$ we have simulated the same flow but taking as friction coefficient at the lateral walls $\wh{\mu_w} = 2\mu_w$. It makes the friction force increase (similar effect is obtained by decreasing the channel width) and the solid volume profile becomes constant almost everywhere in the normal direction, decreasing just close to the mixture upper surface (see blue symbols and red dash-dotted lines in Figure  \ref{fig:analitica_vel_pared}(b)).

	\begin{figure}[!t]
	\begin{center}
		\includegraphics[width=0.49\textwidth]{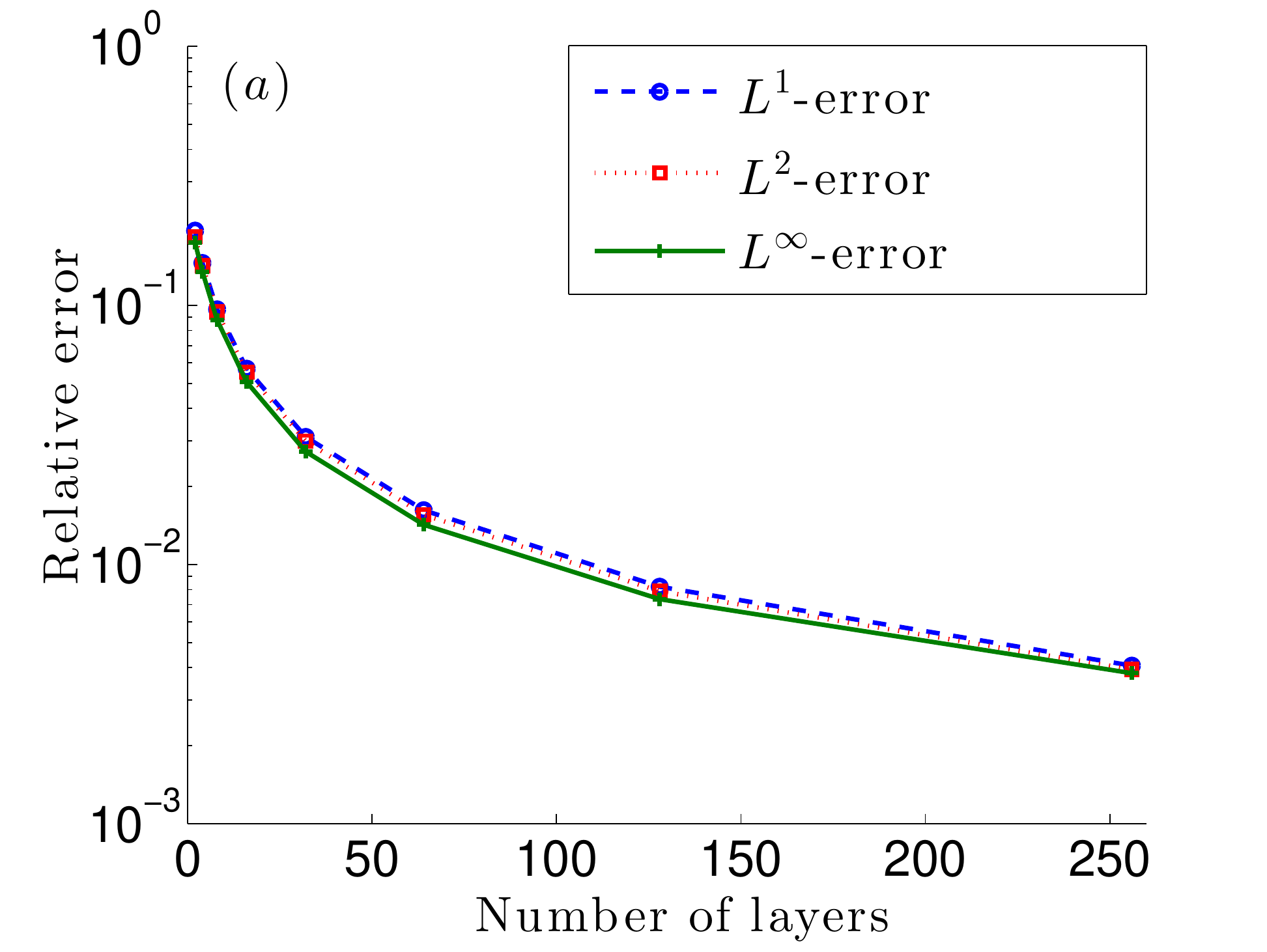}
		\includegraphics[width=0.49\textwidth]{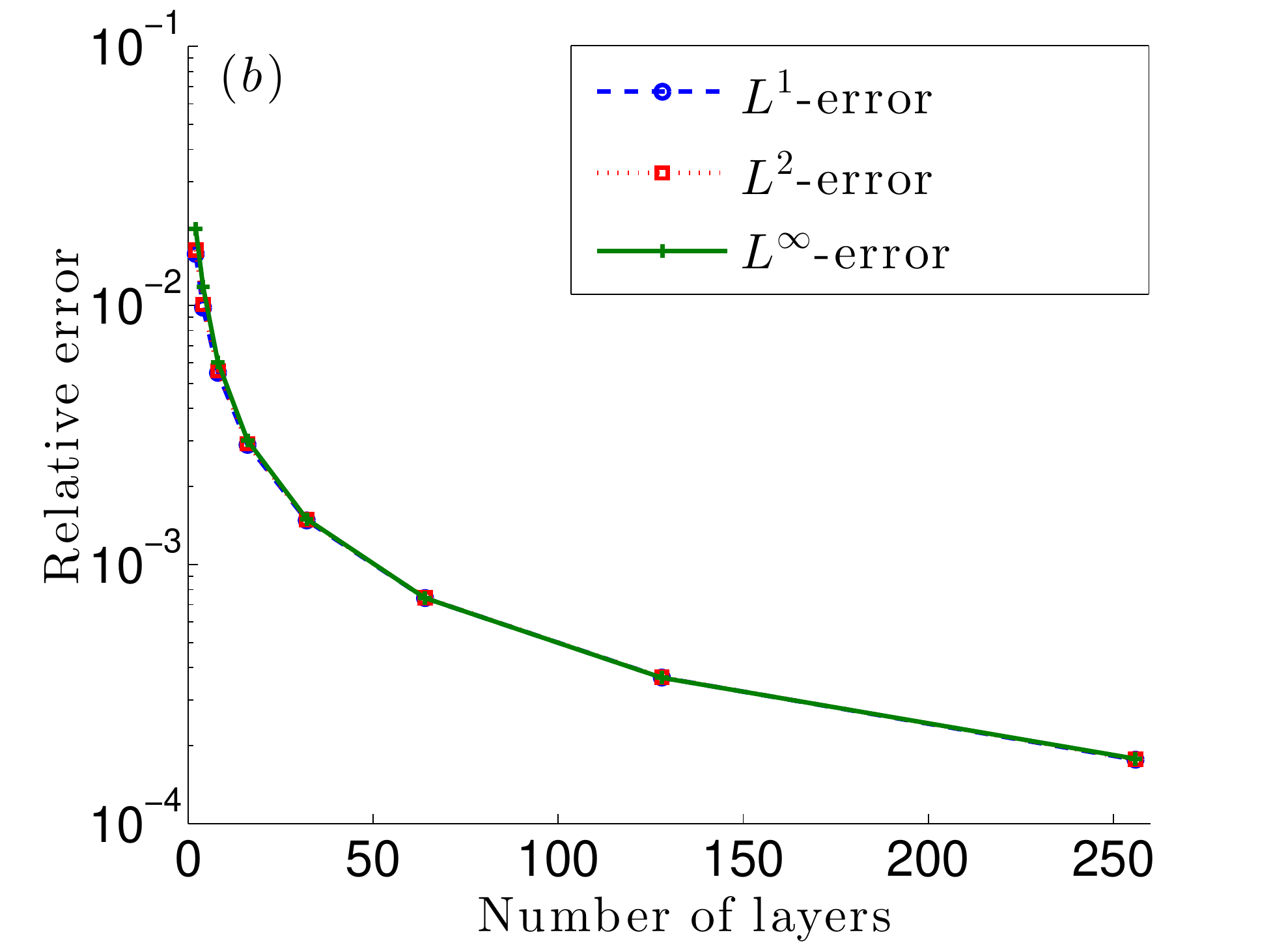}
	\end{center}
	\caption{\label{fig:error_W} \it{Relative errors of (a) the  velocity and (b) the solid volume fraction for the analytical solution with side walls friction and $W=94 d_s$.}}
\end{figure}

In Figure \ref{fig:analitica_vel_pared} we remark that errors between the analytical and the numerical solutions are bigger than the expected ones, specially for the solid volume fraction and a small channel width (see e.g. red dash-dotted lines in Figure \ref{fig:analitica_vel_pared}(b)). This is due to the fact that we have used $20$ layers to reproduce solutions with a strong variations in the normal direction. In order to clarify that, we have performed a convergence test for $W = 94d_s$, also removing the term $\b_\a\left(u_\a-v_\a\right)$, since it is not considered for the analytical solution. We see that the method is first order accurate in this case, for both, the velocity and the solid volume fraction (see Table \ref{tabla2} and Figure \ref{fig:error_W}). In addition, for the two configurations for which the side wall friction is stronger, namely $W=31 d_s, \ 47d_s$, we have also added the corresponding solutions using $160$ layers  in Figure \ref{fig:analitica_vel_pared}(b) (cyan dash-dotted lines). The analytical solution is properly reproduced in that case.

	\begin{table}[!t]
		\begin{center}
			(a) Velocity\\[4mm]
			\begin{tabular}{ccccccc}
				\hline
				$N (\Delta z = 1/N)$ & $L^1$ - Error & $L^1$ - Order & $L^2$ - Error & $L^2$ - Order & $L^{\infty}$ - Error & $L^{\infty}$ - Order\\
				2 & 1.94$\times$10$^{-1}$ & --    & 1.84$\times$10$^{-1}$ & --      & 1.74$\times$10$^{-1}$ & --\\
				4 & 1.45$\times$10$^{-1}$ & 0.47 & 1.42$\times$10$^{-1}$ & 0.37 & 1.34$\times$10$^{-1}$ & 0.37\\
				8 & 9.65$\times$10$^{-2}$ & 0.59 & 9.44$\times$10$^{-2}$ & 0.59 & 8.78$\times$10$^{-2}$ & 0.61\\
				16 & 5.69$\times$10$^{-2}$ & 0.76 & 5.52$\times$10$^{-2}$ & 0.77 & 5.06$\times$10$^{-2}$ & 0.79\\
				32 & 3.10$\times$10$^{-2}$ & 0.87 & 2.99$\times$10$^{-2}$ & 0.88 & 2.72$\times$10$^{-2}$ & 0.98\\
				64 & 1.61$\times$10$^{-2}$ & 0.94 & 1.55$\times$10$^{-2}$ & 0.94 & 1.42$\times$10$^{-2}$ & 0.93\\
				128 & 8.20$\times$10$^{-3}$ & 0.98 & 7.88$\times$10$^{-3}$ & 0.97 & 7.34$\times$10$^{-3}$ & 0.95\\
				256 & 4.06$\times$10$^{-3}$ & 1.01 & 3.93$\times$10$^{-3}$ & 1.00 & 3.80$\times$10$^{-3}$ & 0.94\\
				\hline
			\end{tabular}
		
		\bigskip
		(b) Solid volume fraction\\[4mm]
		\begin{tabular}{ccccccc}
			\hline
			$N (\Delta z = 1/N)$ & $L^1$ - Error & $L^1$ - Order & $L^2$ - Error & $L^2$ - Order & $L^{\infty}$ - Error & $L^{\infty}$ - Order\\
			2 & 1.57$\times$10$^{-2}$ & --    & 1.63$\times$10$^{-2}$ & --      & 1.97$\times$10$^{-2}$ & --\\
			4 & 9.79$\times$10$^{-3}$ & 0.68 & 1.00$\times$10$^{-2}$ & 0.69 & 1.18$\times$10$^{-2}$ & 0.74\\
			8 & 5.49$\times$10$^{-3}$ & 0.83 & 5.59$\times$10$^{-3}$ & 0.85 & 6.03$\times$10$^{-3}$ & 0.96\\
			16 & 2.89$\times$10$^{-3}$ & 0.92 & 2.92$\times$10$^{-3}$ & 0.93 & 3.01$\times$10$^{-3}$ & 1.00\\
			32 & 1.48$\times$10$^{-3}$ & 0.96 & 1.48$\times$10$^{-3}$ & 0.97 & 1.49$\times$10$^{-3}$ & 1.00\\
			64 & 7.42$\times$10$^{-4}$ & 0.99 & 7.44$\times$10$^{-4}$ & 0.99 & 7.43$\times$10$^{-4}$ & 1.01\\
			128 & 3.66$\times$10$^{-4}$ & 1.01 & 3.67$\times$10$^{-4}$ & 1.02 & 3.65$\times$10$^{-4}$ & 1.02\\
			256 & 1.76$\times$10$^{-4}$ & 1.05 & 1.77$\times$10$^{-4}$ & 1.05 & 1.78$\times$10$^{-4}$ & 1.03\\
			\hline
		\end{tabular}
			\caption{\footnotesize  \it{Order of the error for (a) the velocity and (b) the solid volume fraction of the analytical solution with side walls friction, $W=94d_s$, and removing the friction term $\overline{\b_\a}\left(u_\a-v_\a\right)$.}}
			\label{tabla2}
		\end{center}
	\end{table}

\subsection{Comparison with other models and laboratory experiments}\label{se:comp_model}
In this section we compare the results of the PGM model with the results of the model in Bouchut {\it et al.} (2016) \cite{bouchut:2016} and Pailha and Pouliquen (2009) \cite{pailha:2009} (hereafter B-2016 and PP-2009 respectively). However, before that, we evaluate in the next subsection the appropriate model that we propose to use in practice, since the approximation of the solid pressure entrains serious difficulties in the multilayer framework, as we mentioned before.

\subsubsection{Choice of the model}\label{sec:choice}
In this section the simulated cases, as well as the physical and rheological parameters, take the same values as in the previous section. We focus on the dense case for both the high and low viscosity cases because it is the more complex case, since the convergence to the steady solution is slower and it differs far more from other analyzed models, as it is shown in following subsections.

\begin{figure}[!t]
	\begin{center}
		\includegraphics[width=0.49\textwidth]{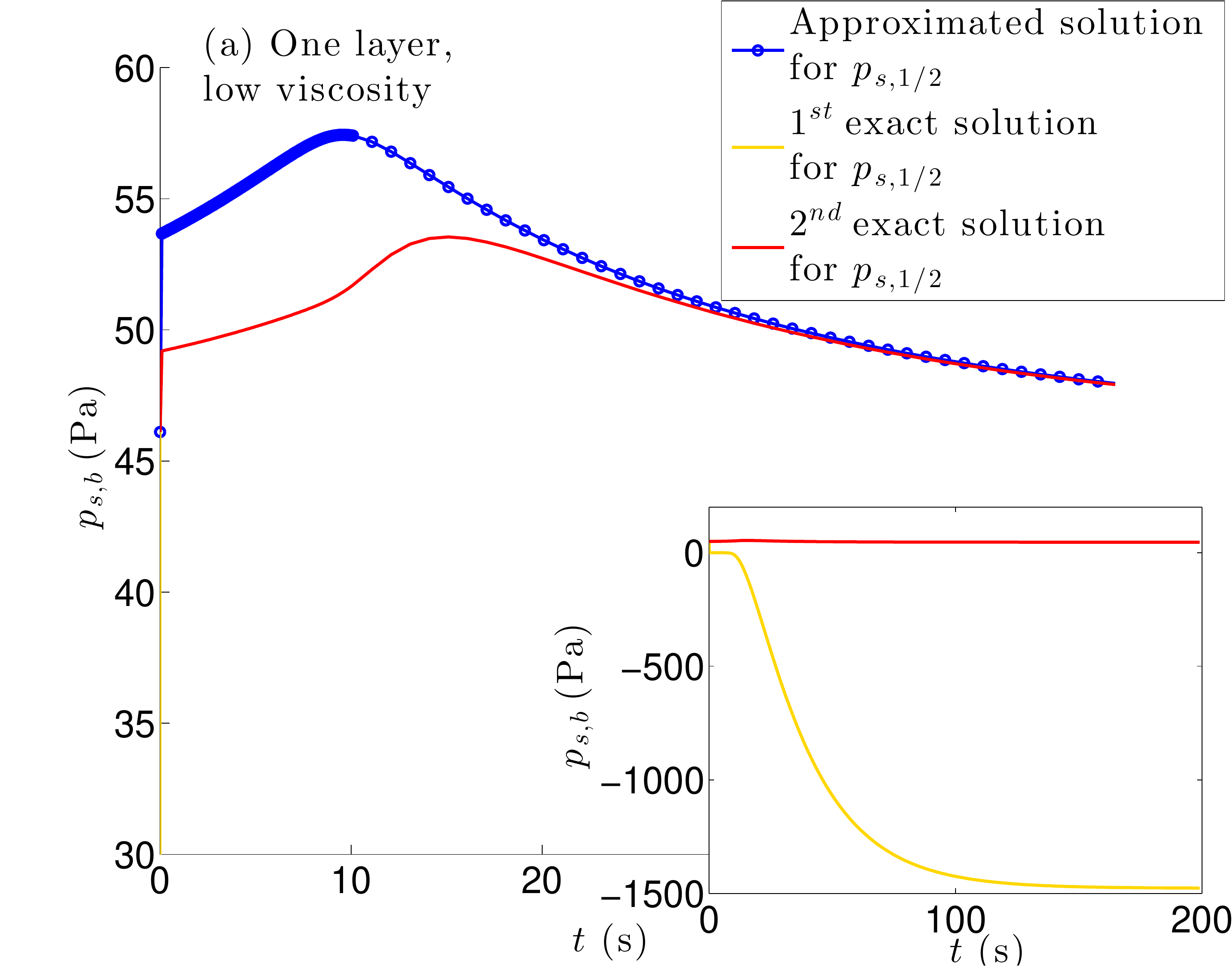}
		\includegraphics[width=0.49\textwidth]{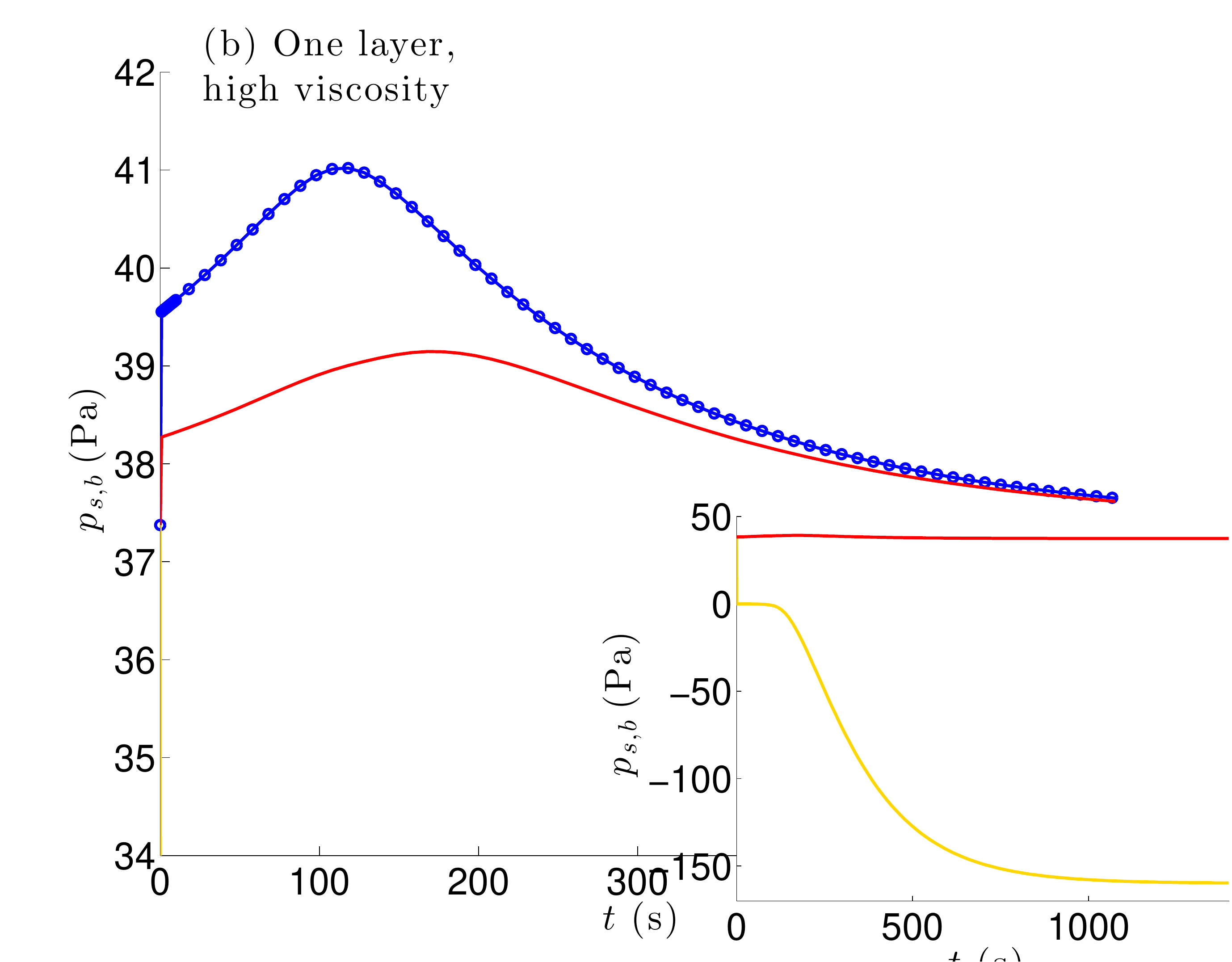}\\
		\includegraphics[width=0.49\textwidth]{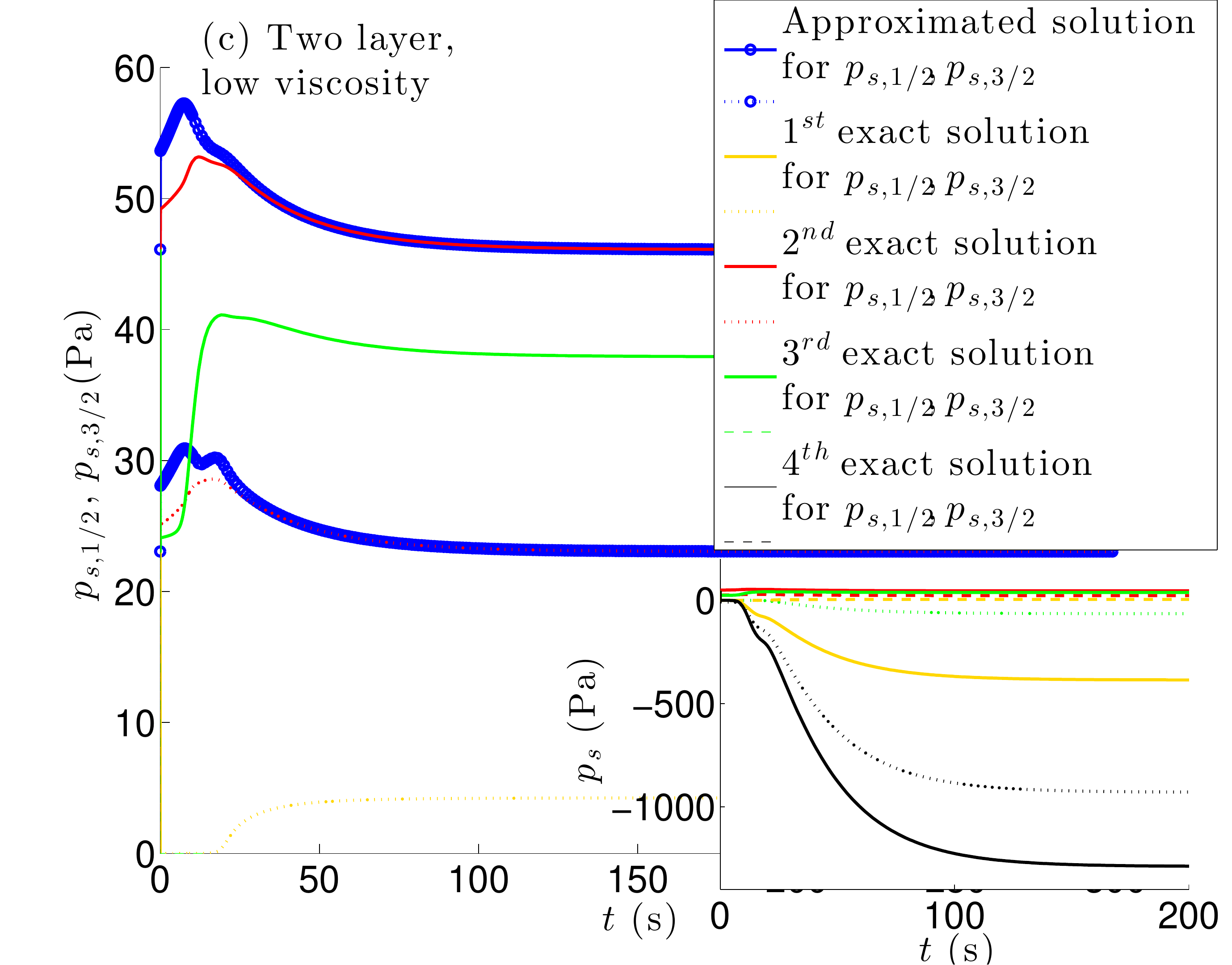}
		\includegraphics[width=0.49\textwidth]{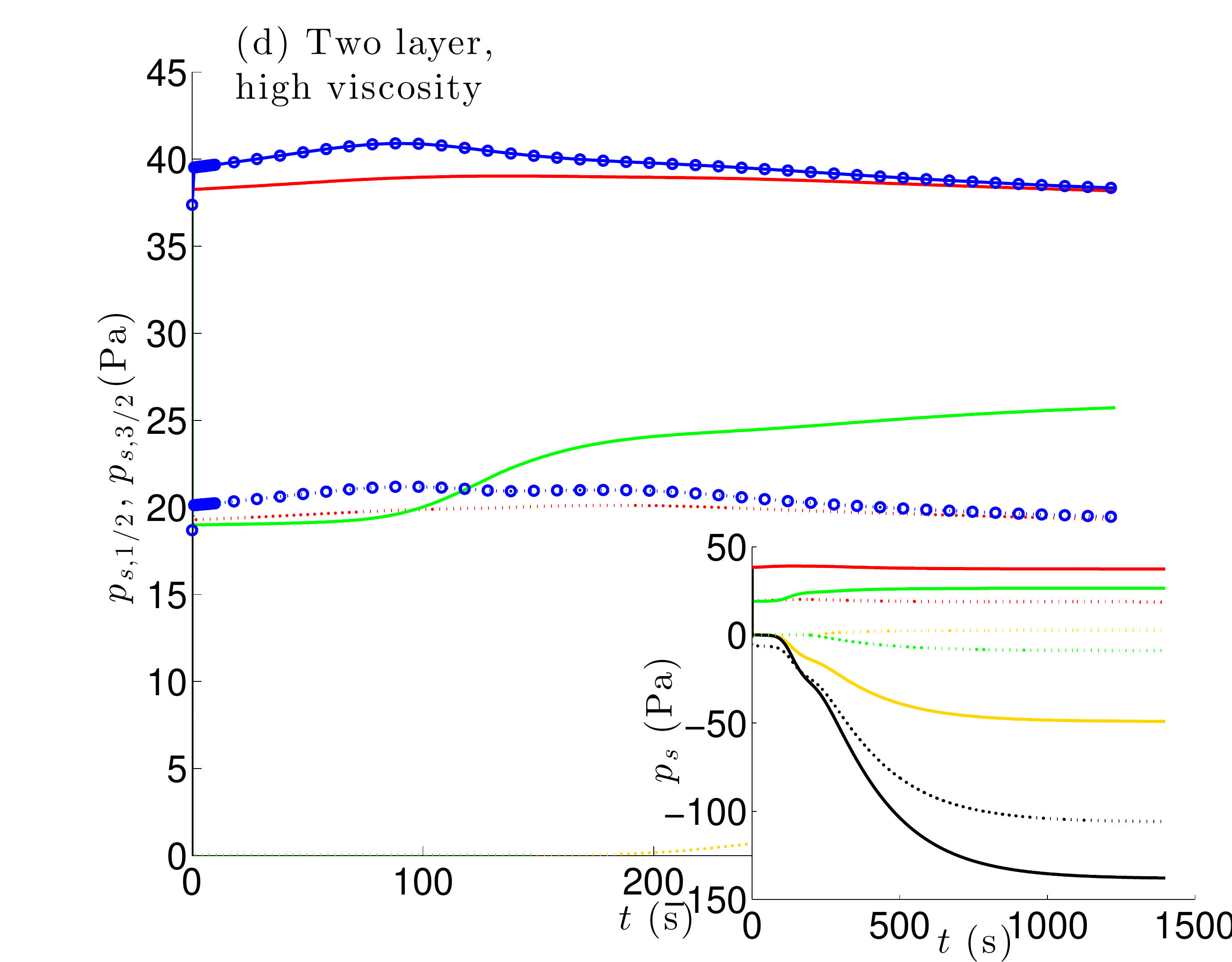}
	\end{center}
	\caption{\label{fig:aprox_ps} \it{Time evolution of exact (solid lines) and computed (blue symbols) solid pressures for the case of a single layer ((a) and (b)) and two layers ((c) and (d)) in the multilayer system, in the low ((a) and (c)) and high ((b) and (d)) viscosity case with dense initial configuration. Dotted lines in the two layer case are the pressures at the internal interface $z=z_{3/2}$, while solid lines are the pressures at the bottom $z=z_{1/2}$. Inset figures are the long time exact solutions.} }
\end{figure}

Firstly, we check that the proposed linearization to compute the pressure gives good results. Note that equations \eqref{eq:eq_pressure} lead to a non-linear system with $N$ equations and unknowns, where each of them is a quadratic equation, and the system is fully coupled due to the excess pore pressure. Therefore, for a configuration with $N$ vertical layers, there exist $2^N$ vector solutions of the system. In particular, for $N=1$, at each time step, once the rest of variables are computed, the pressure at the bottom, $p^n_{s,\frac12}$, is a root of the quadratic equation resulting from \eqref{eq:eq_pressure} with $N=1$. That equation can be exactly solved, obtaining two different roots for $p_{s,\frac12}^n$. However, one of them becomes negative and therefore there is only one solution with a physical meaning. In Figure \ref{fig:aprox_ps}(a)(b) we show that the proposed method in \ref{se:approx_pressure} gives as result an approximation to this appropriate solution. Analogously, for $N=2$, system \eqref{eq:eq_pressure} has $4$ different solutions for the vector $\left(p_{s,\frac12}^n,p_{s,\frac32}^n\right)$ at each time step, corresponding to the pressure at the interfaces $z_\frac12$ and $z_{\frac32}$. Again, the system obtained by \eqref{eq:eq_pressure} with $N=2$ can be exactly solved, where just one of the roots is relevant,  since the others become negative. Figure \ref{fig:aprox_ps}(c)(d) shows that the proposed method approximates this solution. 
Nevertheless, we did not managed to calculate an accurate approximation of this solution at short times where the variation and effect of the excess pore pressure are strong.

\begin{figure}[!t]
	\begin{center}
		\includegraphics[width=0.49\textwidth]{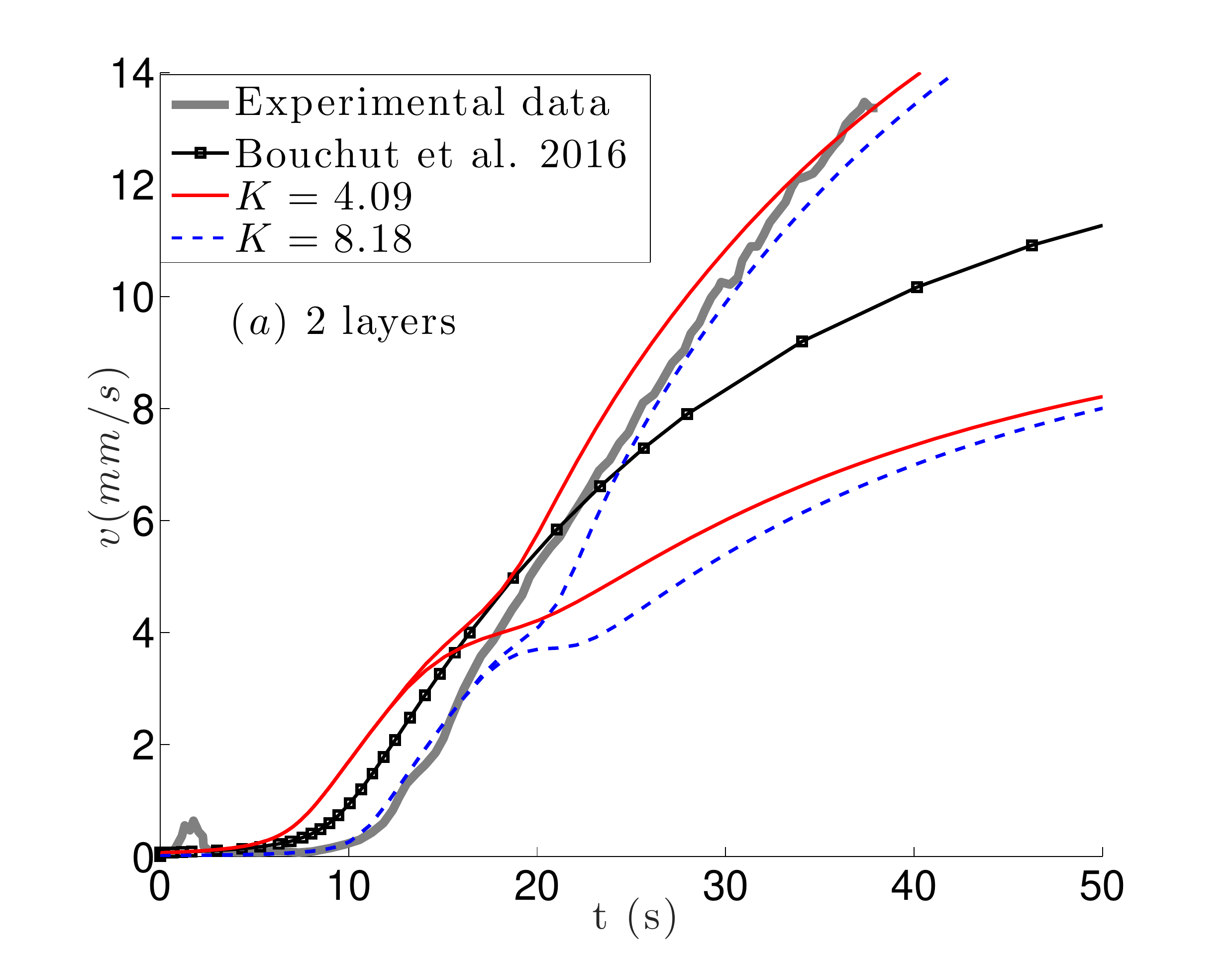}
		\includegraphics[width=0.49\textwidth]{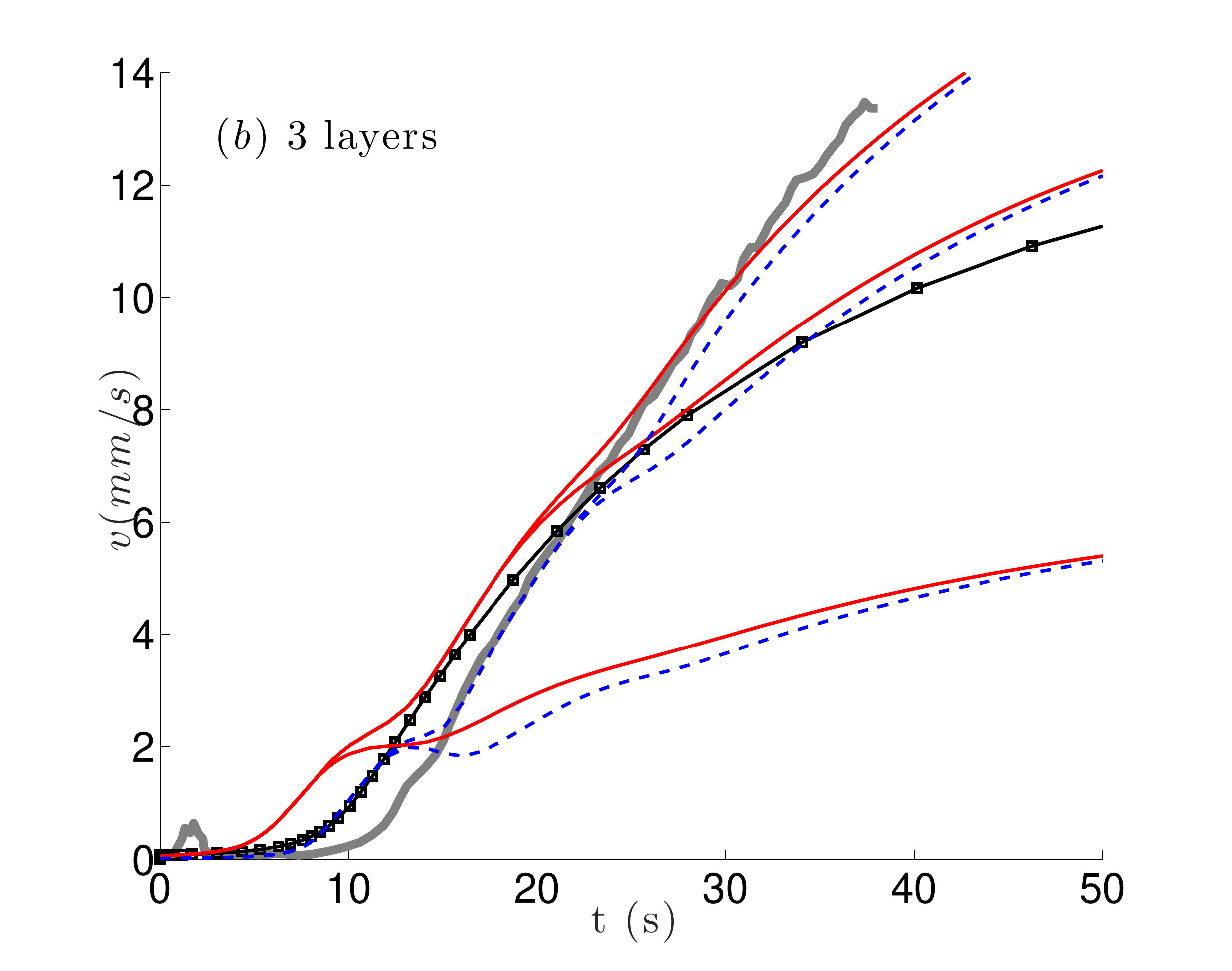}
		\includegraphics[width=0.49\textwidth]{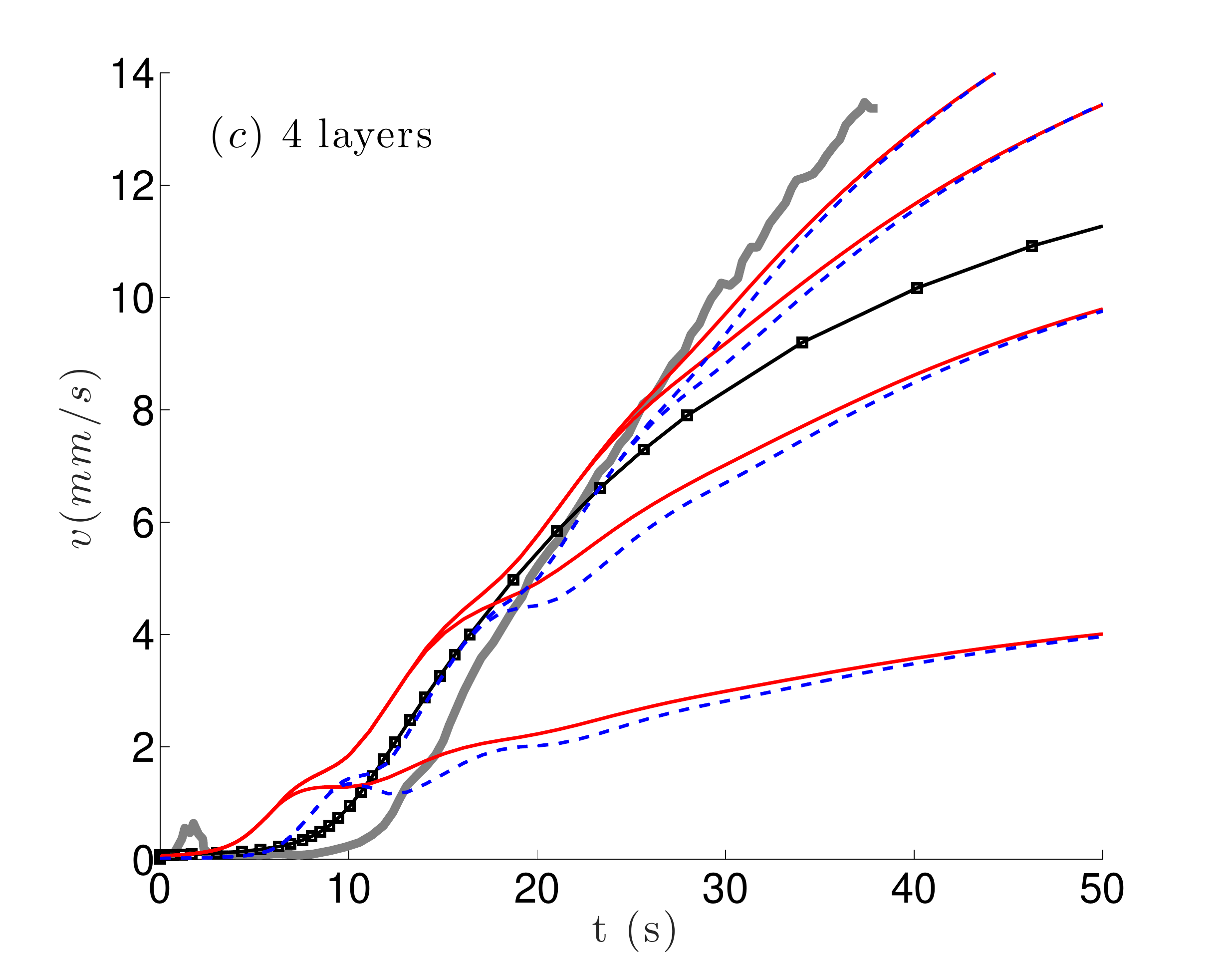}
		\includegraphics[width=0.49\textwidth]{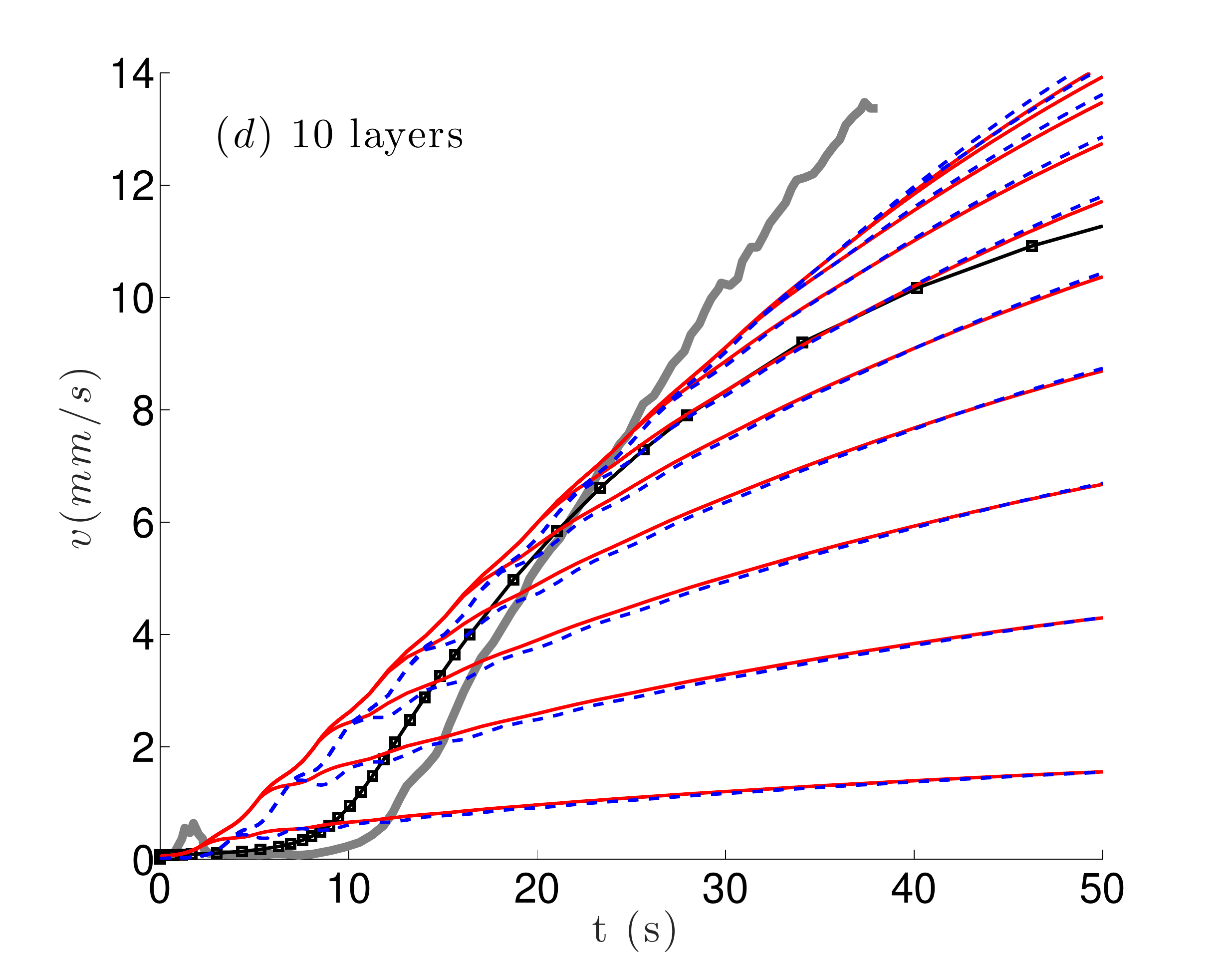}
	\end{center}
	\caption{\label{fig:multi_vel_low} \it{Low viscosity case with dense initial condition. Time evolution of the velocity of each layer for the multilayer model with $2,3,4$ and $10$ layers. Solid red lines are computed with dilatation constant $K=4.09$ and dashed blue lines are with $K=8.18$. The black-squared solid line is the solution of the single-layer model B-2016, and grey lines is the velocity measured at the surface of the mixture in laboratory experiments.}}
\end{figure}

\begin{figure}[!ht]
	\begin{center}
		\includegraphics[width=0.49\textwidth]{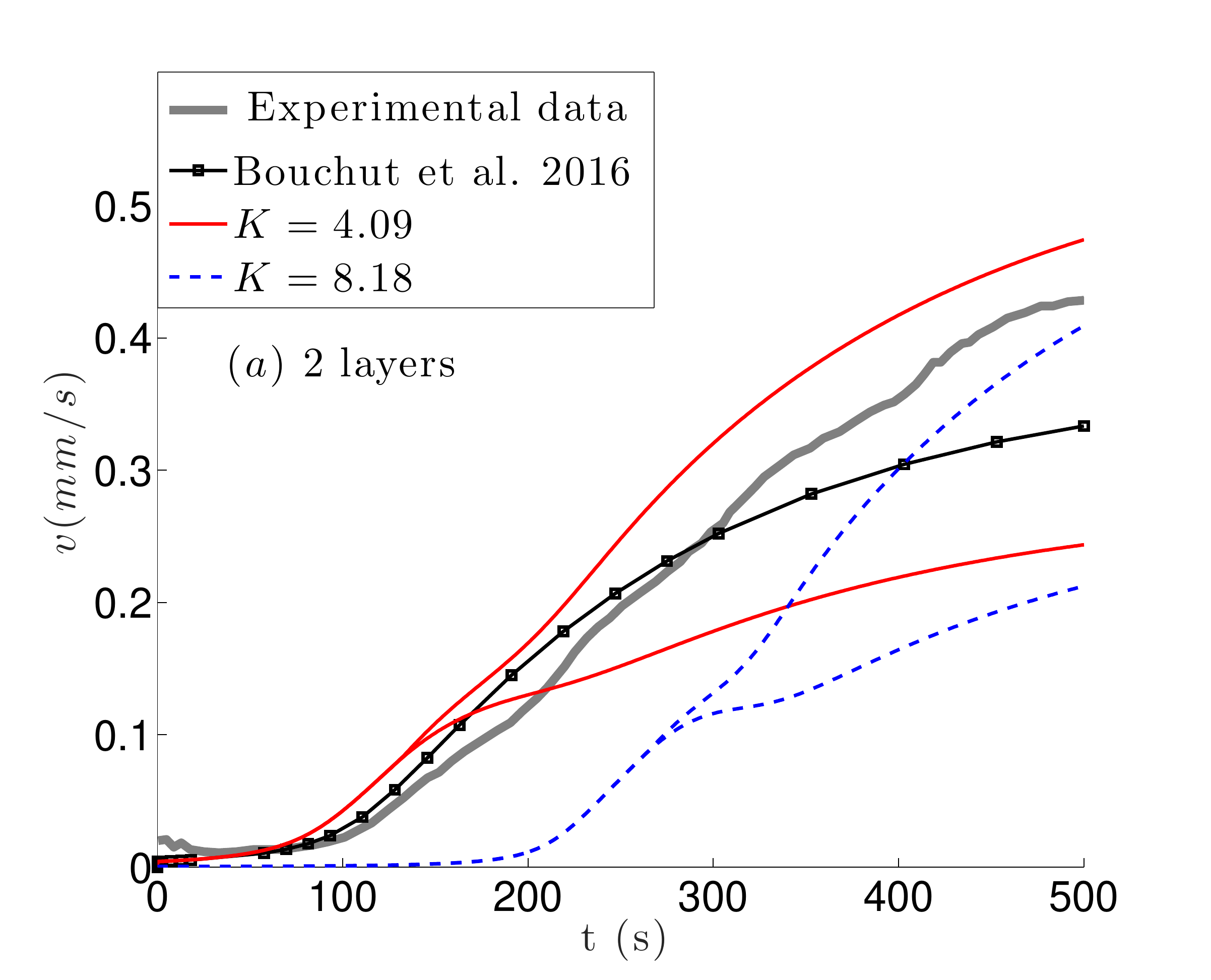}
		\includegraphics[width=0.49\textwidth]{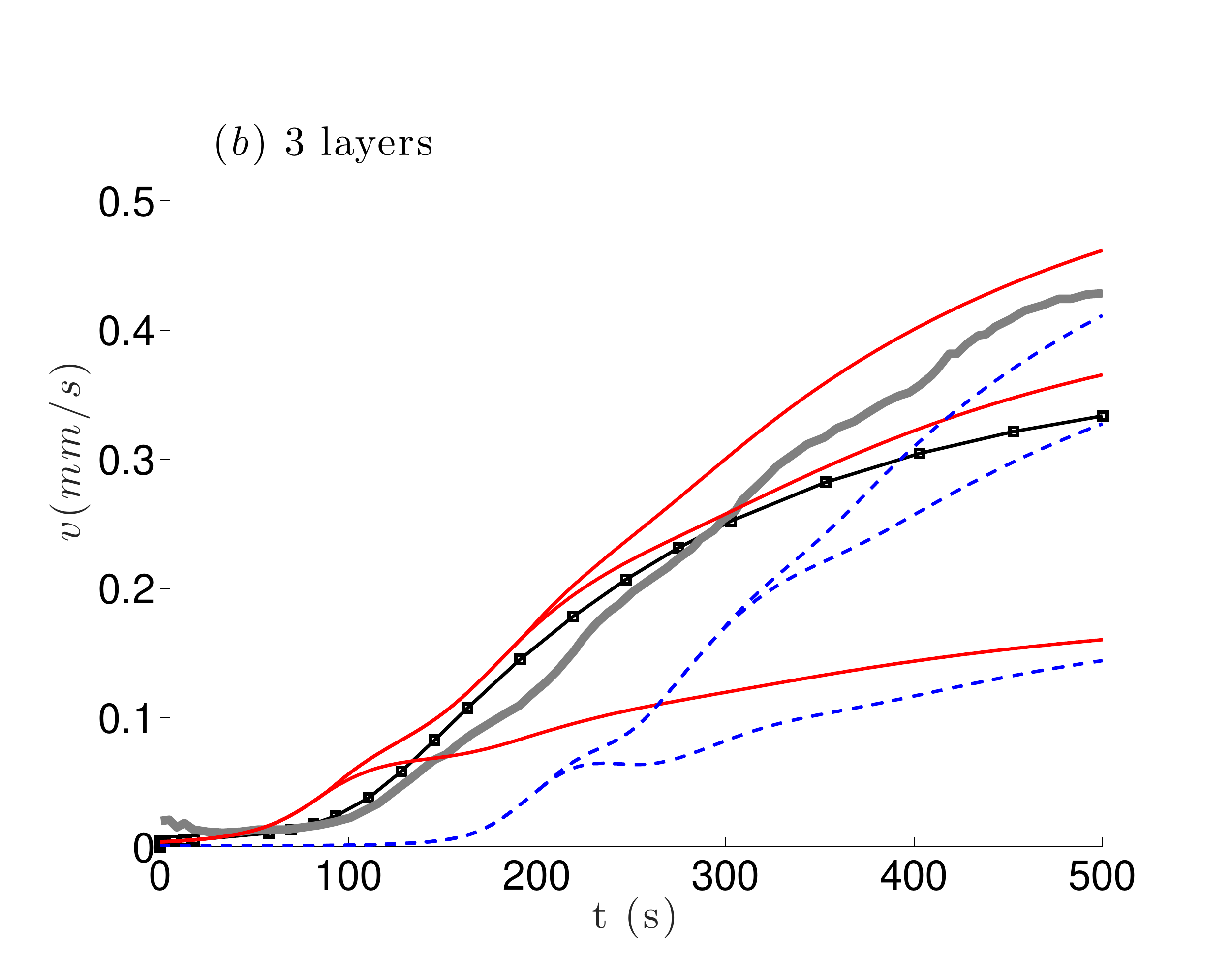}
		\includegraphics[width=0.49\textwidth]{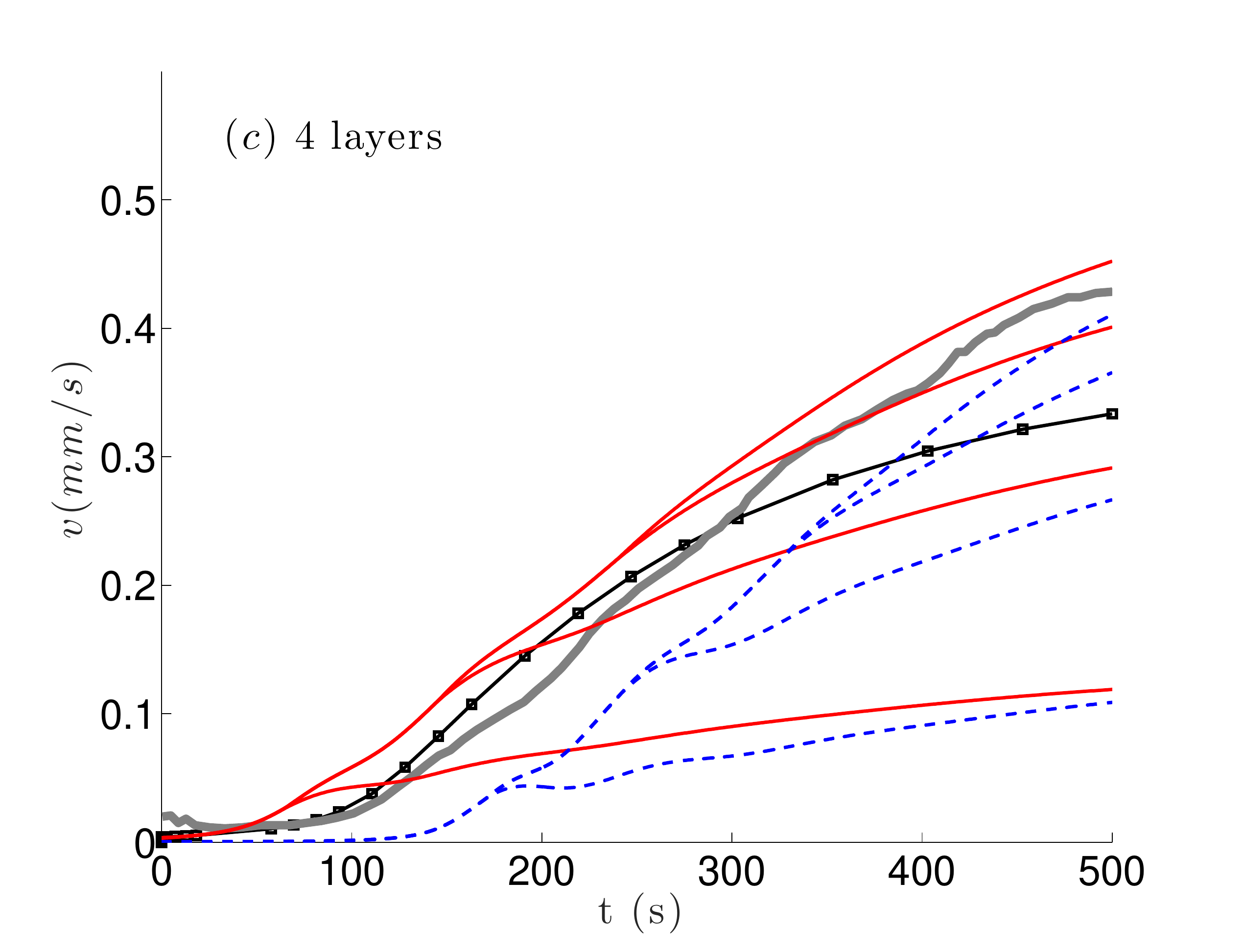}
		\includegraphics[width=0.49\textwidth]{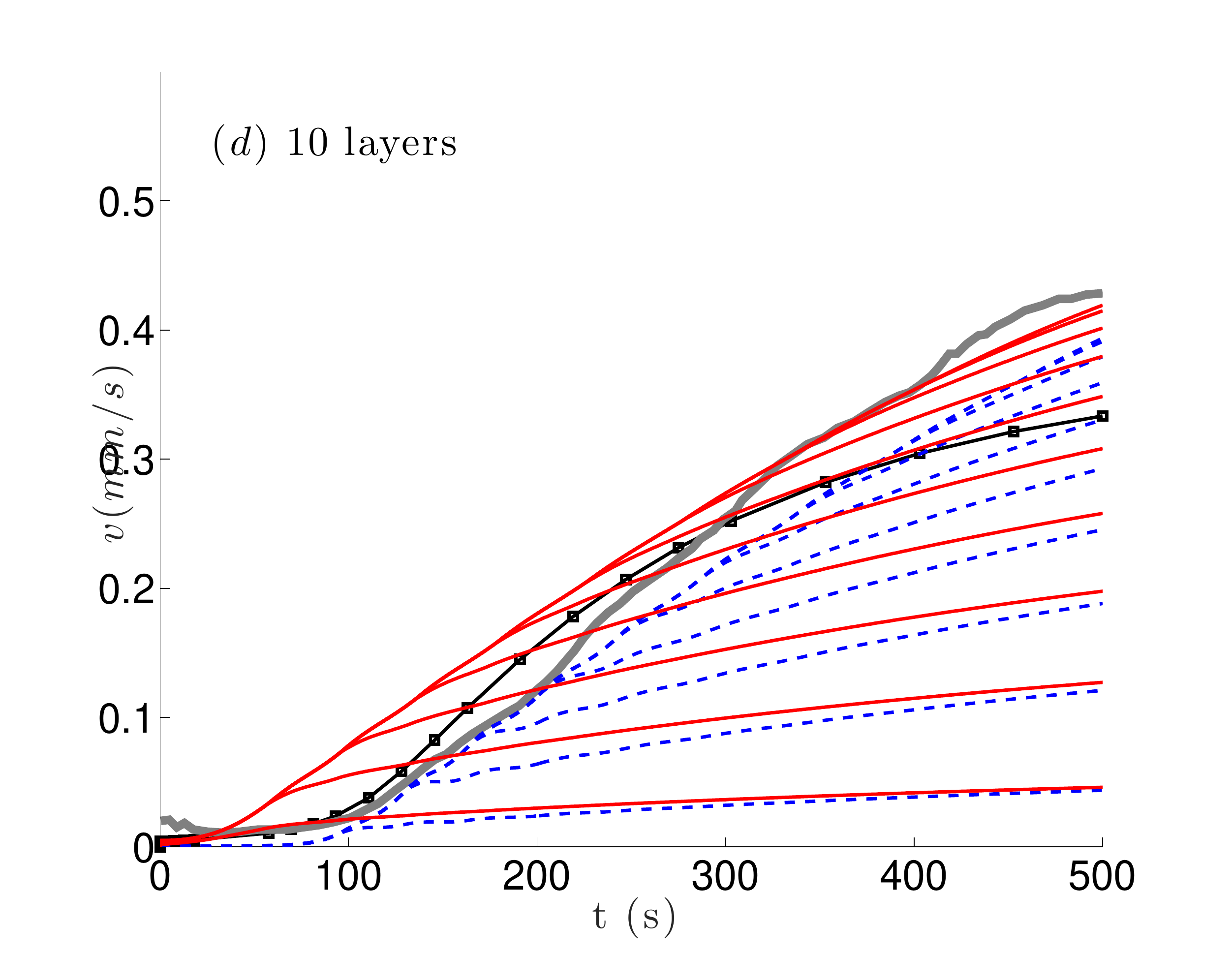}		
	\end{center}
	\caption{\label{fig:multi_vel_high} \it{High viscosity case with dense initial condition. Time evolution of the velocity of each layer for the multilayer model with $2,3,4$ and $10$ layers. Solid red lines are computed with dilatation constant $K=4.09$ and dashed blue lines are with $K=8.18$. The black-squared solid line is the solution of the single-layer model B-2016, and grey lines is the velocity measured at the surface of the mixture in laboratory experiments.}}
\end{figure}

Our results show that it is very difficult to calculate the excess pore pressure when we have a large number of layers because the system is fully coupled, i.e. each pressure depends on all the pressures in the other layers. 
In figures \ref{fig:multi_vel_low} and \ref{fig:multi_vel_high} we see a comparison of the velocity at short times with the multilayer model with $2,3,4$, and $10$ layers, and the velocity measured at the surface of the mixture in the laboratory experiments in \cite{pailha:2009}, for the low and high viscosity cases and the dense initial configuration.


We observe that the model with $2$ or $3$ layers better reproduces the experimental results than the model with 10 layers, in particular at the beginning, when the velocity is very small. The 10 layers model predicts that the velocity grows up too fast in both cases, with low and high viscosity. We see that the single-layer model B-2016 better predicts the time when the velocity starts to grow up, while the model with 2,3 layers captures the maximum (i. e. surface) velocity much better than the single layer model. 

We conclude that the lack of accuracy at short times when using a large number of layers is due to the approximation of the pressure detailed in subsection \ref{se:approx_pressure} and the fact that the non-linear system to solve is strongly coupled, i.e., the errors approximating the pressure in each layer accumulates within the whole domain. A more accurate method to approximate the pressure should be developed if more than 2-3 layers are used. As a result, we will mainly use the 2-layer model in the following.

Figures \ref{fig:multi_vel_low} and \ref{fig:multi_vel_high} show the influence of the dilatation constant $K$ (see \eqref{eq:phieq}), which has been calibrated for single-layer models. If we compare the results obtained with two values of $K$ by showing also the results obtained with $K' = 2K$ (dashed blue lines), we observe a strong sensitivity of the results to this parameter. This is the case in particular for the high viscosity system (see the case with 10 layers in Figure \ref{fig:multi_vel_high}) for the velocity behaviour in particular at the beginning when the velocity starts to grow up. These results suggest that this constant should be calibrated for 3D models (or at least for multi-layer cases).

Therefore, in the following we keep only the model with two layers (Preserving Granular Mass two-layers model, denoted PGM-2 model hereafter) because it is the simplest case giving an appropriate solution, although we could choose also the one with 3 layers. In appendix \ref{Apend_C}, we write this particular case for the reader that is interested on using this particular model, and not on the general multilayer case. In the following subsection we finally compare this two-layer model with previous models in the literature and experimental data.

\subsubsection{PGM-2 model vs other models}\label{se:numtest_PGM2}
We compare here the results of the proposed two-layer model (PGM-2 model) with the results of models B-2016 and PP-2009. These depth-averaged models only compute the averaged velocity and solid volume fraction, while the PGM-2 model allows us to compute two values of these quantities in the normal direction. In particular, in the configuration considered here, the velocity in the bottom layer is lower than the velocity in the top layer, further called the maximum velocity. As previously, the physical and rheological parameter are fixed in tables \ref{tabla_bifasico} and \ref{tabla_bifasico2}.

\begin{figure}[!b]
	\begin{center}
		\includegraphics[width=0.8\textwidth]{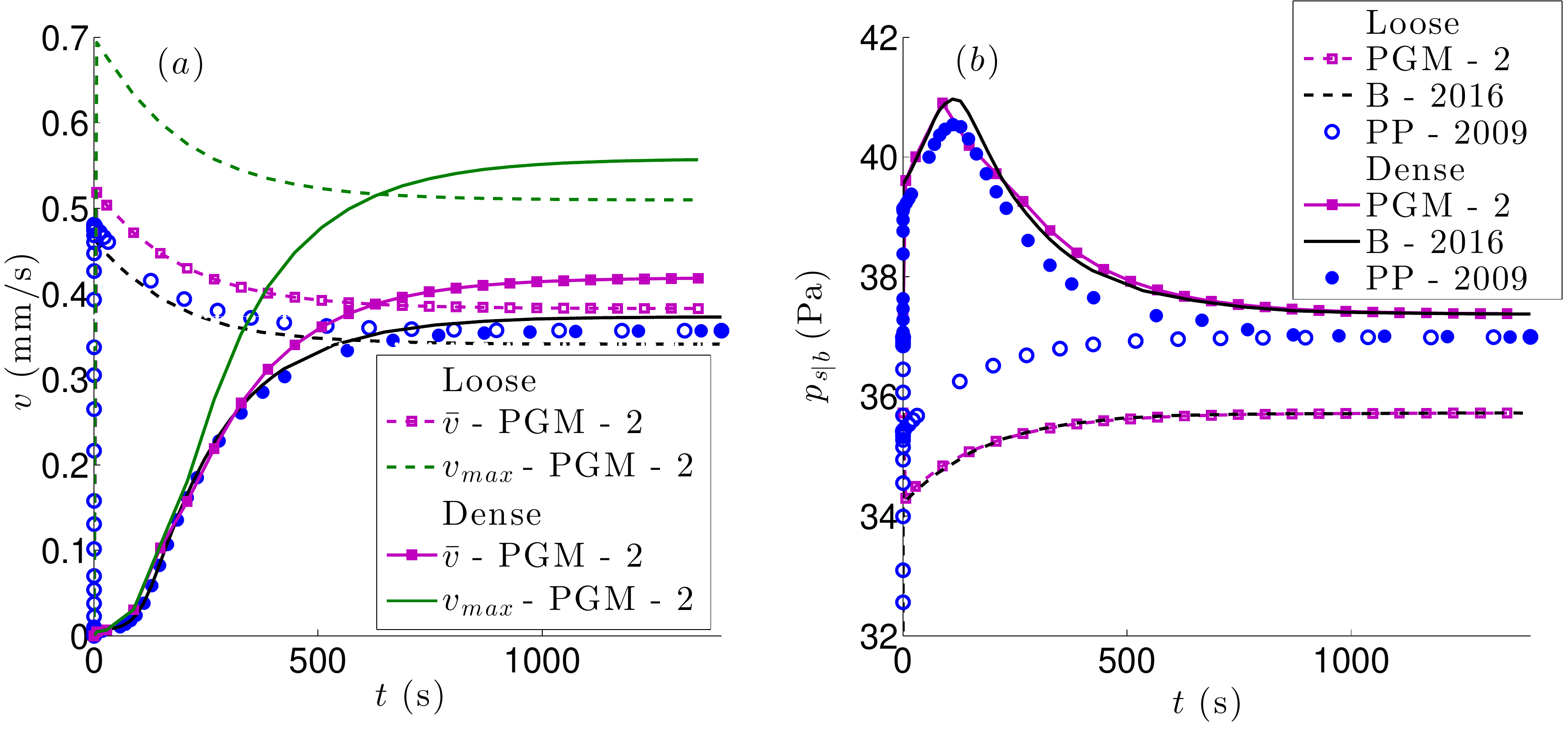}
		\includegraphics[width=0.8\textwidth]{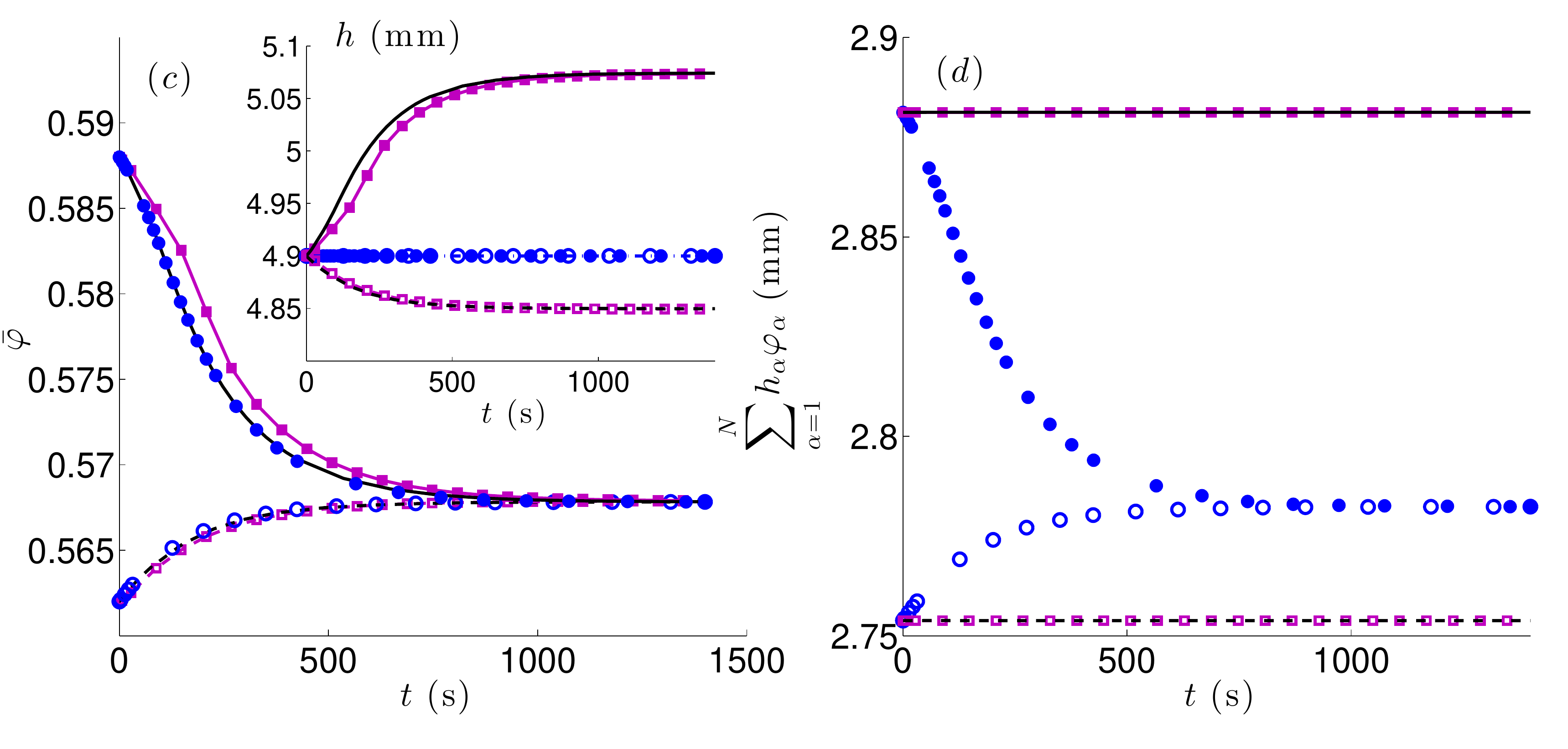}
		\caption{\label{fig:high_vel_pres_conc} \it{Time evolution of (a) velocity, (b) pressure at the bottom, (c) averaged solid volume fraction, (d) solid granular mass in the high viscosity case.}}
	\end{center}
\end{figure}

      \begin{figure}[!ht]
 	\begin{center}
 		\includegraphics[width=0.8\textwidth]{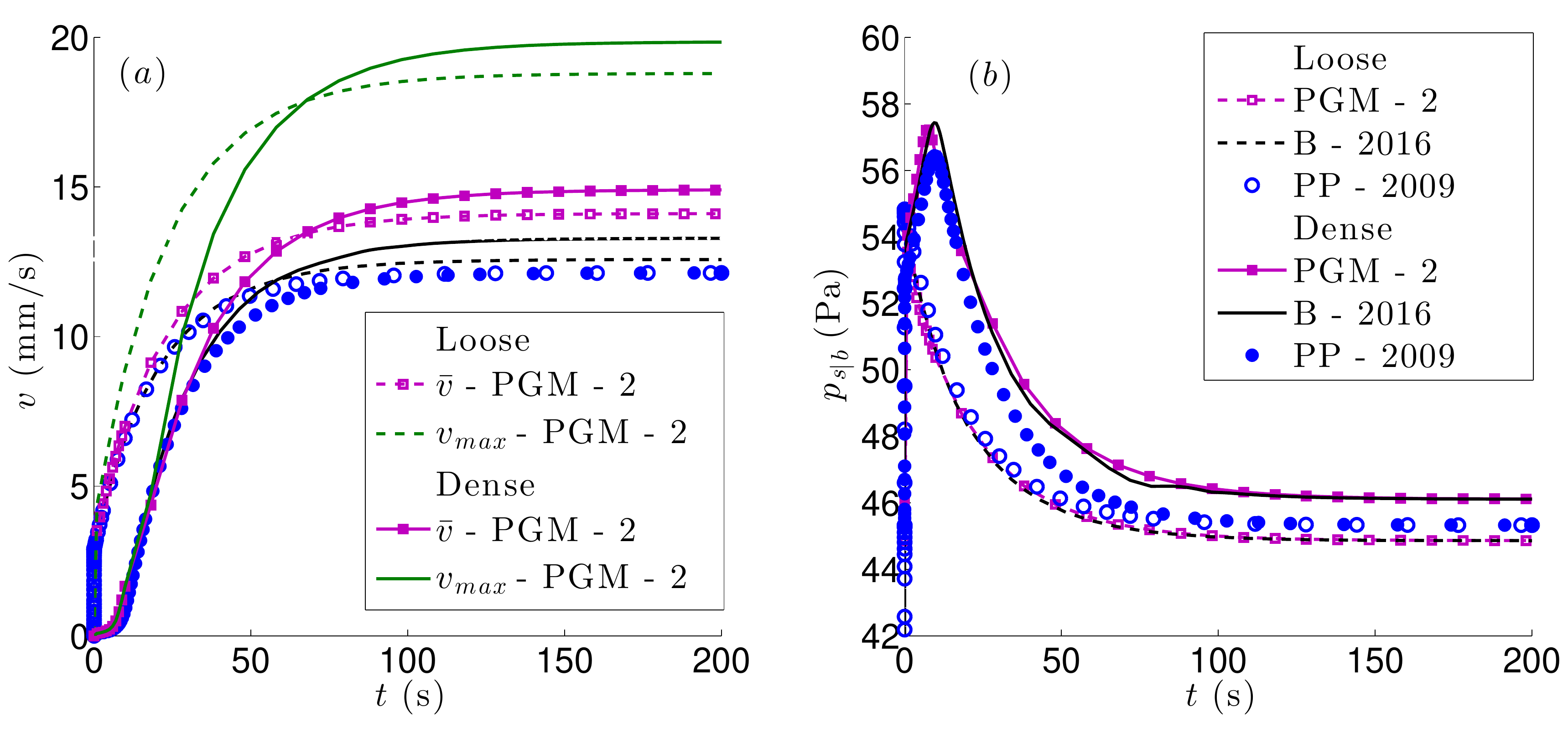}
 		\includegraphics[width=0.8\textwidth]{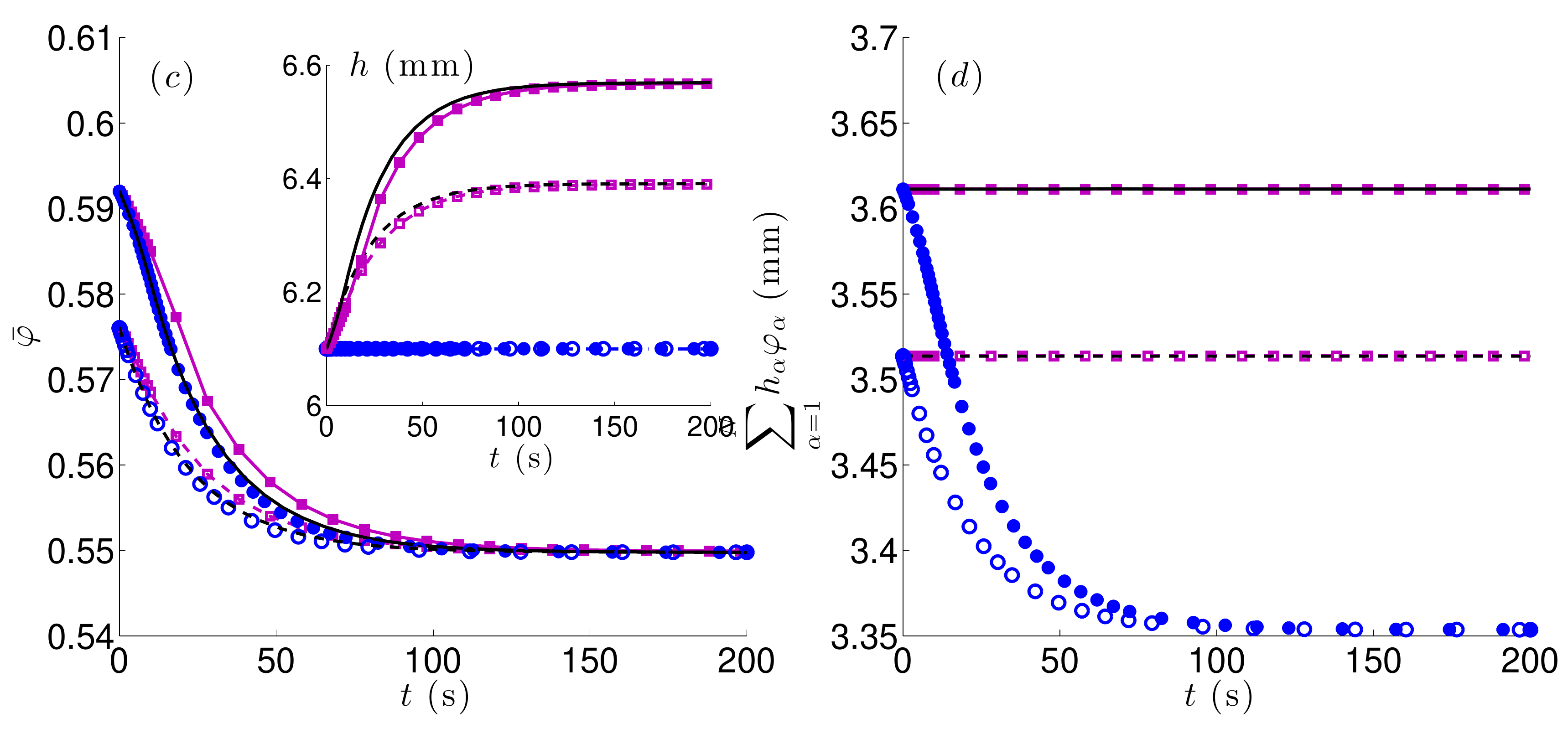}

 	\caption{\label{fig:low_vel_pres_conc} \it{Time evolution of (a) velocity, (b) pressure at the bottom, (c) averaged solid volume fraction, (d) solid granular mass in the low viscosity case.}}
 \end{center}
 \end{figure}
 \begin{figure}[!ht]
 	\begin{center}
 		\includegraphics[width=0.8\textwidth]{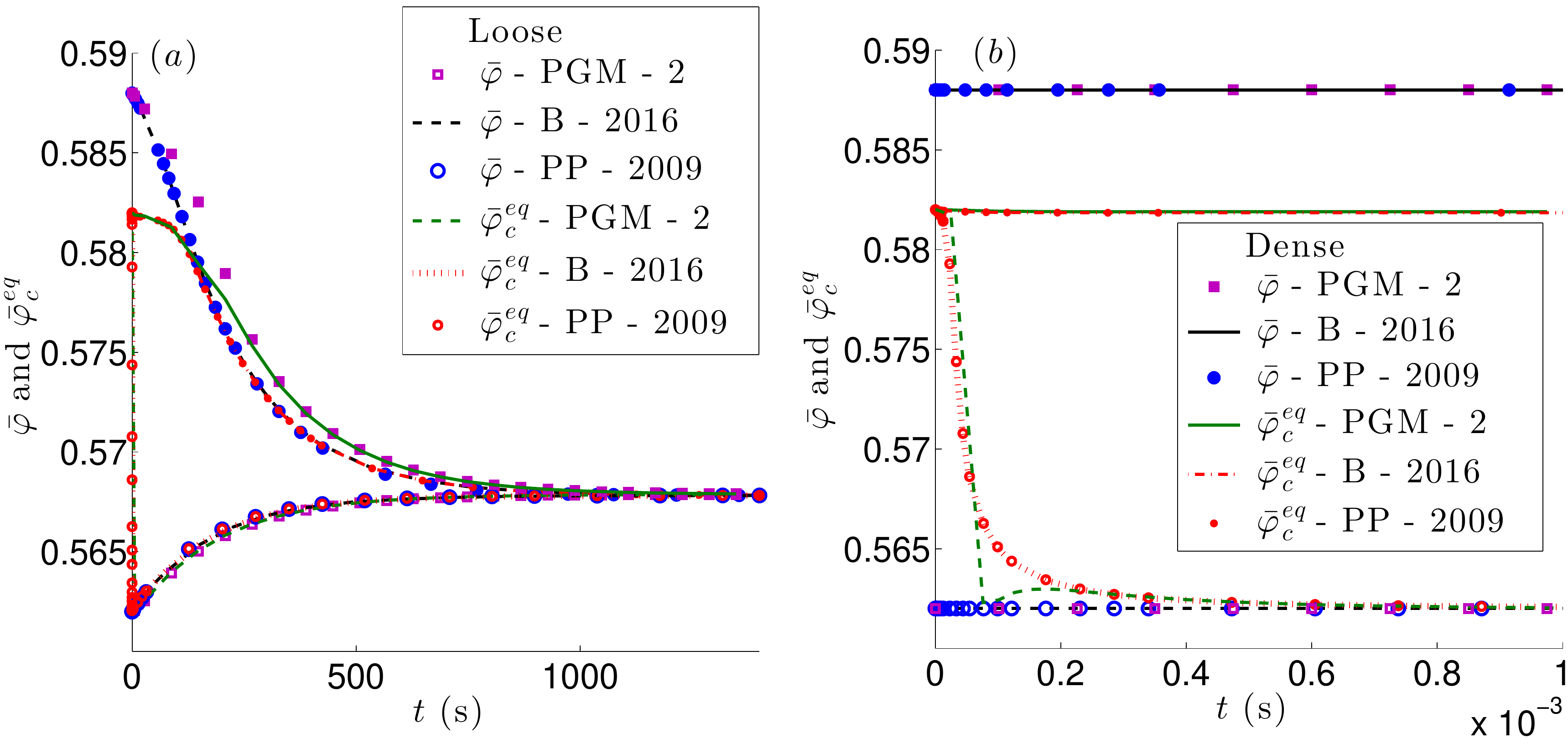}
 		\caption{\label{fig:high_conc_eq} \it{(a) Time evolution of the averaged volume solid fraction and equilibrium concentration for the high viscosity case and (b) time evolution at short times.}}
 	\end{center}
 \end{figure}
 \begin{figure}[!ht]
 	\begin{center}
 		\includegraphics[width=0.8\textwidth]{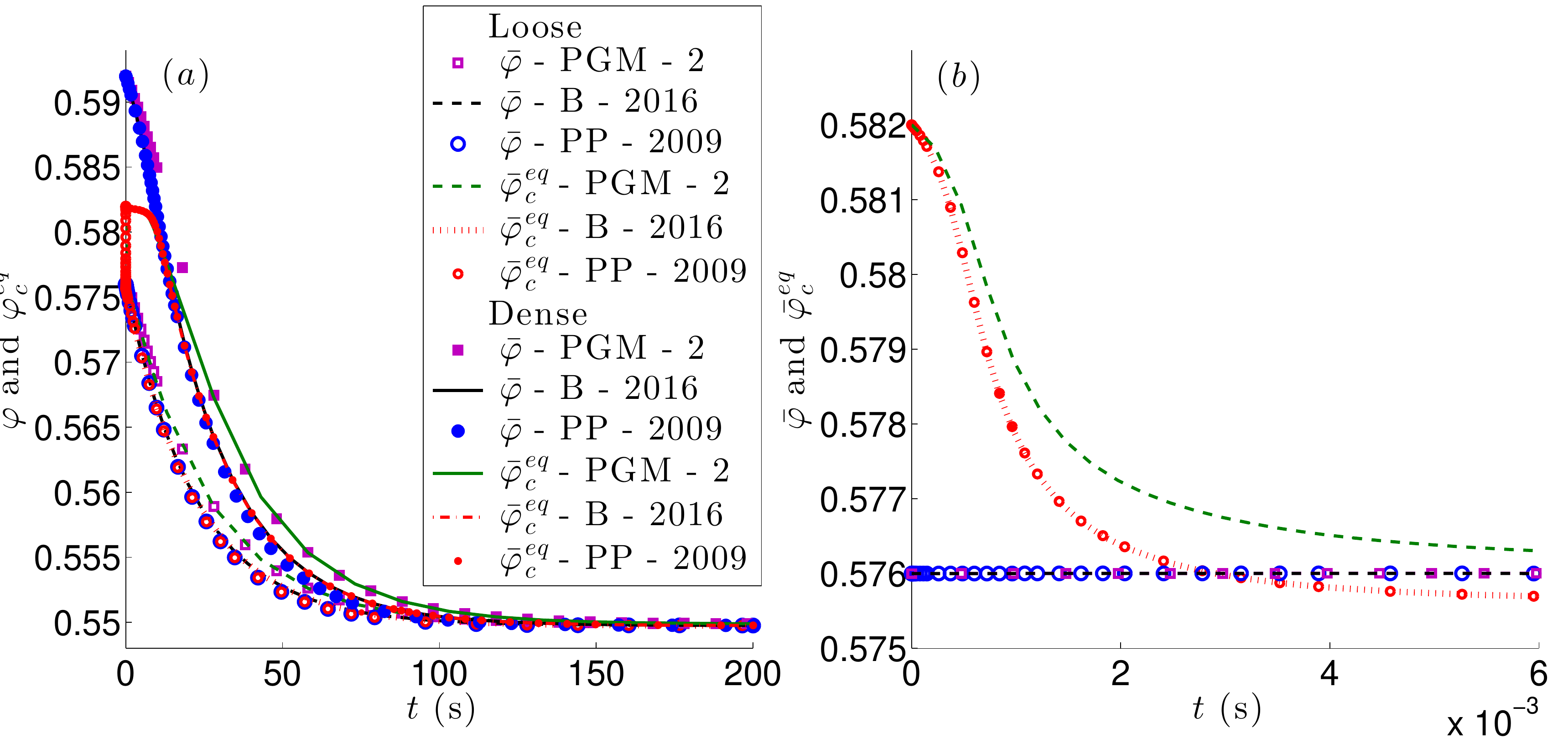}
 	\end{center}
 	\caption{\label{fig:low_conc_eq} \it{(a) Time evolution of the averaged volume solid fraction and equilibrium concentration for the low viscosity case and (b) time evolution at short times.}}
 \end{figure}

\begin{figure}[!ht]
	\begin{center}
		\includegraphics[width=0.8\textwidth]{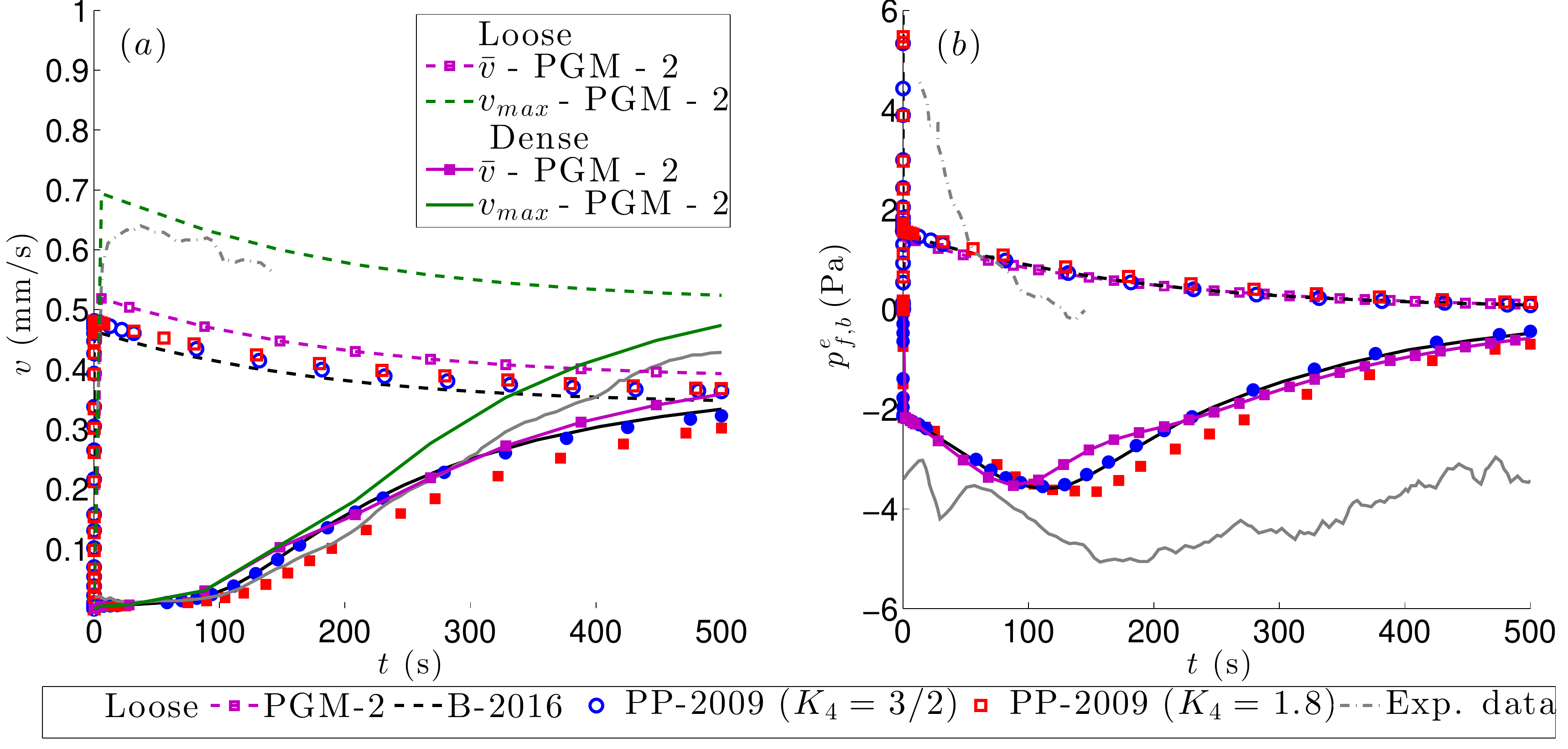}
		\includegraphics[width=0.8\textwidth]{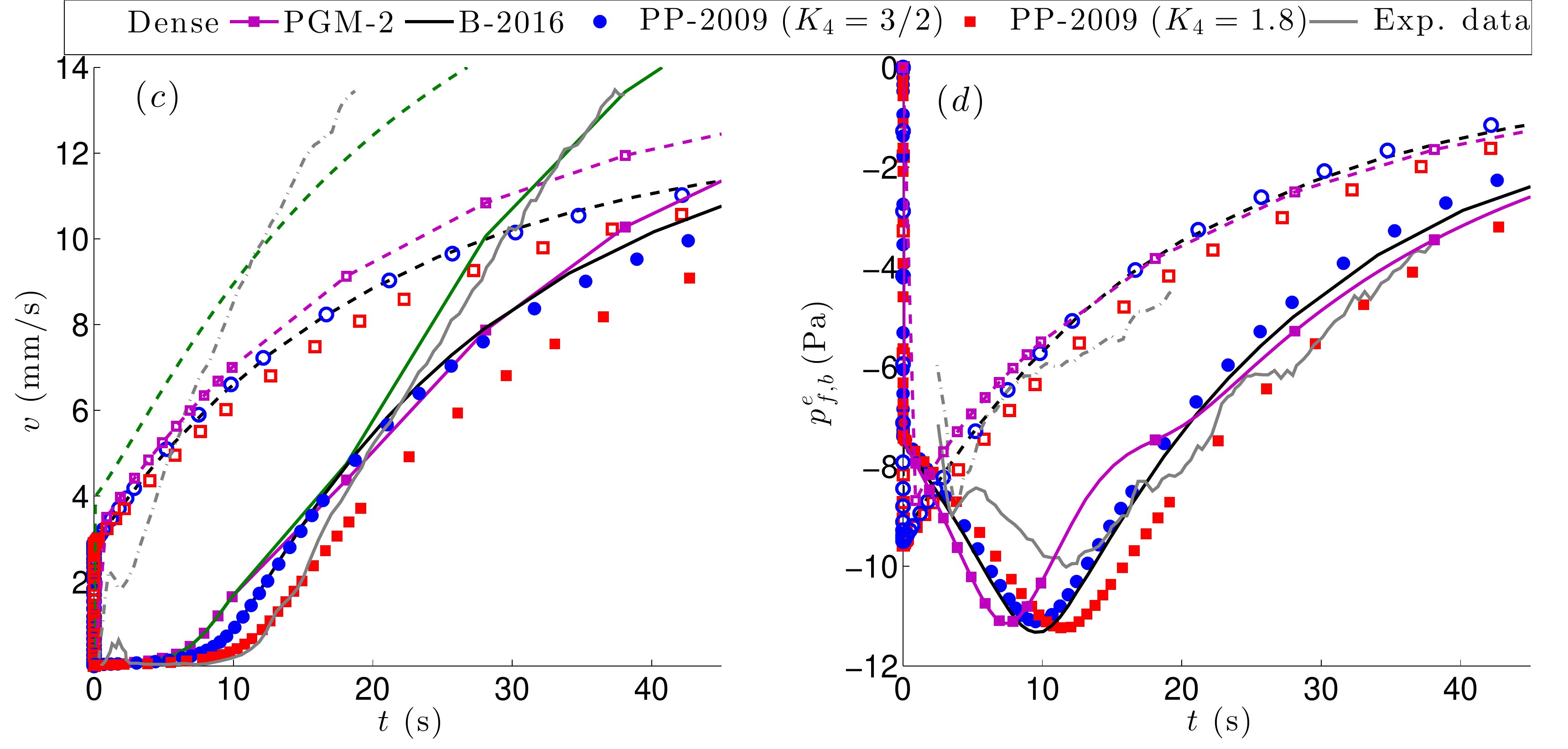}
	\end{center}
	\caption{\label{fig:datos_lab} \it{(a),(c) Time evolution of the averaged velocity; (b),(d) the excess pore pressure at the bottom in the high viscosity case ((a),(b)) and the low viscosity case ((c),(d)). Grey lines are the experiment data.}}
\end{figure}

\begin{figure}[!ht]
	\begin{center}
		\includegraphics[width=0.8\textwidth]{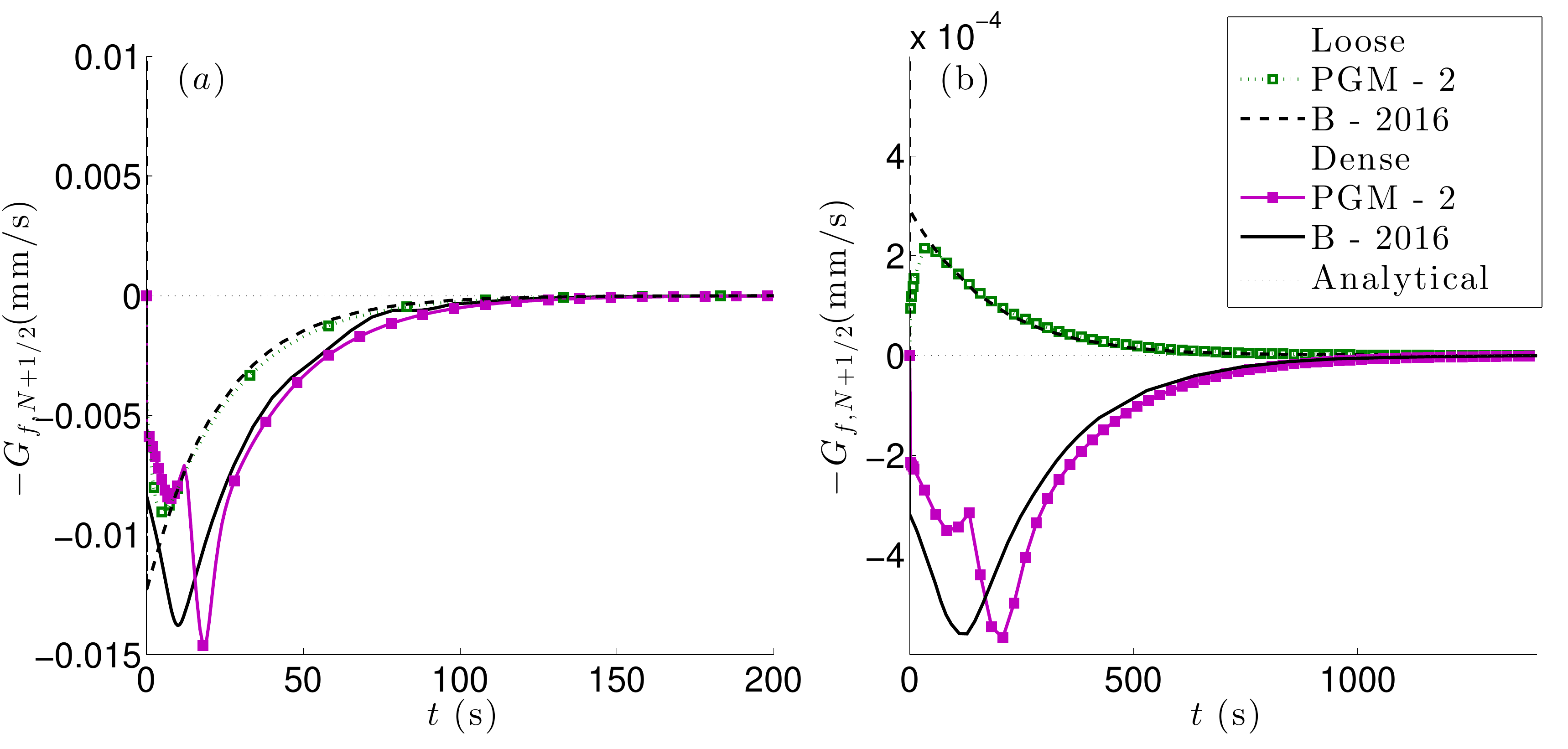}
	\end{center}
	\caption{\label{fig:gf12} \it{Time evolution of the fluid transference at the interface between the mixture and the upper fluid layer in the (a) low and (b) high viscosity cases, with dense/loose initial condition.}}
\end{figure}

\begin{figure}[!ht]
	\begin{center}
		\includegraphics[width=0.8\textwidth]{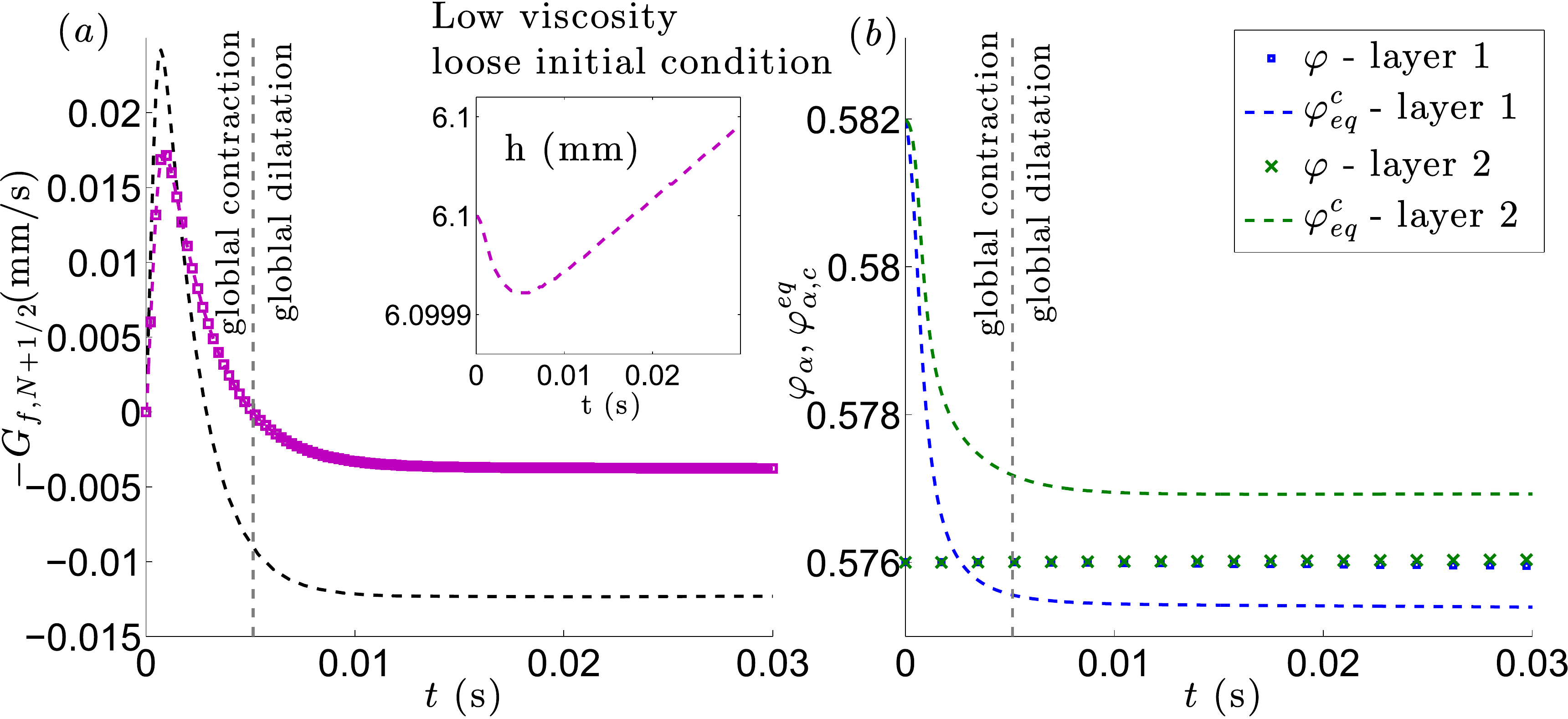}
	\end{center}
	\caption{\label{fig:gf12_low_loose} \it{Time evolution of the (a) fluid transference at the interface between the mixture and the upper fluid layer, and (b) the solid volume fraction at short times for the low viscosity case and loose initial condition.}}
\end{figure}


In figures \ref{fig:high_vel_pres_conc} and \ref{fig:low_vel_pres_conc} we see the time evolution of the maximum and the averaged velocities, the pressure at the bottom, the concentration, the solid mass and the height for all the models in both, the loose and dense initial configuration, for the high and low viscosity cases. All the models give similar results at the steady state although the time evolution can differ. We see that the results of the proposed PGM-2 model and the  B-2016 model are very close for the pressure, averaged solid volume fraction and height, whereas the velocities are slightly different. The behaviour is very similar, for example, the steady velocity for the dense initial configuration is greater that the one for the loose configuration. However, with the 2-layers model the obtained maximum velocity in the top layer is much closer to the velocity measured at the surface of the mixture in the experiments (see Figure \ref{fig:datos_lab}). Therefore, the main advantage of this model with respect to the previous depth-averaged models B-2016 and PP-2009 is the approximation of the velocity field.

Both PGM-2 and B-2016 models preserve the granular mass $\left(\sum_{\a=1}^{N}h_\a\vp_\a\right)$, contrary to the PP-2009 model for which the total height ($h$) is preserved (see figures \ref{fig:high_vel_pres_conc}(c)(d) and \ref{fig:low_vel_pres_conc}(c)(d)). Note that a multilayer model preserving the total height has been also obtained in subsection \ref{se:closure_GN12}.
Imposing mass conservation leads to different velocity, pressure, and solid mass in steady state for initially loose and dense cases (PGM-2 and B-2016 models), while the steady state reached with PP-2009 model does not depend on the initial dense/loose configuration. These results are the expected following the analysis of the  proposed model and the one presented in subsection \ref{se:closure_GN12}.

Figures \ref{fig:high_conc_eq} and \ref{fig:low_conc_eq} show the convergence of the solid volume fraction to the equilibrium concentration in the high and low viscosity cases. This convergence is much slower in the dense configuration than in the loose one, as commented before. In fact, we can see in Figure \ref{fig:high_conc_eq}b that for the loose configuration the solid volume fraction $\bar{\vp}$ goes to $\bar{\vp}_{c}^{eq}$ at very short times ($t\sim 10^{-3}$).

In Figure \ref{fig:datos_lab} we show the velocity and excess pore pressure with models PP-2009, B-2016 and PGM-2, and also the laboratory experiments in Pailha {\it et al.} 2009 \cite{pailha:2009}, in both the high and low viscosity case. Looking at the averaged velocities computed with the depth-averaged models, we see that they cannot predict the velocity observed at the mixture surface in the experiments. However, the maximum velocity computed by the proposed two-layers model has the order of magnitude of the surface velocity in the experiments. On the other hand, having two layers instead one layer does not significantly change the calculated excess pore pressure. As expected, this excess pore pressure goes to zero after some time in all models. 

Finally, in Figure \ref{fig:gf12} we see the evolution of the fluid transference, $G_{f,N+\frac12}$, at the interface between the mixture layer and the upper fluid layer. We recall that $G_{f,N+\frac12}$ have the opposite sign to $\mathcal{V}_f$ in previous work \cite{bouchut:2016} (see \eqref{eq:gn12nuf}). For the high viscosity case ( Figure 19(b)) we see very clearly the dynamics. When the mixtures starts to flow in the initially dense case, we have $\vp>\vp_c^{eq}$, leading to a positive dilatation angle $\tan\psi = K\left(\vp-\vp_{c}^{eq}\right) > 0$ and therefore to dilatation. As a consequence the fluid is sucked into the mixture ($\mathcal{V}_f=-G_{f,N+\frac12}<0$) and the height of the mixture increases (see inset in Figure \ref{fig:high_vel_pres_conc}c). On the contrary, in the initially loose case, $\vp<\vp_c^{eq}$, leading to $\tan\psi = K\left(\vp-\vp_{c}^{eq}\right) < 0$ and therefore to contraction. The fluid is thus expelled from the mixture ($\mathcal{V}_f=-G_{f,N+\frac12}>0$) and the mixture height decreases.

Focusing now on the low viscosity case, we observe the same behaviour for the initially dense case, but not for the loose case. Figure \ref{fig:gf12_low_loose} shows the loose case at very short times, where the material contracts at the beginning as expected, i.e. the fluid is expelled ($-G_{f,N+\frac12}>0$) and the height decreases. However we see that after some time $\vp_2 < \vp_{c,2}^{eq}$ (contraction) in the top layer (layer 2) whereas $\vp_1 > \vp_{c,1}^{eq}$ (dilatation) in the bottom layer (layer 1).
 Moreover, it makes that the fluid transference at the top changes its sign and becomes negative at time $t\approx 5.1\times 10^{-3}$, and therefore the fluid is sucked and the height increases. That is, a different contraction/dilatation situation could occur in each layer in the multilayer configuration.

Physically, it could mean that when the fluid is expelled at the beginning, the lower part is the first one that remain with less fluid (because each layer is expelling fluid toward the upper layer, but receiving it from the layer below), and therefore it becomes to a dense situation. After that, the lower part dilates and the upper part contracts, going to a globally dilatation situation, as we can see in Figure \ref{fig:gf12_low_loose}.

\section{Conclusions}\label{se:conclusions}
In this work we deal with dilatancy effects in granular models of fluidized flows. A multilayer extension of the two-phase model proposed by Bouchut {\it et al.} (2016) \cite{bouchut:2016} has been presented. It is derived from a dimensional analysis and the multilayer approach following an analogous procedure to \cite{fernandezNieto:2016}. This model is in principle able to recover the structure of the fluid in the direction normal to the topography. This is essential to improve the description of friction and viscous terms when considering a complex rheology compared to depth-averaged single-layer models, as showed in \cite{fernandezNieto:2016,fernandezNieto:2018}. In particular, it allows to recover the normal velocity profiles, but also the normal profiles of the solid volume fraction and pore fluid pressure.

The proposed two-phase model behaves as the B-2016 \cite{bouchut:2016} model, preserving the total solid mass. In addition, a model that behaves as the PP-2009 model \cite{pailha:2009} $-$preserving the total mixture height$-$ has been also obtained for a particular closure relation for the solid mass transference at the mixture upper interface. That 
closure relation allows us to obtain either a model (PH) preserving the total height (and maybe loosing solid mass) or a model (PGM) preserving the total solid mass (the height vary depending on the contraction/dilatation of the solid). This last model is the physically meaningful model that we further compare with laboratory experiments.

The main difficulty of the proposed two-phase model is the excess pore fluid pressure that appears in particular in the solid pressure, which is no more hydrostatic even with the thin layer approximation. This pressure varies strongly in time when starting from non-equilibrium conditions due to dilatation or contraction of the granular medium. It further goes to zero when reaching steady states. Dealing with this strongly varying, non-linear term is shown to be a real numerical challenge to approximate the model when variations in the normal direction are considered. For this reason, we only deal with uniform flows in the numerical tests in this work, although the model is deduced for general flows, and simulation of $x$-dependent flows could be part of a forthcoming work.

Even in the case of uniform flows, where the model is reduced to a system of ODEs, the numerical approximation is not easy and a specific numerical scheme is necessary. As consequence of this pore excess pressure term, the pressure cannot be easily computed, even in the single-layer case where the pressure at the bottom is one of the two roots of a quadratic equation. A numerical scheme based on the first-order Taylor polynomial has been proposed to approximate the pressure. 

We have validated the model by comparing with the analytical solution for two-phase uniform flows. In this case we have chosen the PH model because computing the analytical solution in that case is simpler than for the PGM model, since the height is directly obtained. We have seen that the steady solution does not depend on the dense/loose initial value of the solid volume fraction for the PH model (also PP-2009 model), where the solid mass varies in time, in contrast with the results obtained with the PGM model (also B-2016 model), where the solid mass is preserved and the steady states for the velocity and pressure depend on the initial solid volume fraction.

For these tests, a constant profile of solid volume fraction has been obtained. Furthermore, an analytical solution for confined flows, where side wall friction notably changes the dynamics of the flow, has been deduced from the resolution of a integro-differential equation. In that case, the velocity profile goes from a Bagnold profile to a S-shaped profile, very similar to the case of dry flows (see \cite{fernandezNieto:2018}). It is interesting that the profile of solid volume fraction is not constant in this case, but linear, and when a flow/no-flow transition is observed in the velocity profile, then the solid volume fraction becomes constant in the no-flow (i. e. rigid) zone. In addition, we see that including side wall friction as in Jop {\it et al.} \cite{jop:2005} is a good approximation for shallow flows. An interesting feature of this approach is that is allows to give an approximation of flow/no-flow transition which is obviously not the case for depth-averaged single-layer models. Furthermore, for the analytical solutions (with and without side wall friction) we have also shown convergence tests depending on the number of vertical layers. In particular, for solutions with a strong vertical structure (the case of a narrow channel width) the analytical solution is perfectly recovered when increasing the number of layers.

We have shown that the approximation of the pressure with the proposed method is acceptable, in the sense that it approximates the good solution from the $2^{N}$ possible solution vectors for the pressure. However, this approximation is not accurate enough at short times and the error notably increases with the number of layers in the normal direction. We have seen that the model with 2 or 3 layers gives reasonable results when compared with experimental data, allowing to make the quantities (velocity, volume fraction, pore pressure, etc.) vary in the normal direction. The pore pressure estimates however is getting less and less accurate when increasing the number of layers due to the accumulation of the numerical error for this full coupled models. This problem will be encountered also if solving the complete Jackson's equation for any 3D method. Another conclusion of this work is that the dilatation constant should be fitted for 3D models, since it has a strong influence on the results, in particular when using a large number of layers.

The model with two layers (PGM-2 model) has been chosen for further comparison with experimental results. We conclude that a more accurate method to compute the pressure is necessary in order to approximate the solution at short times. An interesting alternative could be to use an adaptive multilayer method, which uses few layers at short times giving a good approximation of the solid pressure, and increases the number of layers when the excess pore pressure becomes small. That would allow us to get a better approximation of the velocity profile at latter times and will be investigated in the future.

We compare the results of the PGM-2 model with the model of Bouchut {\it et al.} (2016) \cite{bouchut:2016} (B-2016) and Pailha and Pouliquen (2009) \cite{pailha:2009} (PP-2009). We see that our results are similar to the B-2016 model, except for the velocity. Actually, these previous depth-averaged models were not able to predict the velocity measured at the mixture surface, while the maximum velocity calculated at the top layer with the PGM-2 model is closer of this observed velocity. More differences are found when comparing the proposed model with the PP-2009 model as expected, since this model preserves the total height instead of the mass and therefore the steady states of the loose/dense initial condition are the same, in contrast with models B-2016 and PGM-2 models.

\section*{Acknowledgements}
This research has been partially supported by the Spanish Government under grants MTM2015-70490-C2-2-R and RTI2018-096064-B-C22 with the participation of FEDER, by the ERC contract ERC-CG-2013-PE10-617472 SLIDEQUAKES.

\appendix

\section{Explicit expression of the excess pore pressure}\label{Apend_B}
In this appendix we detail the computation of the excess pore pressure in equation \eqref{eq:pres_fluid}
\begin{equation}
\label{eq:ap_pres_fluid_e}
p_{f,\a}^e = p_{f,\a+\frac12}^e  \,+\, \varepsilon^{k+2} \dfrac{\b_\a}{\v1pa}\dint_{z}^{z_{\a+\frac12}} \left(u^z-v^z\right)\,dz',
\end{equation}
with
\begin{equation}
\label{eq:ap_pres_fluid_e_a12}
p_{f,\a+\frac12}^e = p_{f,\a+1}^{e}\left(z=z_{\a+\frac12}\right) = \varepsilon^{k+2}\dsum_{\gamma = \a+1}^N \dfrac{\b_{\gamma}}{\left(1-\vp_\gamma\right)}\dint_{z_{\gamma-\frac12}}^{z_{\gamma+\frac12}}  \left(u^z-v^z\right)\,dz'.
\end{equation}
Firstly, we integrate the dilatancy equation \eqref{eq:nondim_dil_3D} between $z_{\a-\frac12}$ and $z\in\left(z_{\a-\frac12},z_{\a+\frac12}\right)$ as it is made for the calculation of the vertical velocity, resulting
\begin{equation}
\label{eq:ap_vert_vel_sol}
 v^z_{\a}(z) \;=\; (v^z_{\a-\frac{1}{2}})^+ \;-\; (z-z_{\a-\frac{1}{2}})\left(\p\!_{x}v_{\a} - \dfrac{1}{\varepsilon}\Phi_\a\right).
\end{equation}
We can also integrate equation \eqref{eq:incomp_mixt} regarding the incompressibility of the mixture, obtaining
\begin{equation}
\label{eq:ap_vert_vel_fl}
\begin{array}{ll}
 \vpa u^z_{\a}(z) + \v1pa u^z_{\a}(z) &=\; \v1pa (u^z_{\a-\frac{1}{2}})^+  \,+\, \vpa (v^z_{\a-\frac{1}{2}})^+ \\[4mm]
 & -\, \left(z-z_{\a-\frac12}\right) \p\!_{x} \left(\vpa v_{\a}+\v1pa u_{\a}\right).
\end{array}
\end{equation}
Now, by considering equation \eqref{eq:ap_vert_vel_fl} minus equation \eqref{eq:ap_vert_vel_sol} and dividing the result by $\v1pa$, we get

\begin{equation*}
u^z_{\a}(z)-v^z_{\a}(z) \;=\; (u^z_{\a-\frac{1}{2}})^+ - (v^z_{\a-\frac{1}{2}})^+ \,-\, \dfrac{\left(z-z_{\a-\frac12}\right)}{\v1pa} \bigg(  \p\!_{x} \Big(\v1pa \left(u_{\a}-v_\a\right)\Big) + \dfrac{1}{\varepsilon}\Phi_\a\bigg).
\end{equation*}
\noindent Once we have the explicit expression for the difference of the vertical velocities, we have to compute the integrals in \eqref{eq:ap_pres_fluid_e}-\eqref{eq:ap_pres_fluid_e_a12}. Tanking into account that
$$
\dint_{z}^{z_{\a+\frac12}}\left(z'-z_{\a-\frac12}\right) dz' = \dfrac{h_\a^2 - \left(z-z_{\a-\frac12}\right)^2}{2},
$$
we obtain that
\begin{equation}
\label{eq:ap_pe_alfa}
\begin{array}{ll}
p_{f,\a}^e &=\, p_{f,\a+\frac12}^e  \,+\, \varepsilon^{k+2} \dfrac{\b_\a}{\v1pa} \left(z_{\a+\frac12}-z\right)\left((u^z_{\a-\frac{1}{2}})^+ - (v^z_{\a-\frac{1}{2}})^+\right)\\[4mm]
&-\, \varepsilon^{k+2} \dfrac{\b_\a}{\v1pa}\dfrac{h_\a^2 - \left(z-z_{\a-\frac12}\right)^2}{2\v1pa}\bigg(  \p\!_{x} \Big(\v1pa \left(u_{\a}-v_\a\right)\Big) + \dfrac{1}{\varepsilon}\Phi_\a\bigg).
\end{array}
\end{equation}
In this work we are interested on the case of a strong friction between the fluid and granular phases. This case is more suitable for natural context applications as discussed in \cite{bouchut:2016}, and moreover it is simpler than the moderate friction choice, involving only first-order derivatives in the momentum equations. Therefore we take $k=-1$ in previous equation, leading to
$$
\begin{array}{l}
p_{f,\a}^e \,=\, p_{f,\a+\frac12}^e  \,+\, \varepsilon \dfrac{\b_\a}{\v1pa} \left(z_{\a+\frac12}-z\right)\left((u^z_{\a-\frac{1}{2}})^+ - (v^z_{\a-\frac{1}{2}})^+\right)
\,-\, \b_\a\dfrac{h_\a^2 - \left(z-z_{\a-\frac12}\right)^2}{2\v1pa^2}\Phi_\a + \mathcal{O}(\varepsilon).
\end{array}
$$
Using the mass transference terms \eqref{eq:G} we obtain
$$
(u^z_{\a-\frac{1}{2}})^+ - (v^z_{\a-\frac{1}{2}})^+ = \dfrac{G_{s,\a-\frac12}}{\vpa} - \dfrac{G_{f,\a-\frac12}}{\v1pa} + \left(u_\a - v_\a\right)\p_x z_{\a-\frac12},
$$
and using the expression for the solid and fluid mass transference terms \eqref{eq:GsGf}, we get
$$
\varepsilon \dfrac{\b_\a}{\v1pa} \left(z_{\a+\frac12}-z\right)\left((u^z_{\a-\frac{1}{2}})^+ - (v^z_{s,\a-\frac{1}{2}})^+\right) = \dfrac{-\b_\a}{\vpa\v1pa^2}\left(z_{\a+\frac12}-z\right)\mathlarger{\dsum}_{\gamma=1}^{\a-1} \vp_{\gamma}h_{\gamma}\Phi_{\gamma} \,+\, \mathcal{O}(\varepsilon).
$$
Note that here we have used the nondimensional expression of the mass transference terms, whose leading order is $1/\varepsilon$, corresponding to the term involving the dilatancy function. As a result, we get the following explicit expression for the excess pore  pressure

\begin{equation*}
\label{eq:ap_press_fluid_e_exp}
p_{f,\a}^e = p_{f,\a+\frac12}^e - \b_\a\dfrac{h_\a^2 - \left(z-z_{\a-\frac12}\right)^2}{2\v1pa^2}\Phi_\a \,-\, \dfrac{\b_\a}{\vpa\v1pa^2}\left(z_{\a+\frac12}-z\right)\mathlarger{\dsum}_{\gamma=1}^{\a-1} \vp_{\gamma}h_{\gamma}\Phi_{\gamma} \,+\, \mathcal{O}(\varepsilon),
\end{equation*}
with
$$p_{f,\a+\frac12}^e = p_{f,\a+1}^{e}\left(z=z_{\a+\frac12}\right) = \mathlarger{\mathlarger{\dsum}}_{\xi=\a+1}^N\dfrac{-\b_\xi\,h_{\xi}}{\vp_\xi\left(1-\vp_\xi\right)^2}\,\left(\mathlarger{\dsum}_{\gamma = 1}^{\xi-1}\vp_{\gamma}h_\gamma\Phi_\gamma \,+\, \dfrac{\vp_{\xi}h_\xi\Phi_\xi}{2}\right) \,+\, \mathcal{O}(\varepsilon).
$$

\bigskip

In this case, we consider the moderate friction regime by fixing $k=0$ in \eqref{eq:ap_pe_alfa}, the excess pore  pressure results
\begin{equation*}
\label{eq:ap_pe_alfa_debil}
\begin{array}{ll}
p_{f,\a}^e &=\, p_{f,\a+\frac12}^e  \,+\, \varepsilon^{2} \dfrac{\b_\a}{\v1pa} \left(z_{\a+\frac12}-z\right)\left(\dfrac{G_{s,\a-\frac{1}{2}}}{\vpa}-\dfrac{G_{f,\a-\frac{1}{2}}}{\v1pa} + \left(u_\a-v_\a\right)\p_x z_{\a-\frac{1}{2}}\right)\\[4mm]
&-\, \varepsilon^{2} \dfrac{\b_\a}{\v1pa}\dfrac{h_\a^2 - \left(z-z_{\a-\frac12}\right)^2}{2\v1pa}\bigg(  \p\!_{x} \Big(\v1pa \left(u_{\a}-v_\a\right)\Big) + \dfrac{1}{\varepsilon}\Phi_\a\bigg),
\end{array}
\end{equation*}
where $p_{f,\a+\frac12}^e = p_{f,\a+1}^e\left(z=z_{\a+\frac12}\right)$, and $G_s$, $G_f$ are given by \eqref{eq:GsGf}.
Note that this expression is more complicated because it involves second order derivatives in the momentum equations.

\section{Preserving Granular Mass two-layers model (PGM-2 model)}\label{Apend_C}
This appendix is devoted to presenting the two-layer model, which we have used in section \ref{se:numtest_PGM2} to compare our results with previous models in the literature: Bouchut {\it et al.} (2016) \cite{bouchut:2016} and Pailha and Pouliquen (2009) \cite{pailha:2009}, for readers that are interested on using this model without look at the detail of the multilayer case.

Thus, the PGM-2 model is written as ($G_{f,\frac12}=G_{s,\frac12}=0$, $k_b = 0$ and $k_i=0$)
\begin{equation*}
	\label{eq:PGM2_uniform}
	\left\{
	\begin{array}{l}
	\p_t h = G_{f,\frac52}, \\[4mm]
	\p_{t}\vp_1 = -\,\vp_1\Phi_1, \\[4mm]
	l_{1}\,\r_s\,\p_{t}\left(h\vp_1 v_{1}\right) = l_1\,h\,\overline{\b_1}\left(u_1- v_1\right)\,-\,\left(\r_s-\r_f\right)g\sin\theta l_1 h \vp_1  \nonumber\\[2mm]
	\quad+ K_{s,\frac{1}{2}} - K_{s,\frac{3}{2}}\; +\;\r_s G_{s,\frac{3}{2}}\dfrac{v_{2} + v_{1}}{2}, \\[4mm]
	l_{1}\,\r_f\,\p_{t}\left(h\left(1-\vp_1\right) u_{1}\right) = \,-\,l_1\,h\,\overline{\b_1}\left(u_1 - v_1\right)\nonumber\\[2mm]
	\quad - K_{f,\frac{3}{2}}\; +\;\r_f G_{f,\frac{3}{2}}\dfrac{u_{2} + u_{1}}{2},\\[4mm]
	l_{2}\,\r_s\,\p_{t}\left(h\vpa v_{2}\right) = l_2\,h\,\overline{\b_2}\left(u_2 - v_2\right)\,-\,\left(\r_s-\r_f\right)g\sin\theta l_2 h \vp_2  \nonumber\\[2mm]
	\quad+ K_{s,\frac{3}{2}} \;-\; \r_s G_{s,\frac{3}{2}}\dfrac{v_{2} + v_{1}}{2}, \\[4mm]
	l_{2}\,\r_f\,\p_{t}\left(h\left(1-\vp_2\right) u_{2}\right) = \,-\,l_2\,h\,\overline{\b_2}\left(u_2 - v_{2}\right)\nonumber\\[2mm]
	\quad+ K_{f,\frac{3}{2}}\; +\;\r_f G_{f,\frac{5}{2}}u_{2} \;-\; \dfrac12\r_f G_{f,\frac{3}{2}}\dfrac{u_{2} + u_{1}}{2},
	\end{array}\right.
\end{equation*}
where
$$
\overline{\b_\a} = \dfrac{150\vpa^2}{d_s^2\left(1-\vpa\right)}\eta_f, \quad \mbox{for } \a=1,2,
$$
the mass transference terms are
$$
G_{f,\frac52} = \dfrac{l_1\vp_1\Phi_1+l_2\vp_2\Phi_2}{l_1\vp_1+l_2\vp_2}h, \quad G_{s,\frac32} = l_1\left(\vp_1G_{f,\frac52}-\vp_1h\Phi_1\right),\quad G_{f,\frac32} = l_1\left(\left(1-\vp_1\right)G_{f,\frac52} + \vp_1 h \Phi_1\right),
$$
and the dilatancy function is
\begin{equation*}
\Phi_\a = \dot{\gamma}_\a K\left(\vp_\a-\vp_{c,\a}^{eq}\right),\quad\text{with}\quad \vp^{eq}_{c,\a} = \vp_c^{stat} - K_2\,\dfrac{\eta_f\dot{\gamma}_\a}{p_{s,\a-\frac12}}, \quad\mbox{for } \a=1,2
\end{equation*}
with
$$
\dot{\gamma}_1 = \dfrac{\lambda\abs{v_{1}}}{l_1 h},\quad \dot{\gamma}_2 = \dfrac{\abs{v_2-v_{1}}}{l_2 h},
$$
with $\lambda=1$ (friction) or $\lambda=2$ (no slip).
The viscous terms are
$$
\begin{array}{lll}
K_{s,\frac{3}{2}} = - \dfrac12\eta_{s,\frac{3}{2}}\, \dfrac{v_2 - v_1}{l_1 h},  & &
K_{s,\frac{1}{2}} = -\r_s g\,\cos\theta\,h\,\left(\mu\big(I_{\frac12}\big)+\tan\psi_{1/2}\right)\dfrac{v_{1}}{\abs{v_{1}}},\\
K_{f,\frac{3}{2}} = - \eta_{f}\,\dfrac{u_2 - u_1}{l_1 h}, &  & K_{f,\frac{1}{2}} = 0,
\end{array}
$$
with $\eta_{s,\frac32}$ and $I_{\frac12}$ defined by \eqref{eq:approx_eta_interfaz} and \eqref{eq:def_I_interfaz}. Finally, the solid pressure at the continuous level are written
$$
\begin{array}{l}
p_{s,\frac12} = p_{s,\frac32} + \left(\r_s-\r_f\right)g\cos\theta l_1 h\vp_1 \,+\, \dfrac{\overline{\b}_1\,l_1 h}{\vp_1\left(1-\vp_1\right)^2}\, \dfrac{\vp_{1}l_1 h\Phi_1}{2},\\[4mm]
p_{s,\frac32} = \left(\r_s-\r_f\right)g\cos\theta l_2 h\vp_2 \,+\, \dfrac{\overline{\b}_2\,l_2 h}{\vp_2\left(1-\vp_2\right)^2}\, \left(\vp_{1}l_1 h\Phi_1 + \dfrac{\vp_{2}l_2 h\Phi_2}{2}\right),
\end{array}$$
where $\Phi_1$ and $\Phi_2$ depend on $p_{s,\frac12}$ and $p_{s,\frac32}$ and cannot be explicitly computed. They are computed as in section \ref{se:approx_pressure}.
\bibliographystyle{plain}
\bibliography{Biblio}

\end{document}